\documentclass[aps,12pt,a4paper]{revtex4-2}
\usepackage[warn]{mathtext}
\usepackage[utf8]{inputenc}
\usepackage{booktabs}
\usepackage{graphicx}
\usepackage[english]{babel}
\usepackage{wrapfig}
\usepackage{float}
\usepackage{hyperref}
\usepackage{physics}
\usepackage{amsmath,amssymb,amsthm,mathtools}
\usepackage{verbatim}
\usepackage{fixmath}
\usepackage{braket}
\usepackage{xcolor}
\theoremstyle{definition}

\theoremstyle{plain}

\theoremstyle{plain}

\newcommand{\eq}[1]{(\ref{#1})}
\newcommand{\fig}[1]{Fig.~\ref{#1}}

\newcommand{\be}{\begin{equation}}
\newcommand{\ee}{\end{equation}}

\newcommand\disp{\displaystyle}

\newcommand{\la}{\langle}
\newcommand{\ra}{\rangle}

\parskip=\medskipamount

\begin{document}

\title{DSSYK, open ASEP/TASEP and 2d dilaton gravity at strong coupling}

\author{Alexander Gorsky$^{1,2,4}$, Elizaveta Kovalenko$^{3}$, and Sergei Nechaev$^{5,2}$}

\affiliation{
$^1$Institute for Information Transmission Problems RAS, 127051 Moscow, Russia \\
$^2$Laboratory of Complex Networks, Center for Neurophysics and Neuromorphic Technologies, Moscow, Russia \\
$^3$Moscow Institute of Physics and Technology, Institutsky lane 9, Dolgoprudny 141700, Russia \\
$^4$Phystech School of Applied Mathematics and Computer Science, Moscow Institute for Physics and Technology, Dolgoprudnyi, Russia \\
$^5$LPTMS, CNRS -- Universit\'e Paris Saclay, 91405 Orsay Cedex, France}

\date{\today}

\begin{abstract}

We discuss the triality between stochastic growth models in the KPZ universality class, the DSSYK model, and low dimensional dilaton gravity. The  TASEP model, which is our primary focus, is identified with the strong-coupling limit of two-dimensional quantum gravity. The Hamiltonian of DSSYK and the transfer matrix of ASEP coincide up to an additive constant, and we derive the exact expression for the resolvent of the ASEP transfer matrix on a finite interval. The connection of corresponding cumulants with the Tutte polynomials $T(1,q)$ of crossing graphs dual to chord diagrams implies the relationship with the $Q=0$ Potts model coupled to gravity. We further exploit the realization of all these models in terms of different versions of discrete and continuous random walks on two-dimensional manifolds, which parameterize the Hilbert spaces of respective theories. The representation of TASEP  as the growth model of heaps provides a new insight into the strong-coupling gravity, suggesting its interpretation as an emerging phenomenon. We also discuss some new dualities between ASEP, $q$-TASEP, and heap models.  

\end{abstract}

\maketitle

\tableofcontents

\section{Introduction}

Integrable probabilistic models in both continuum and discrete versions, KPZ-- and ASEP--like families, respectively, play a very important role in the variety of growth problems \cite{corwin2012kardar, derrida1998exactly}. The rich algebraic structure of these models allows one to use powerful combinatorial methods based on representations of underlying groups for their study. The asymptotic behavior of these models fixes the so-called Kardar--Parisi--Zhang (KPZ) universality class. Due to the common algebraic framework, it is not surprising that stochastic models are naturally closely linked with the theory of integrable models \cite{Crampe_2014, blythe2007nonequilibrium}.

From a broad perspective, there are two major families of integrable probability models: the Macdonald process \cite{borodin2014macdonald} and the  higher-spin vertex model \cite{borodin2016stochastic}. These models are at the top of their respective hierarchies -- see \cite{borodin2016lectures, corwin2014macdonald} for reviews. In the first family, the two-parameter $(q,t)$-Macdonald process family degenerates to the $q$-TASEP model at $t=0$, and in the limit $q\to 0$, it reduces to TASEP. In the second family, the higher spin-vertex model degenerates to ASEP, involving the parameter $q'$, and then to TASEP as $q'\to 0$. Thus, the two families meet at the level of KPZ in the continuum and at the level of the Schur process in the discrete setting.

There are several dualities relating various stochastic models \cite{gorsky2022dualities}. The multispecies inhomogeneous probabilistic models involve two sets of parameters: inhomogeneities that yield the local clocks for the update and the boundary conditions. The familiar bispectral duality in the world of integrable models interchanges them within each family. Moreover, there is the counterpart of Matsuo-Cherednik duality, or $QQ$-duality, which connects the probabilistic models from two different families. The identification of familiar dualities in the integrability framework via the probabilistic description has been done in \cite{gorsky2022dualities}. For instance, the inhomogeneous multispecies TASEP is dual to the Goldfish probabilistic process. In this study, we will not discuss the top models in the families, while starting to consider models at the levels $q$-TASEP and ASEP.

Generically, two types of boundary conditions are imposed on probabilistic models: periodic and open boundary conditions, the latter representing interactions with the reservoir in non-equilibrium systems. For open boundary conditions, one can formulate a steady state in which the probability distribution of configurations is time-independent, while a nonzero current may persist \cite{derrida2007non}. In the last decade, substantial progress has been made in the study of open KPZ and ASEP models. Recent developments from probabilistic and algebraic perspectives are reviewed in \cite{corwin2022some}, while a combinatorial description is presented in \cite{williams2022combinatorics}.

In \cite{corwin2018open,bryc2023markov} it was shown that the steady state for an open ASEP in the weakly asymmetric limit reduces to the steady state for an open KPZ. Moreover, it turns out \cite{corwin2018open} that the very probabilistic process on the interval can be formulated in terms of two processes. For an open KPZ, these processes were identified as the Brownian motion plus the Liouville stochastic model -- see \cite{barraquand2021steady, corwin2024stationary}, and the corresponding matrix product representation in terms of the $q$-boson algebra has been developed in \cite{barraquand2023stationary}. The central role of the $q$-Hermite or Askey-Wilson polynomials in the exact solution of the model has been clarified in \cite{uchiyama2004asymmetric}. The relation of the TASEP to heaps models and Lee-Yang zeros has been discussed in \cite{tamm2022steady}. An attempt has been made to develop the matrix model representation for TASEP in \cite{eynard2009matrix}.

A seemingly unrelated line of development concerns low-dimensional holography, which relates large-$N$ quantum mechanics to two-dimensional gravity coupled to a scalar field. In particular, the SYK model \cite{sachdev1993gapless, kitaev2015simple}, describing a quantum dot of a large number of Majorana or complex fermions with random all-to-all four-fermion interactions, is dual at low energies to JT gravity, which describes the near-horizon dynamics of certain higher-dimensional black holes. The effective low-energy Schwarzian action in the SYK model \cite{bagrets2016sachdev} describes the Goldstone mode associated with the spontaneous breaking of the reparametrization symmetry in Euclidean time. On the JT gravity side, the Schwarzian action describes an effective degree of freedom associated with fluctuations of the UV boundary cutoff \cite{maldacena2016conformal, kitaev2019statistical}. A matrix-model representation of JT gravity was developed in \cite{saad2019jt}, where it was argued that the matrix appearing in the matrix integral can, to some extent, be interpreted as a random Hamiltonian for boundary theory. Some deformations of this matrix model and aspects of the ensemble interpretation have been analyzed in \cite{witten2020matrix, stanford2019jt, johnson2021jackiw}. 

More general holographic duality has been found in the double scaled limit of the SYK model (DSSYK) with $p$-fermion coupling instead of the four-fermion interaction subject to conditions $\lambda=\frac{p^2}{N}= \mathrm{const}$, $N\rightarrow \infty, p\rightarrow \infty$. The boundary DSSYK model is dual to $\sin$--dilaton gravity with a specific $\sin \Phi$ dilaton potential and the spectral densities coincide at all energies \cite{blommaert2025entropic, blommaert2025wormholes, heller2025krylov}. On the gravity side, we have the geometry with varying curvature in this case. The representation by the ETH matrix model for DSSYK has been discussed in \cite{jafferis2023jackiw,jafferis2023matrix,okuyama2023discrete}. The potential in this one-matrix ETH matrix model involves Chebyshev polynomials, and if the matter fields are added, it becomes a two-matrix model. In the limit $q=e^{-\lambda}\rightarrow 1$ the DSSYK reduces to the SYK model. The DSSYK effectively interpolates between the JT gravity  at the left spectral edge and the dS gravity at the right spectral edge \cite{verlinde2024syk,sekino2025double}.

As shown in \cite{berkooz2019towards}, DSSYK is exactly solvable in the planar limit using the technique of the chord diagram and the transfer matrix. At infinite $N$, the chord basis is equivalent to the Krylov basis for $\ket{O}$. The Krylov basis seems to be the most suitable way to treat holography in the DSSYK model \cite{fu2025toward, aguilar2025building, nandy2025tridiagonal, heller2025sitter, aguilargutierrez2026probingchaosintegrabilitytransition, ambrosini2025operator, miyaji2025finite}. The elements of the basis are represented by the $q$-Hermite polynomials, and the $q$-boson generator algebra acts as the creation or elimination of chords. The chord basis also naturally emerges as the Lanczos tridiagonalization of the $q$--Gaussian matrix ensemble \cite{bozejko1997q}. The review of the solution of DSSYK and the description of the chord basis can be found in \cite{berkooz2024cordial}. 

The chord basis was also identified in the gravitational dual. In \cite{lin2022bulk}, it was shown that the open chord-number states $\ket{n}$ appearing in the transfer-matrix formalism provide a basis for the bulk Hilbert space, with the bulk geodesic length discretized in units of the DSSYK coupling, $l=\lambda n$. In the triple-scaling limit, the bulk theory reduces to JT gravity, while the chord-number basis becomes the phase space parametrized by the geodesic length and its conjugate momentum. Correspondingly, the DSSYK transfer matrix reduces to the Hamiltonian of Liouville quantum mechanics. An alternative description of the bulk geometry is provided by non-commutative $AdS_2$ \cite{berkooz2023quantum}.

It was possible to add the End-of-World (EOW) brane in the gravity dual. In JT gravity, the EOW brane results in a Morse potential \cite{gao2022effective} with the Whittaker function as the solution. In the DSSYK case the single EOW brane yields the modification of the $q$-Hermite polynomial to the big $q$-Hermite polynomial and to Al-Salam-Chihara polynomials for the two boundary EOW branes \cite{okuyama2023endworldbranedouble}. The tensions of the EOW branes provide the parameters of polynomials. Remarkably, the EOW branes can be interpreted as coherent states for the $q$-oscillator algebra \cite{okuyama2023endworldbranedouble}

Recently, the evaluation of the partition function and moments of DSSYK has been related to another counting problem \cite{gaiotto2025syk}. It was rigorously shown that the DSSYK partition function at infinite temperature coincides with the Schur half-index of $\mathcal{N}=2$ $SU(2)$ pure SYM theory with $n_F=0$, which counts a particular class of half-BPS line operators \cite{gadde2013gauge}. Similarly, the moments of the DSSYK Hamiltonian are given by a generalized Schur half-index, with the insertion of the DSSYK Hamiltonian corresponding to the insertion of a Wilson line. This relation was generalized to $SU(2)$ SYM theories with matter in the fundamental and adjoint representations in \cite{berkooz2025comments}. Generalizations of DSSYK were identified in which $q$ -Hermite polynomials are replaced by more general Askey--Wilson polynomials. The connection between the Schur index and the counting of BPS states on the Coulomb branch was discussed in \cite{cordova2016schur, lewis2025schur}.

The underlying reason of this new duality is the relation of the chord algebra in DSSYK with the Skein algebra generated by Wilson and t'Hooft line operators in 4d SYM theory. This Skein algebra has also been linked to the algebra acting on the Hilbert space of $SL(2,C)$ CS theory in $D\times I$ geometry, where $D$ is the disk with fixed boundary Wilson loops and an irregular singularity at one point, which is doubled, yielding a pair of irregular singularities. The interplay between the modification of the DSSYK Hamiltonian and the insertion of the EOW branes has been discussed in \cite{aguilar2026q}. In particular, the difference between the counting problems for the fundamental and adjoint matter in terms of the EOW branes has been clarified.

Another new line of research suggests that JT gravity can be thought of as the worldsheet theory for the topological string in the 3d Calabi-Yau geometry defined by the spectral curve of JT and $\sin$--dilaton gravity \cite{altland2026quantum}. This viewpoint suggests another perspective on the counting problem and the possibility of relating JT gravity as the version of Kahler gravity \cite{altland2026quantum}. This line of reasoning is, to some extent, parallel to old attempts to form quantum geometry or space-time foam from elementary building blocks in the topological string framework \cite{iqbal2008quantum, okounkov2003quantumcalabiyauclassicalcrystals}.

How are the stochastic models and the DSSYK--bulk gravity dual pair related? The most naive argument is based 
on the Cole--Hopf transform, which connects the height distribution in the KPZ equation to the partition function of the continuous polymer in the random medium. The "polymer representation" of the SYK model and JT gravity has been discussed in \cite{kitaev2019statistical, kitaev2018soft, stanford2020finitecutoffjtgravityselfavoiding}. In \cite{gorsky2024kpz} we have made a few modest steps towards the identification of the KPZ universality class in the context of 2D JT gravity. In particular, it was argued \cite{gorsky2024kpz} that the KPZ regime can be identified  numerically in terms of the Krylov basis for JT gravity, which is the oscillator basis in the $q\rightarrow 1$ limit. The Gauss--KPZ transition was clearly seen if we artificially introduce the cut-off in the Krylov space or for a system with a finite number of degrees of freedom.

The interplay between stochastic models and holography in a more general context has been discussed in \cite{pelliconi}. It was found that in the particular regime the gravity can be considered as a mesoscopic system and the origin of stochastic interpretation is the correlation of the wormhole states in the Hilbert space. These arguments were supported by the direct relation between the stochastic averages of the correlators in the bulk $AdS_3$ gravity and their boundary CFT counterparts. In the boundary CFT, the randomness follows from the OPE coefficients. The results of \cite{pelliconi} suggest a new view on holographic duality and there is no immediate need for the averaging over boundary theories. 

As noted in the resent paper \cite{watanabe2025jt} there is a duality between the steady states of the open ASEP and KPZ with fixed boundary flows on the one hand and DSSYK and JT-gravity with the end-of-world (EOW) brane boundary conditions on the other hand. The DSSYK transfer matrix up to the constant coincides with the ASEP transfer matrix in the chord basis, the parameter $q$ in the DSSYK can be identified with the asymmetry parameter in the ASEP, and the identity of the correlators in two models has been explicitly demonstrated. The boundary  parameters in open KPZ define the tensions of the EOW branes in the $q\rightarrow 1$ limit where the JT/KPZ duality occurs. The connection between the DSSYK--ASEP duality on one side and the correspondence between the strange correlators \cite{Vanhove_2018} in the Levin--Wen string-net models \cite{Levin_2005} and the Turaev--Viro state-sum model on the other side has been discussed in \cite{okuyama2026asepdssykdualitystrangecorrelator}.

In this work, we take some preliminary steps toward a unified view of DSSYK, 2d/3d gravity, and stochastic models in the KPZ universality class. Our approach draws on several common elements. First, we make explicit use of the underlying algebraic structure shared by these models. In stochastic models, this algebraic structure underlies both the matrix-model representation and the transfer-matrix approach. In DSSYK, the transfer matrix in the chord representation naturally realizes the relevant algebraic structure, whereas in the gravity framework, the corresponding role is played by the phase structure in the length representation.

The third unifying perspective is provided by the path--integral interpretation of the stochastic models. Remarkably, the probability distributions of configurations in both TASEP and ASEP can be represented as sums over weighted two-dimensional lattice paths. For TASEP, such a path representation of the probabilities was developed in \cite{derrida2004asymmetric}, while the corresponding representation for ASEP was established in \cite{barraquand2021steady, barraquand2023stationary}. In the ASEP case, the paths carry special weights that have a natural interpretation in terms of the direct sum of discrete Liouville theory and free Brownian motion \cite{barraquand2021steady, barraquand2023stationary, corwin2022some}. This construction generalizes the Motzkin-path representation previously developed for the TASEP partition function in \cite{derrida1993asep}. In the KPZ scaling limit of ASEP, $q\to 1$, the two-dimensional lattice approaches a continuum manifold, and the corresponding path counting is governed by the continuum Liouville measure.

On the other hand, the DSSYK model admits a statistical interpretation in terms of path counting for a spinning particle on a hyperbolic disk \cite{kitaev2018soft, stanford2020finitecutoffjtgravityselfavoiding}, or, equivalently, for a particle moving in a constant magnetic field. The KPZ scaling limit in this representation was discussed in \cite{gorsky2021lifshitz}. From a gravity perspective, the DSSYK Hamiltonian is realized as a Wilson line in $SL(2,\mathbb{C})$ Chern--Simons theory, representing a probe particle. The Hilbert space of the Chern--Simons theory can in turn be parametrized by a two-dimensional lattice of charges. In particular, it was shown in \cite{okuyama2023endworldbranedouble, watanabe2025jt} that the boundary states of ASEP coincide with the coherent states in the chord basis, which are identified with EOW branes.

\begin{figure}[ht]
\centering
\includegraphics[width=12cm]{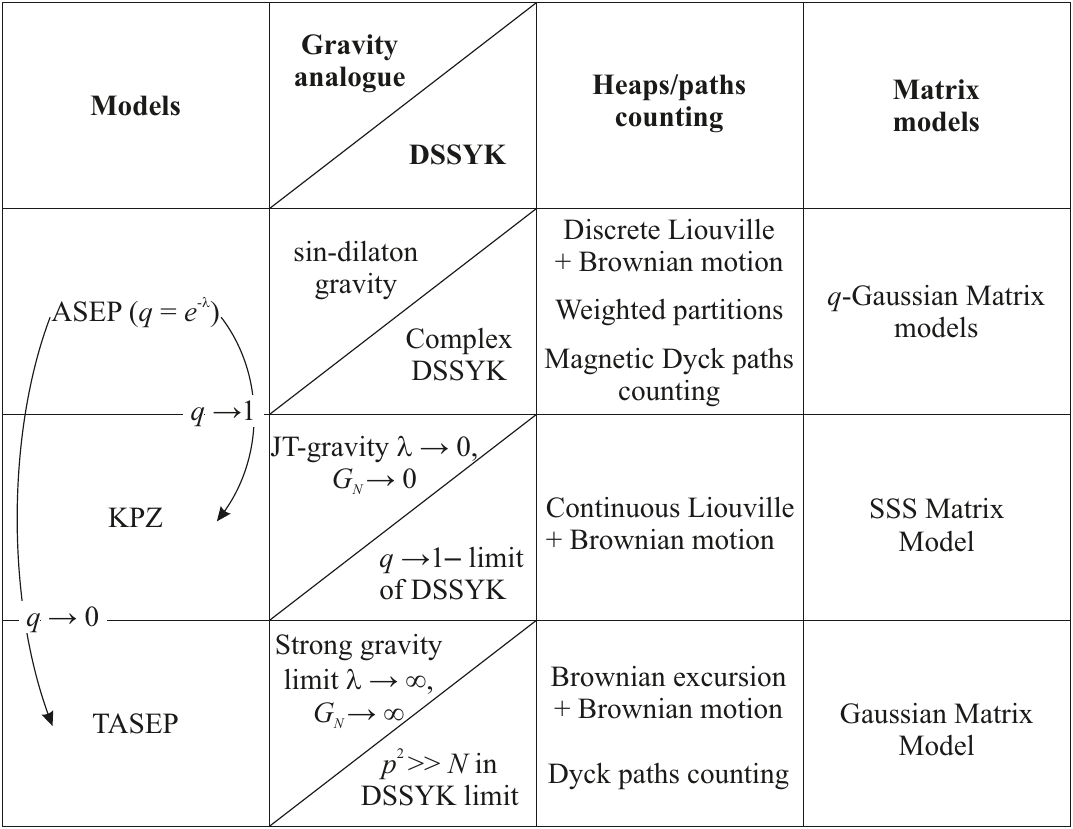}
\caption{Schematic connection between different models discussed in the paper.}
\label{fig:flowchart}
\end{figure}

In this work, we explore the triality among: (i) stochastic growth models of the KPZ--, ASEP--, and TASEP--type, (ii) the DSSYK model, and (iii) 2d/3d gravity. We identify open TASEP with the strong-coupling limit of 2d $\sin$--dilaton gravity. By exploiting the growth interpretation of the partition functions of open TASEP and $q$-TASEP, we obtain new perspectives on strong-coupling gravity. We briefly discuss how the stochastic nature common to all these models can be captured within a matrix-model representation. Furthermore, we exploit the realization of these models in terms of various versions of discrete and continuous random walks on two-dimensional manifolds, which provide a parameterization of the state spaces of the respective stochastic models. The interpretation of TASEP as a heap growth model offers a new perspective on strong-coupling gravity and suggests that the strong-coupling regime may be understood as an emergent phenomenon. We also discuss a new duality between ASEP and $q^{-1}$-heaps, which provides a useful growth interpretation of the ASEP model. To facilitate an understanding of the connections between the different models discussed throughout the paper, in \fig{fig:flowchart} we provide a flowchart summarizing these relationships.

The paper is organized as follows. In Section \ref{sect:2}, we define the ASEP stochastic model on an interval and describe its steady state using the matrix-product ansatz. We derive exact expressions for the resolvents of the transfer matrices for arbitrary $q$, and discuss the limits $q\to 1$, corresponding to the KPZ regime, and $q\to 0$, corresponding to TASEP. In Section \ref{sect:3}, we explore the relation between DSSYK and $\sin$--dilaton gravity through their common algebraic structure and chord basis. We discuss the representation of cumulants in terms  of crossing graphs and $T(1,q)$ Tutte polynomials. Section \ref{sect:4} is devoted to the duality between stochastic models, DSSYK, and dual gravity. In Section \ref{sect:5}, we examine the emergence of stochastic terms in matrix models in the double-scaling limit and review several aspects of the $q$-Gaussian probability measure. In Section \ref{sect:6}, we investigate various aspects of the growth interpretation of ASEP/TASEP models. In particular, we formulate new dualities between $q$-TASEP and ASEP and particular generalizations of ballistic deposition models. In Discussion, we comment on directions for future research, while in conclusion we briefly summarize the main results of our study. The appendices \ref{sect:app1}--\ref{sect:app4} contain technical details of the derivations presented in the main text.

\section{ASEP model}
\label{sect:2}

\subsection{Definition}

\begin{figure}[ht]
\centering
\includegraphics[width=12cm]{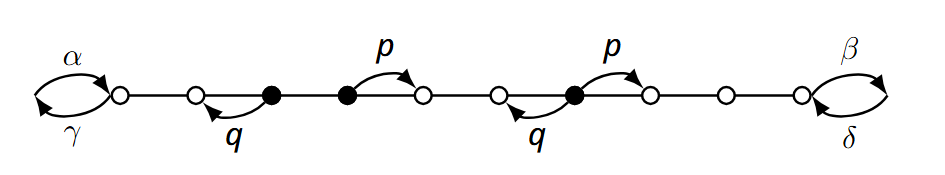}
\caption{ASEP on a 1D lattice with open boundaries. Black dots denote occupied sites, empty circles correspond to holes; bulk hopping rates are $p$ (to the right) and $q$ (to the left); boundary rates are: injection/removal $\alpha,\gamma$ (left) and $\delta,\beta$ (right).}
\label{fig:asep}
\end{figure}

Let us briefly recap the dynamics of the one-dimensional asymmetric simple exclusion process (ASEP) with open boundary conditions. The review of the combinatorial description of the ASEP-like models can be found in \cite{wood2020combinatorial}, the exact solution of the steady-state was first found in \cite{derrida1993asep}. The standard setting of the model is as follows: during every infinitesimal time $dt$, any given site of a finite $N$-site lattice is either occupied exactly by one particle ($\tau_i=1$), or is empty ($\tau_i=0$). If the neighboring site is empty, the particle at the site $2\leq i \leq N$ jumps to the left or to the right with probabilities $qdt$ and $pdt$. The left and right edges of the lattice are connected to the reservoirs: particles enter the system from the reservoirs at the left and right edges with probabilities $\alpha dt$ and $\delta dt$, and are removed from it with probabilities $\gamma dt$ and $\beta dt$, as shown in \fig{fig:asep}.

The ASEP model provides a discrete regularization of the KPZ equation. The stationary measure for KPZ on an interval has been derived in \cite{corwin2024stationary,bryc2023markov, barraquand2021steady, barraquand2023stationary} through an appropriate scaling limit and the matrix-product ansatz \cite{enaud2004large}. The stationary probability measure of ASEP admits a representation in terms of asymmetrically weighted random paths on $\mathbb{Z}^2$. In the KPZ scaling limit, this representation reduces to a combination of two stochastic processes: Brownian motion in one direction and Liouville quantum mechanics in the other \cite{barraquand2021steady}. This perspective will be useful for establishing connections between ASEP, DSSYK, and 2d gravity.

The evolution equation for the probability distribution $P_N(\tau_1,...,\tau_N,t)$, where $\tau_i \in \{0,1\}$ can be written as 
\begin{multline}
\frac{d}{dt}P_N(\tau_1, \dots, \tau_N,t) = \sum_{\sigma_1} (h_1)_{\tau_1\sigma_1}P_N(\sigma_1,\tau_2,\dots,\tau_N,t) + \\ \sum_{i=1}^{N-1}\sum_{\sigma_i\sigma_{i+1}} (h)_{(\tau_i\tau_{i+1}) (\sigma_i\sigma_{i+1})} P_N(\dots,\sigma_i,\sigma_{i+1},\dots,t) + \sum_{\sigma_N} (h_N)_{\tau_N\sigma_N}P_N(\tau_1,\dots,\tau_{N-1}, \sigma_N,t)
\label{master}
\end{multline}
Here $h$ are the matrices for the transition rates:
\be
h_1 = \begin{pmatrix} -\alpha & \gamma \\ \alpha & -\gamma \end{pmatrix} \quad h_N = \begin{pmatrix} -\delta & \beta \\ \delta & -\beta  \end{pmatrix}
\ee
where the basis is $\{0,1\}$. That means, for example, $\{(0,\tau_2,\dots,\tau_N);(1, \tau_2,\dots,\tau_N)\}$ for $h_1$. In the basis $\{(0,0); (0,1); (1,0);(1,1) \}$ the transition matrix $(h)$ can be written as
\be
h = \begin{pmatrix}
0 & 0 & 0 & 0 \\
0 & -q & p & 0 \\
0 & q & -p & 0\\
0 & 0 & 0 & 0
\end{pmatrix}    
\ee 

We are interested in the steady-state $\frac{d}{dt}P_N(\tau_1,...,\tau_N,t)=0$. The corresponding solution yields the operator representation \cite{derrida1993asep, wood2020combinatorial}. Recall briefly the main steps of the construction. At the left boundary, one has
\be\label{left balance}
\sum_{\sigma_1} \dots =  \begin{cases}
-\alpha P_N(0, \tau_2, \dots) + \gamma P_N(1,\tau_2,\dots) \quad \tau_1=0\\
\alpha P_N(0, \tau_2, \dots) - \gamma P_N(1,\tau_2,\dots) \quad \tau_1=1
\end{cases}
\ee
Note that the two expressions in \eqref{left balance}, corresponding to $\tau_1=0$ and $\tau_1=1$, differ only by a sign, so we may combine these two cases by a factor $(2\tau_1-1)$. Now assume that there exists a coefficient $x_L$ independent of $\tau_2,\ldots,\tau_N$, such that: 
\be
\sum_{\sigma_1} \dots = (2\tau_1-1)x_LP_{N-1}(\tau_2,\dots \tau_N).
\label{left}
\ee 
A similar result in bulk may be obtained by considering separately the four possible values of $(\tau_i,\tau_{i+1})$. The states $(0,0)$ and $(1,1)$ do not give a contribution, whereas for $(0,1)$ and $(1,0)$ the two terms occur with the hopping rates $p$ and $q$. Assuming that there exists such $x$ that these contributions can be expressed through the corresponding probabilities of the site $(N-1)$ obtained by removing the site $i$ or the site $i+1$, the four cases can be written as follows:
\be
\sum_{\sigma_i\sigma_{i+1}}\dots = -(2\tau_i-1)x P_{N-1}(\dots,\tau_{i-1},\tau_{i+1},\dots)+(2 \tau_{i+1} -1 )xP_{N-1}(\dots,\tau_i,\tau_{i+2},\dots)
\label{middle}.
\ee
At the right boundary, analogously to \eq{left}, one has 
\be
\sum_{\sigma_N} \dots = -(2\tau_N-1)x_RP_{N-1}(\tau_1,\dots,\tau_{N-1}) 
\label{right}
\ee
If such coefficients with $x=x_L=x_R$ exist, then $P_N$ given by recurrent equations \eqref{left}-\eqref{right} automatically satisfy the balance equation \eqref{master}. So, the problem is reduced to solving equations \eqref{left}-\eqref{right}.

Defining $X_k= \tau_K D+(1-\tau_k)E$ where $D$ and $E$ are some matrices, consider the ansatz \cite{derrida1993asep}: 
\be
P_N(\tau_1,...,\tau_N) = \bra{W} \prod_{k=1}^N X_k\ket{V}
\label{ansatz}
\ee
where $L_i= \bra{W}X_1,\dots,X_{i-1}$ and $R_i= X_{i+2},\dots,X_N\ket{V}$; thus $P_N(\dots,01,\dots)=L_iEDR_i$, $P_N(\dots,10,\dots) = L_iDER_i$. Substituting \eq{ansatz} into \eq{middle} we get conditions on $D$ and $E$. For example, for $(\tau_i,\tau_{i+1})=(0,1)$ one has $L_i(pDE-qED)R_i=xL_i(D+E)R_i$. The case $(\tau_i,\tau_{i+1})=(1,0)$ gives the same condition. From the middle part, we obtain the equation that defines the algebra of operators $D$ and $E$:
\be
pDE-qED=x(D+E)    
\label{res}
\ee
 Similarly, the boundary condition is obtained at the right and left ends, substituting \eq{ansatz} into \eq{left} and \eq{right}
\be
\begin{array}{c}
\disp -\alpha\bra{W}E+\gamma\bra{W}D = -x\bra{W}, \medskip \\
\disp -\delta E \ket{V} + \beta D\ket{V} = x \ket{V}
\end{array}
\label{boundary}
\ee
Here we consider $x=x_R=x_L$ to satisfy \eq{master}.

Note that \eq{res} is identical to $DE-qED=E+D$. Actually, taking a common notation $x=p$, up to the redefinitions $p\neq 0$, $\frac{q}{p}=r$ we have $DE-rED = E + D$. Later we used "$q$" instead of "$r$", as an asymmetry parameter. Similarly, replacing $\tilde \alpha = \frac{\alpha}{p}, \tilde \beta= \frac{\beta}{p}, \tilde \gamma = \frac{\gamma}{p}, \tilde \delta =\frac{\delta}{p}$ in the boundary equations and omitting the tilde for the sake of notation hereafter, we obtain the ASEP matrix algebra.
\be\label{ASEP alg}
\begin{array}{c}
\disp DE-qED=D+E, \medskip\\
-\alpha\bra{W}E+\gamma\bra{W}D = -\bra{W}, \medskip \\
\disp -\delta E \ket{V} + \beta D\ket{V} =  \ket{V}
\end{array}
\ee
\subsection{Matrix product state ansatz for ASEP}

The deformation of the algebra of infinite-dimensional operators $D$ and $E$ in the form $DE-qED=D+E$ implies the symmetrized ASEP transfer matrix $T=D+E$ \cite{wood2020combinatorial}:
\be
T=\frac{1}{1-q} \left(\begin{array}{ccccc}
    2 & \sqrt{1-q^1} & 0 & 0 & \cdots\\
    \sqrt{1-q^1} & 2 &\sqrt{1-q^2} & 0 & \\
    0 & \sqrt{1-q^2} & 2 & \sqrt{1-q^3}\\
    0 & 0 & \sqrt{1-q^3} & 2 \\
    \vdots & & & & \ddots
\end{array}\right)
\ee
For the left and right boundaries  one gets 
\be
\begin{array}{l}
\disp \bra{W} = \kappa \left(1, \frac{P_1}{\sqrt{1-q}},\frac{P_2}{\sqrt{(1-q)(1-q)^2}}, \frac{P_3}{\sqrt{(1-q)(1-q)^2 (1-q)^3}}, \dots \right) \medskip \\
\disp \ket{V} = \kappa \left(1, \frac{Q_1}{\sqrt{1-q}},\frac{Q_2}{\sqrt{(1-q)(1-q)^2}}, \frac{Q_3}{\sqrt{(1-q)(1-q)^2 (1-q)^3}}, \dots \right)^{\top}
\end{array}
\label{braket}
\ee
where $P_n$ and $Q_n$ satisfy the recursions:
\be
\begin{array}{ll}
\disp P_{n+1}=AP_n+B(1-q^n)P_{n-1}, & P_0=1, \quad P_1 = A \medskip \\
\disp Q_{n+1}=CQ_n+D(1-q^n)Q_{n-1}, &  Q_0=1, \quad Q_1 = C
\end{array}
\label{PQ}
\ee
and the coefficients $A,B,C,D$ are as follows 
\be
\begin{cases} \disp A=\frac{1-q-\alpha+\gamma}{\alpha} \medskip \\ \disp B=\frac{\gamma}{\alpha} \end{cases}; \quad 
\begin{cases} \disp C=\frac{1-q-\beta+\delta}{\beta} \medskip \\ \disp D=\frac{\delta}{\beta} \end{cases}
\label{CD}
\ee
Thus, the most general forms of $\bra{W}$ and $\ket{V}$ are
\be
\begin{array}{ll}
\disp \bra{W} = \kappa \left(1, \frac{A}{\sqrt{1-q}},\frac{A^2+B(1-q)}{\sqrt{(1-q)(1-q)^2}}, \frac{A^3+AB[(1-q)+(1-q^2)]}{\sqrt{(1-q)(1-q)^2 (1-q)^3}}, \dots \right) \medskip \\
\disp \ket{V} = \kappa \left(1, \frac{C}{\sqrt{1-q}},\frac{C^2+D(1-q)}{\sqrt{(1-q)(1-q)^2}}, \frac{C^3+CD[(1-q)+(1-q^2)]}{\sqrt{(1-q)(1-q)^2 (1-q)^3}}, \dots \right)^{\top} 
\end{array}
\label{bra-ket}
\ee
where $\kappa$ is set by the condition $\braket{W|V}=1$. However, we suppress the general normalization factor $\kappa$ of the boundary vectors in subsequent derivations. The described representation for ASEP with open boundaries was introduced in \cite{uchiyama2004asymmetric}.

Truncate the infinite matrix $T$ and consider the  $n\times n$ matrix $T(n)$. The boundary vectors become finite, $\bra{W} \to \bra{W^{(n)}}$ and $\ket{V} \to \ket{V^{(n)}}$. Using ansatz \eq{ansatz}, the ASEP partition function $Z_N(n)$ can be written as
\be
Z_N(n) = \bra{W^{(n)}}T^N(n)\ket{V^{(n)}}
\label{partition}
\ee
Define the vector $\textbf{Z}_N=(Z_{N,1},Z_{N,2}, \dots, Z_{N,n})^{\top}$ that satisfies the recursion
\be
\textbf{Z}_{N+1}=T(n)\textbf{Z}_N, \quad \textbf{Z}_{N=0}=\ket{V^{(n)}}
\label{rec-part}
\ee
Equation \ref{rec-part} can be solved using the generating function method. For the generating function
\be
\textbf{Y}(s)=(Y(s,1), Y(s,2),\dots, Y(s,n))^{\top}=\sum_{N=0}^{\infty} \textbf{Z}_N\, s^N
\label{W'}
\ee
the recursion \eq{rec-part} becomes the set of linear algebraic equations
\be
\left(T(n)-\frac{1}{s}I\right)\textbf{Y}(s)=-\frac{1}{s}\textbf{Z}_0  
\label{gen}
\ee
where $I$ is the identity matrix.

\subsection{Statistical properties of ASEP}

During the past two decades, the orthogonal-polynomial approach to ASEP-type models has become increasingly popular \cite{sasamoto1999one}. In particular, it is known \cite{uchiyama2004asymmetric} that ASEP with open boundaries is closely related to continuous $q$-Hermite polynomials, and the corresponding generating function was previously obtained in integral form \cite{uchiyama2004asymmetric,blythe2000exact}. Here, we present an alternative expression for the ASEP generating function in terms of a double sum over continuous $q$-Hermite polynomials. To our knowledge, this representation is new; in addition, it is particularly transparent and useful for analyzing the limiting cases $q\to 0$ and $q\to 1$. We leave the details of the derivation to Appendix \ref{sect:app1} and state here the resulting expression for the ASEP generating function $\Omega(s)$:
\begin{multline} 
\Omega_n^{ASEP}(s) = \sum_{N=0}^{\infty} \bra{W^{(n)}}T^N(n)\ket{V^{(n)}}s^N = \frac{(-1)}{s} \frac{1-q}{H_n(x|q)} \sum_{k=1}^{n}(-1)^k \frac{(i\sqrt{B})^{k-1}H_{k-1}(a|q)}{\sqrt{(q;q)_{k-1}}} \times \\ \bigg[\sqrt{(q;q)_{k-1}}  H_{n-k}^{(k)}(x|q)\sum_{j=1}^{k} \frac{(-i\sqrt{D})^{j-1}}{(q;q)_{j-1}}H_{j-1}(y|q) H_{j-1}(x|q)+ \\ \frac{H_{k-1}(x|q)}{\sqrt{(q;q)_{k-1}}}\sum_{j=k+1}^{n} (-i\sqrt{D})^{j-1} H_{j-1}(y|q) H_{n-j}^{(j)}(x|q) \bigg] 
\label{GenASEP}
\end{multline}
Recall that $x=1-\frac{1-q}{2s}, \quad y = \frac{C}{2i\sqrt{D}}, \quad a=\frac{A}{2i\sqrt{B}}$.

%\subsubsection{Phase transitions in ASEP}

For completeness, we comment here on the phase behavior in ASEP. The analysis of phase transitions in the two- and four-boundary-rate ASEP has been presented in numerous previous works; see, for example, \cite{uchiyama2004asymmetric, wood2020combinatorial, sandow1994partially}. Employing a representation of the boundary vectors in terms of a series of continuous $q$-Hermite polynomials, we significantly simplify the derivation for ASEP with four boundaries compared to the approach based on the full integral expression \cite{uchiyama2004asymmetric}. Since the discussion of phase transitions is not the main focus of the present paper and is included mainly to illustrate the effectiveness of the method, we add the corresponding material in the Appendix \ref{sect:app3}. Here, we retain only the explicit form of the generating function in the limit $n \to \infty$, which will be required for subsequent analysis.

To analyze phase transitions in the full ASEP model, we take a limit $n \rightarrow \infty$ in the generating function $\Omega^{ASEP}(s)$ defined in \eq{GenASEP}:
\be
\Omega^{ASEP}(s) = \sum_{N\geq0}\bra{W}T^N\ket{V} s^N = \bra{W}(I-sT)^{-1}\ket{V}
\ee
Let $\phi_n(\theta) = \frac{(-1)^n}{\sqrt{(q;q)_n}}H_n(\cos(\theta)), \quad 0\leq \theta \leq \pi$. Using the transfer matrix $T$ in $\phi_n(\theta)$ and using the recursion for continuous $q$-Hermite polynomials, we get 
\be
(T \phi(\theta))_n = \frac{2}{1-q}\phi_n + \frac{1}{1-q}\left(\sqrt{1-q^{n+1}}\phi_{n+1}+\sqrt{1-q^n} \phi_{n-1}\right) = \frac{2-2\cos(\theta)}{1-q} \phi_n(\theta)
\ee
which allows us to rewrite the generating function as follows:
\be
\Omega(s) = \int_0^{\pi} \frac{\widehat{W}(\theta)\widehat{V}(\theta)}{1-s\lambda(\theta)} d\mu_q(\theta); \quad 
\widehat{W}(\theta) = \sum_{n\geq0} w_n \phi_n(\theta), \quad \widehat{V}(\theta)=\sum_{n\geq0} v_n \phi_n (\theta)
\label{spec}
\ee
where 
\be
d\mu_q(x) = \frac{1}{2\pi}\frac{y(x\mid q)}{\sqrt{1-x^2}}dx, \quad y(x \mid q)=\left|\left(e^{2 i \theta}; q\right)_{\infty}\right|^2
\ee 
is the spectral measure of continuous $q$-Hermite polynomials, and $w_n$, $v_n$ are components of the boundary vectors \eq{braket}. The analysis of the analytic structure of the corresponding expressions for different sets of boundary parameters is given in the Appendix \ref{sect:app3}.

\subsection{Particular solution of ASEP in terms of Touchard-Riordan $q$-Catalans}

The method of solving the generic ASEP model using a generating function is valid for any boundary vector. Let us make a specific choice: 
\be
\bra{W^{(n)}} = \bra{e_1} = (1,0,0,\dots,0), \quad \ket{V^{(n)}} = \ket{e_1} = (1,0,0,\dots,0)^{\top}
\label{eq:braket-bridge}
\ee
%which could be obtained formally from generic vectors $\bra{W^{(n)}}$ and $\ket{V^{(n)}}$ by sending $\alpha\to \infty$ and $\beta\to \infty$ in \eq{boundary}.
This particular case has a rich algebraic structure, since it establishes a bridge between ASEP and specific models of growth discussed in forthcoming sections. So, we work with the following partition function:
\be
Z_N(n)=\bra{W^{(n)}}T^N(n)\ket{V^{(n)}} = \bra{e_1}T^N(n)\ket{e_1}
\ee
and the vectors $\textbf{Z}_N=(Z_{N,1}, Z_{N,2}, \dots, Z_{N,n})^{\top}$,  $\textbf{Z}_{N+1} = T(n)\textbf{Z}_N$ and $\textbf{Z}_{N=0}=\ket{V^{(n)}}=\ket{e_1}$. The generating function $\textbf{Y}(s) = (Y(s,1), Y(s,2), \dots, Y(s,n))^{\top}=\sum_{N=0}^{\infty}\mathbf{Z_N}\,s^N$ defined in \eq{gen} can be represented in the determinantal form (see \eq{W} of Appendix \ref{sect:app1} for details):
\be
Y(s,k) = \frac{\text{det}B'(k)}{\text{det}(T(n)-\frac1sI)} = \frac{v'_{n,k}}{u_n}
\label{eq:determ}
\ee
where $B'(k)$ is obtained from $\left(T(n)- \frac{1}{s} I\right)$ by replacing the $k$-th column by the vector $-\frac{1}{s} \ket{e_1} = (-\frac1s, 0, 0,\dots,0)^{\top}$, while the denominator $u_n$ remains unchanged. Repeating the derivation provided in Appendix \ref{sect:app1}, one gets the following explicit expression of $v'_{n,k}$, which is much simpler than the generic case due to the simpler structure of the boundary vectors:
\be
v'_{n,k} = \frac{(-1)^k}{s(1-q)^{n-1}} \sqrt{(q;q)_{k-1}} H^{(k)}_{n-k}(x|q), \quad x= 1-\frac{1-q}{2s}.
\ee
Therefore
\be
Y(s,k) = (-1)^k \frac{1-q}{s} \frac{\sqrt{(q;q)_{k-1}}H_{n-k}^{(k)}(x|q)}{H_n(x|q)}
\ee
In particular, we are interested in the specific case $k=1$ that corresponds to the selection of the vector $\ket{V^{(n)}} = \ket{e_1} = (1,0,0,\dots,0)^{\top}$. The generating function by definition is:
\be
\Omega_n(s)=\sum_{N=0}^{\infty} \bra{W^{(n)}} T^N(n) \ket{V^{(n)}}s^{N} \equiv Y(s,1) = -\frac{1-q}{s} \frac{H_{n-1}^{(1)}(x|q)}{H_n(x|q)}, \quad x=1-\frac{1-q}{2s}
\label{eq:ASEP-bridge}
\ee
On the other hand (see \eq{spec}), 
\be
\Omega(s) = \int_0^{\pi} \frac{\widehat{W}(\theta)\widehat{V}(\theta)}{1-s\lambda(\theta)}\, d\mu_q(\theta)
\end{equation}
Substituting the new boundary vectors $\bra{e_1}$ and $\ket{e_1}$ for $\widehat{W}$ and $\widehat{V}$ defined in \eq{eq:braket-bridge}, we obtain:
\be
\Omega(s) = \int\frac{d\mu_q(\lambda)}{1-s\lambda} = \sum_{N \geq 0 }s^{N} \int \lambda^Nd\mu_q(\lambda)
\ee
Recalling that $\lambda = \frac{2(1-x)}{1-q}$ and $x=\cos(\theta)$, we get from the previous Section the following result:
\be
\bra{e_1}T^N\ket{e_1} = \left(\frac{2}{1-q}\right)^N \int_{-1}^1 (1-x)^N d\mu_q(x) =  \left(\frac{2}{1-q}\right)^N\sum_{r=0}^N \binom{N}{r} (-1)^r \int_{-1}^1 x^r\, d\mu_q(x)
\label{eTe}
\ee
where $m_r = \int x^r d\mu_q(x)$ is exactly the $r$-th moment of the continuous $q$-Hermite polynomial. It is known (see, for example, \cite{ismail1987combinatorics}) that all $m_r$ with odd $r$ are zero, while for even $r$ they admit expression by crossing $q$-Catalan numbers, which are also called Touchard–Riordan polynomials:
\be
m_{2n} = \Big(\frac{1-q}{4}\Big)^n \sum_{\pi \in \mathcal{P}(2n)} q^{\text{cr}( \pi)} = \Big(\frac{1-q}{4}\Big)^n C^T_n (q)
\label{C^T_q}
\ee
where the sum runs over all pairings of $\{1,2,\dots,2n\}$ and $\text{cr}(\pi)$ is the number of crossings in the definition of $C^T_n(q)$ (see \cite{riordan1975distribution}).

Substituting \eq{C^T_q} into \eq{eTe}, we arrive at the following expression:
\begin{multline}
\bra{e_1}T^N\ket{e_1} =  \left(\frac{2}{1-q}\right)^N \sum_{k=0}^{\left[\frac{N}{2}\right]}\binom{N}{2k} \Big(\frac{1-q}{4}\Big)^k \sum_{\pi \in \mathcal{P}(2k)} q^{\text{cr}( \pi)}=\\  =\left(\frac{2}{1-q}\right)^N \sum_{k=0}^{\left[\frac{N}{2}\right]}\binom{N}{2k} \Big(\frac{1-q}{4}\Big)^k C^T_k(q)
\label{ASEP-Catalans0}
\end{multline}

Therefore, the ASEP generating function for arbitrary values of $q$ in the limit $\alpha\to\infty$, $\beta\to\infty$ can be expressed via Touchard-Riordan $q$-Catalan crossing numbers, $C^T_k(q)$, as follows
\be
\Omega(s) = \sum_{N=0}^{\infty} \left(\frac{2s}{1-q}\right)^N \sum_{k=0}^{\left[\frac{N}{2}\right]}\binom{N}{2k} \Big(\frac{1-q}{4}\Big)^k C^T_k(q)
\label{ASEP-Catalans}
\ee

\subsubsection{Generating function: resolvent in the limit $q\to 1$}

Let us consider the  example of nontrivial scaling $q\to 1$
taking into account simultaneously limits $q\rightarrow 1$ and $n\rightarrow \infty$ in the ASEP generating function \eqref{GenASEP} in the scaling regime $\lim\limits_{\substack{q\to 1 \\ n\to\infty}} n(1-q) = \lambda$ where $\lambda \in (0, \infty)$ (other modes are trivial). In what follows, we use the notation $\varepsilon = 1-q$. Recall the explicit expression of the generating function (see \eqref{Omega ASEP} in Appendix \ref{sect:app1})
\begin{equation}\label{Omega limit}
\Omega_n^{ASEP} (s) = \frac{\varepsilon}{s}\sum_{k=1}^{n}\sum_{j=1}^n
(-1)^{k+j+1}\frac{P_{k-1}Q_{j-1}}{(q;q)_{m-1}} \frac{H_{m-1}(x|q)H_{n-M}^{(M)}(x|q)}{H_n(x|q)}
\end{equation}
where $m = \min(k,j), \; M=\max(k,j)$ and we denoted $P_{k-1} = (i\sqrt{B})^{k-1}H_{k-1}(a|q), \quad Q_{j-1}=(i\sqrt{D})^{j-1}H_{j-1}(y|q)$. Introducing scaled variables 
\begin{equation}
\left\{\begin{array}{ll}
\kappa = (k-1)\varepsilon, & \eta = (j-1)\varepsilon, \quad 0\leq\kappa, \eta \leq \lambda \medskip \\ m-1 = \min(k-1,j-1), & \mu = (m-1) \varepsilon = \min(\kappa,\eta) \medskip \\ 
M-1=\max(k-1,j-1), & \nu = \max(\kappa, \eta)       
\end{array}
\right.
\end{equation}
we derive the leading asymptotics for each term of \eqref{Omega limit}:
\begin{enumerate}
    \item The term $(q;q)_{m-1} = \prod\limits_{l=1}^{m-1}(1-q^l)$. Using the decomposition $1-q^l = 1-e^{-l\varepsilon}+O(\varepsilon)$ we obtain
    \begin{equation}
        \log(q;q)_{m-1}=\sum_{l=1}^{m-1}\log(1-q^l) = \frac1\varepsilon\sum_{l=1}^{m-1}\varepsilon\log(1-e^{-l\varepsilon}) \sim \frac{1}{\varepsilon} \int_{0}^{\mu}\log(1-e^{-t})dt
    \end{equation}
Therefore, the total contribution of this term is
\begin{equation}
        (q;q)_{m-1} \sim \exp \Big(\frac{1}{\varepsilon} \int_{0}^{\mu}\log(1-e^{-t})dt\Big)
\end{equation}
    \item The term $P_{k-1}$. The total contribution is:
\begin{equation}
    P_{k-1} \sim \exp\Big(\frac{1}{\varepsilon} \int_0^{\kappa} \log u_{A}(t) dt\Big)
\end{equation}
where $u_A(t) = \frac{A_0 \pm \sqrt{A_0^2+4B_0(1-e^{-t})}}{2}$ with $A_0 = \frac{\gamma-\alpha}{\alpha}, \quad B_0 = \frac{\gamma}{\alpha}$. The branch should be chosen such that $u_A(0)=A_0$.
    \item Similarly, contribution of the term $Q_{j-1}$:
\begin{equation}
    Q_{j-1} \sim \exp\Big(\frac{1}{\varepsilon} \int_0^{\eta} \log u_{C}(t) dt\Big)
\end{equation}
where $u_C(t) = \frac{C_0 \pm \sqrt{C_0^2+4D_0(1-e^{-t})}}{2}$ with $C_0 = \frac{\delta-\beta}{\beta}, \quad D_0 = \frac{\delta}{\beta}$. The branch should be chosen such that $u_C(0)=C_0$.
    \item The term $\frac{H_{m-1}(x|q)H_{n-M}^{(M)}(x|q)}{H_n(x|q)}$. The total contribution is:
\begin{equation}
    \frac{H_{m-1}(x|q)H_{n-M}^{(M)}(x|q)}{H_n(x|q)} \sim \exp\Big(-\frac1\varepsilon \int_{\mu}^{\nu} \log v(t)dt\Big)
\end{equation}
where $v(t) = 1+e^{-\frac{t}{2}}$.
\end{enumerate}
The four exponential contributions, collected together, produce a single functional like an action $\Phi(\kappa,\eta)$. The corresponding generating function becomes the following.
\begin{equation}
  \Omega_n^{ASEP}(s) \sim -\frac{\varepsilon}{s}\sum_{k,j} \exp\Big(\frac{1}{\varepsilon}(\Phi(\kappa,\eta)+i\pi(\kappa+\eta)\Big)
\end{equation}
Where $\kappa = (k-1)\varepsilon, \quad \eta = (j-1)\varepsilon, \quad 0\leq\kappa,\eta \leq \lambda$ and 
\begin{equation}
    \Phi(\kappa,\eta) = \int_0^{\kappa}\log u_A(t)dt + \int_0^{\eta}\log u_C(t)dt - \int_0^{\mu}\log(1-e^{-t})dt-\int_{\mu}^{\nu}\log v(t)dt
\end{equation}
Here $\mu = \min(\kappa,\eta), \quad \nu = \max(\kappa, \eta)$.

The leading exponential growth is governed by the maximum of $\Phi(\kappa,\eta)$:
\begin{equation}
\varepsilon \log|\Omega_n(s)^{ASEP}| \longrightarrow \sup_{0\leq\kappa,\eta\leq \lambda} \Re \Phi(\kappa,\eta) 
\end{equation}
For vanishing boundary rates $\gamma = \delta = 0$, this supremum can be found explicitly:
\begin{equation}
    \left|\Omega_n(s)^{ASEP}\right| \to \exp\left(\frac1\varepsilon\left(\frac{\pi^2}{6} - \mathrm{Li}_2(e^{-\lambda})\right) \right)
\end{equation}
The example considered demonstrates that there are several nontrivial 
scalings for weakly asymmetric ASEP.
\subsubsection{Generating function in the limit $q\to 0$ -- TASEP regime}

By sending $q\to 0$ we block backward jumps and restore the TASEP limit which corresponds to the non-deformed algebra of operators $D$ and $E$: $DE = D+E$. Assuming also $\gamma=0, \delta=0$, we have 
\be
\left[\begin{array}{ll}
B = \frac{\gamma}{\alpha} = 0, & A=\frac{1-\alpha}{\alpha} \medskip \\ 
D=\frac{\delta}{\beta}=0, & C=\frac{1-\beta}{\beta}
\end{array}\right], \quad
\begin{array}{l}
P_{n+1} = AP_n \Rightarrow P_n = A^n \medskip \\
Q_{n+1}=C Q_n  \Rightarrow Q_n = C^n
\end{array}, \quad
\begin{array}{l}        
H_r(x|q) \to U_r(x), \medskip \\
H^{(M)}_{n-M}(x|q) \to U_{n-M}(x)
\end{array}
\ee
where $U_r(x)$ are the Chebyshev polynomials of the second kind. The generating function \eq{gen-ASEP} in this limit takes the following form
\be
\Omega_n^{(TASEP)}(s) = \frac{1}{sU_n(x)} \sum_{k=1}^n\sum_{j=1}^n(-1)^{k+j+1}A^{k-1}C^{j-1}U_{m-1}(x)U_{n-M}(x)
\label{gen-TASEP0}
\ee
Equation \eq{gen-TASEP0} is essentially simplified in the limit $n\rightarrow \infty$. Define two functions,
\be
\eta(s)=\frac{1-\sqrt{1-4s}}{2}; \quad \rho= - \frac{\eta}{1-\eta} 
\ee
Since $s= \eta(1-\eta)$, we have $x = 1-\frac{1}{2s} = \frac{\rho+\rho^{-1}}{2}$. Thus, the Chebyshev polynomials may be written as
\be
U_n(x) = \frac{(x+\sqrt{x^2-1})^{n+1} - (x-\sqrt{x^2-1})^{n+1}}{2\sqrt{x^2 -1}}  = \frac{\rho^{n+1} - \rho^{-n-1}}{\rho-\rho^{-1}}
\ee

Taking all limits as discussed in Appendix \ref{sect:app2}, in the $n \rightarrow \infty$ limit the generating function $\Omega_n^{TASEP}(s)$ takes the well-known TASEP form
\be
\Omega^{TASEP}(s) = -\frac{\rho}{s}\frac{1}{(1-AC)(1+A\rho)(1+C\rho)}= \frac{\alpha \beta}{\alpha + \beta -1} \frac{1}{(1-\frac{\eta}{\alpha})(1-\frac{\eta}{\beta})}
\ee
where $ \eta = \frac{1-\sqrt{1-4s}}{2}$.

\subsection{Limit to KPZ}
%\subsusbsection{Limits  of evolution equations}
%\subsubsection{Limits of steady sates}

The description of the ASEP steady state can be reformulated in terms of the weighted two-dimensional random walk on $Z^2$ using the parameterization of the states suggested in  \cite{derrida2004asymmetric}
\begin{equation}
    2\tau_i -1=n_i -n_{i-1} +\sigma_i
\end{equation}
where $\sigma_i$ is a random variable defined as follows 
\be
\sigma_i=\begin{cases} \pm 1 & \mbox{with equal probabilities $\tfrac{1}{2}$ when $n_{i-1}=n_n$} \medskip \\ 0 & \mbox{when $n_{i-1}\neq n_i$} \end{cases}
\ee
Let us introduce $m_i=\sum_{j=1}^{i}\sigma_j$. The steady state probability can be written as 
\begin{equation}
    P(\mathbf{n},\,\mathbf{m})\propto \frac{1}{Z_L(q)}\prod_{i-1}^L[n_i]_qP^{SSRW}_{n_0,0}(\mathbf{n},\,\mathbf{m})
\end{equation}
where $P^{SSRW}_{n_0,0}$ is the probability measure of the simple symmetric random walk on $Z^2$ that started t $(n_0,0)$ with equal probability steps in four directions.

To derive the continuum KPZ equation, we perform the following rescaling of the variables:
\begin{equation}
q=e^{-\lambda},\quad i=x\lambda^{-2}, \quad L=4M\lambda^{-2}.
\end{equation}
The boundary rates are rescaled as
\begin{equation}
\rho_a=\frac{1}{2} +\frac{u}{4}\lambda, \quad  \rho_b=\frac{1}{2} +\frac{v}{4}\lambda.
\end{equation}
We define the height function for ASEP as
\begin{equation}
h^{ASEP}(t,i)= -2N +\sum_{j=1}^{i}(2\tau_j-1),
\end{equation}
where $N$ is the net number of particles that have been added to the system at the first site from the left reservoir. In the stationary regime, one has the following.
\begin{equation}
h^{ASEP}(t,i)-h^{ASEP}(t,0)=n_i-n_0+m_i
\end{equation}
The connection with the polymer representation provides the Cole--Hopf transformation.
\begin{equation}
Z(t,j) =e^{\tfrac{\lambda}{2}h^{ASEP}(t,j) +ct}, \qquad c=1+q-2\sqrt{q}
\end{equation}
Under the above rescaling, the function $Z\left(16\lambda t,\tfrac{4x}{\lambda^2}\right)$ obeys the stochastic heat equation, which also represents the partition function of a polymer in random media.
\begin{equation}
\partial_t Z(t,x)=\partial_{xx}Z(t,x) +\xi(t,x)Z(t,x)
\end{equation}
Meanwhile, the height function $h^{KPZ}(t,x)$ satisfies the continuous KPZ equation on an open segment,
\begin{equation}
\partial_t h^{KPZ}(t,x)= \partial_x^2 h^{KPZ}(t,x) +\left(\partial_x h^{KPZ}\right)^2 + \xi(t,x),
\quad \langle\xi(t,x)\xi(t',x')\rangle =\delta(t-t')\delta(x-x')
\end{equation}

In a steady state, consider the rescaled height function
\begin{equation}
    H(\lambda,x)=\frac{\lambda}{2}(h^{ASEP}(t,4x\lambda^{-2}) - h^{ASEP}(t,0))
\end{equation}
which in the $\lambda\rightarrow 0$ limit reduces to the steady height function of open KPZ. It can be presented as the sum of two stochastic variables that describe the motion in the "n" and "m" directions
\begin{equation}
    H(x)=W(x)+X(x)
\end{equation}
where $W(x)$ is a one-sided Brownian process with boundary conditions: $W(0)=0$ $W(L)$-free. The stochastic field $X(x)=U(x)-U(0)$ is the Liouville process with the conventional path integral measure for Liouville quantum mechanics with particular boundary conditions
\begin{equation}
    dU\exp(-2uU(0) -2vU(L)- \int_{0}^{L}dx[U'^2 + e^{-2U(x)}])
\end{equation}
were $U(L),\,U(0)$ are free. 
It is exactly the representation of the open KPZ in the interval as the sum of the Liouville and Brownian processes found in \cite{barraquand2021steady,barraquand2023stationary,barraquand2025large}.

\subsection{Spin chain representation}

\subsubsection{Homogeneous spin chains}

The ASEP model can be represented as an XXZ spin chain with appropriate boundary conditions, either on an interval or on a circle \cite{sandow1994partially}. Integrable many-body spin systems arise when the Markov operator is interpreted as the Hamiltonian governing the time evolution. In particular, the Markov evolution of the joint probability distribution $P$ for ASEP takes the following form \cite{gwa1992sixvertex}:
\be
\frac{dP}{dt}=M\, P
\label{rr7}
\ee
where the Markov matrix $M$ for a homogeneous periodic lattice of size $L$ in the spin representation is
\be
M=\sum_{i=1}^{L}\left(sS_i^{-}S_{i+1}^{+}+ rS_i^{+}S_{i+1}^{-} + \frac{1}{4}S_i^{z}S_{i+1}^{z}\right) -\frac{L}{4}
\label{rr8}
\ee
Here, $s$ and $r$ are real parameters that parameterize the jump rates to the left and to the right, respectively, and $S^{-}_i,S^{+}_i,S^{z}_i$ are three generators of $SU(2)$ acting on the $i$-th site. Upon the gauge transformation, $M$ is mapped onto the Hamiltonian of $SU(2)$ XXZ Heisenberg (quantum) spin chain
\be
H_{XXZ}= \sum_{i=1}^{L}\left(S_i^{x}S_{i+1}^{x}+ S_i^{y}S_{i+1}^{y} + \frac{\alpha + \alpha^{-1}}{2}S_i^{z}S_{i+1}^{z}\right) +\alpha^{L}S_L^{-}S_1^{+} + \alpha^{-L}S_L^{+}S_1^{-} + \frac{\alpha + \alpha^{-1}}{2}S_L^{z}S_{1}^{z}
\label{rr9}
\ee
where
\be
\alpha=\sqrt{\frac{s}{r}}, \qquad S^{+}_{L+1}=\alpha^{L}S^{+}_{1}, \qquad S^{-}_{L+1}=\alpha^{-L}S^{-}_{1}
\label{rr10}
\ee
To obtain the TASEP, we consider the limit $s=0$, which yields the effective Ising model with particular boundary conditions. For the SSEP one selects $s=r=\tfrac{1}{2}$ and the Hamiltonian of the XXX chain emerges. The stationary solution of \eq{rr8} with $E=0$ corresponds to the steady state whose eigenfunction can be represented in the form of a matrix product. 

Note that the XXZ spin chain for an open ASEP is non-Hermitian because of the boundary terms. This makes the Bethe Anzatz solution of this model more difficult, contrary to the periodic ASEP when the standard BA tools can be applied. Nevertheless, a slight modification of BA provides an exact solution for the spectrum for open ASEP as well \cite{de2005bethe}.

\subsubsection{Inhomogeneous stochastic models}

The homogeneous models allow for a straightforward generalization to the inhomogeneous stochastic ASEP and its TASEP limit when the jump rates for each particle are different \cite{arita2013matrix, ayyer2014inhomogeneous, kim2023schubert}. In the spin chain language, we consider the inhomogeneous  XXZ spin chain with  local spectral parameters. Instead of the local Hamiltonian that governs a Markovian evolution of the probability of the homogeneous models in continuous time, the inhomogeneous models are described in terms of the transfer matrix
\be
T(x|\vec{z})=\Tr_0\Big[R_{0,L}\Big(\frac{x}{z_L}\Big)R_{0,L-1}\Big(\frac{x}{z_{L-1}}\Big)\dots
R_{0,1}\Big(\frac{x}{z_1}\Big)\Big]
\ee
where $R$ is the $R$-matrix of the corresponding model. In the particular gauge, it is considered with respect to unitarity, which is necessary for the probabilistic interpretation. The spectral parameter $x$ is generically a complex variable, and the inhomogeneities $z_i$ are also arbitrary numbers ${\mathbb C}$. The transfer matrix commutes for different values of the spectral parameter $x$ due to the Yang-Baxter equation.

Instead of a continuous time evolution for the homogeneous TASEP only, the discrete-time evolution is available now with the updating rules dictated by the Markov matrix:
\be
P(t+1) = M(x|\mathbf{z})P(t)
\ee
where the Markov matrix $M(x|\mathbf{z})$ is the normalized transfer matrix
\be
M(x|\mathbf{z})= T(x|\mathbf{z})T(z_1|\mathbf{z})^{-1}
\ee
In the limit $z_i\rightarrow 1$ the continuous time TASEP model is restored and the Markov matrix in this case is $M'(1)$.

The analog of the steady state can also be defined in the inhomogeneous case. It corresponds to the state
\be
M\ket{S}=\ket{S}
\ee
If one introduces the vector
\be 
\mathbf{v}(z)=\left(\begin{array}{cccc} z \\ 1 \end{array}\right)
\label{vec}
\ee
which satisfies the associative Faddeev-Zamolodchikov quadratic algebra
\be
R_{12}(z_2/z_1)v_1(z_1)v_2(z_2)= v_2(z_2)v_1(z_1)
\ee
Then the steady state can be written as
\be
\ket{S} =v_1(z_1)v_2(z_2)\dots v_L(z_L)
\ee
The steady state for a generic open lattice with boundary conditions can be presented in a similar matrix product form
\be
P_{\rm steady}\propto \bra{W}A(z_1) \dots A(z_N)\ket{V}
\ee
where $A(z)$ is the vector
\be 
\mathbf{A}(z)=\left( \begin{array}{cccc} E(z) \\ D(z) \end{array}\right)
\label{vec2}
\ee
obeying the quadratic Faddeev-Zamolodchikov algebra
\begin{equation}
    [E(z_1),E(z_2)]=0, \qquad  [D(z_1),D(z_2)]=0, \qquad \frac{z_1}{z_2}D(z_1)E(z_2)= D(z_2)E(z_1)
    \end{equation}
and $z_i$ parameterize a local jump rate.

The inhomogeneous ASEP and TASEP models can be solved using the corresponding Bethe ansatz equations. Let us emphasize that the presence of inhomogeneities allows one to construct non-local conservation laws for these stochastic processes, in close analogy with the non-local conservation laws of inhomogeneous spin chains. In the TASEP model, these conserved quantities can be obtained as residues of the transfer matrix at the positions of the inhomogeneities. These non-local Hamiltonians play a central role in the integrable structure of the stochastic process.

\section{DSSYK and dual dilaton gravity}
\label{sect:3}

\subsection{SYK and DSSYK models}

The conventional SYK model  \cite{sachdev1993gapless,kitaev2015simple} is described as the quantum dot Hamiltonian 
\be
H_{SYK}=i^{p/2}\sum_{i_1\dots i_p} J_{i_1 \dots i_p} \Psi_{i_1} \dots  \Psi_{i_p},\qquad \la J_{i_1\dots i_p}^2 \ra=\frac{J^2}{\lambda C^{N}_{p}}
\ee
with $p=4$ all-to-all interaction term, fermions $\Psi_i$ $i=1\dots N$ can be Majorana or complex.

For Majorana fermions, the low-temperature partition function
\be
Z(\beta) =\int dE\rho(E) e^{-\beta E}
\ee
can be evaluated in the low-energy regime and is dominated by the soft Goldstone mode that emerges from spontaneous breaking of the invariance of time reparameterization $\tau \mapsto f(\tau)$. The low-energy Schwarzian action for this soft mode reads as
\be
I_{SCH}=-C \int_0^{\beta} d \tau \Bigg\{\tan \frac{\pi f(\tau)}{\beta}, \tau \Bigg\}
\ee 
and the corresponding spectral density is 
\begin{equation}
    \rho(E)= C \sinh (2 \pi \sqrt{2CE})
\end{equation}
where $C$ is constant. The Schwartzian action can be presented in the form of Liouville quantum mechanics \cite{bagrets2016sachdev} which we shall use later. 

For complex fermions, a second soft mode, the phase of the fermion, survives at low energies. Thus, the low-energy effective action contains a pair of soft Goldstone degrees of freedom \cite{Gu_2020}: the Liouville mode and the free phase mode
\begin{equation}
S_{Complex}= \int dt \left(c_1 (\partial_t \chi)^2 +c_2 \mathrm{Sch}{f(t)}\right)
\end{equation}
When an effective electric field is added, the two soft modes become coupled. Remarkably, the sum of the free field and the quantum Liouville mode is precisely what appears in the KPZ limit of ASEP in the random-walk representation.

The limit
\be
N\rightarrow \infty, \qquad p \rightarrow \infty,\qquad \lambda= \frac{2p^2}{N} = \mathrm{const}
\ee
defines the DSSYK model. In what follows, it will be useful to introduce the parameter $q=e^{-\lambda}$. The DSSYK model is exactly solvable at all energies in the chord basis \cite{berkooz2019towards}, and its spectral density is given by
\begin{equation}
    \rho(E)dE= \frac{d \theta}{\pi}(q;q)_{\infty} (e^{2i\theta};q)_{\infty}(e^{-2i\theta};q)_{\infty}
\end{equation}
The Liouville Hamiltonian describing the Schwartzian mode becomes the discrete Liouville Hamiltonian
\begin{equation}
    H_{DSSYK}=\sqrt{1-e^{-2x}}\cos p,\qquad [x,p]=\lambda
\end{equation}

Upon averaging over the couplings, the partition function can be written in terms of the $q$-boson algebra
\be
Z=\bra{0}e^{-\beta T}\ket{0}, \qquad T=a^{\dagger} +a , \qquad aa^{\dagger} -qa^{\dagger}a=1
\ee
For the moments in the chord basis, we have
\be
m_k=\left<\Tr(H^k)\right>_J=\bra{0}T^k\ket{0}=\sum_{\text{over $\tfrac{k}{2}$ chords}} q^{\text{\# of chord intersections}}
\ee
The element in the chord  basis $\ket{n}$ in DSSYK is enumerated by the number of open chords, and the effective DSSYK Hamiltonian acts in the chord space as a tridiagonal matrix
\be
H= \frac{J}{\sqrt{\lambda}} (\alpha + \alpha^+)
\ee
where 
\be
\alpha=\sum_{l\ge 0} \sqrt{[l+1]_q}\; \ket{l} \bra{l+1}, \qquad  [l]_q= \frac{1-q^l}{1-q}
\ee
The eigenvalue equation in the chord basis takes the form of the hopping problem
\be
E\Psi_l(E)= \frac{J}{\sqrt{\lambda}} \left(\sqrt{1-q^{l+1}}\, \Psi_{l+1}(E) + \sqrt{1-q^{l}}\, \Psi_{l-1}(E)\right)
\ee
and the spectrum reads as 
\be
E(\theta)= \frac{2J\cos \theta}{\sqrt{\lambda(1-q)}}
\ee
The eigenfunctions are the $q$-Hermite polynomials
\be
\Psi_l(\mu) = \sqrt{(q;q)_{\infty}}(e^{2i\theta};q)_{\infty} \frac{H_l(\mu|q)}{\sqrt{2\pi (q;q)_l}}
\ee
and $\mu=\cos\theta$. 

\subsection{Dual bulk 2D gravity}

The dual $\sin$-dilaton gravity of the DSSYK model has action
\begin{equation}
    S=-\frac{1}{\lambda} \int_M d^2 x \sqrt{g}(\Phi R +2 \sin \Phi) +
     2 \int_{\delta M}dx \sqrt{|\gamma|}  \Big(\Phi_0K + \Phi\Big(K-1\Big)\Big)
\label{eq:action}
\end{equation}
where the second integral corresponds to the boundary contribution. The gravitational coupling constant is defined as
\begin{equation}
    G_N=4\pi \lambda
\end{equation}
hence, $q\rightarrow 1$ corresponds to the weak gravity limit, while the limit $q\rightarrow 0$ gives the strong coupling regime. In the $\lambda \rightarrow 0$ limit at the lower spectral edge $E\rightarrow -E_0$ the action \eq{eq:action} reduces to the JT gravity 
\be    
S_{JT}= \int_M d^2x \sqrt{|g|} \Big(\Phi_0R + \Phi\Big(R+\frac{2}{l^2}\Big)\Big) + 2 \int_{\delta M}dx \sqrt{|\gamma|}  \Big(\Phi_0K + \Phi\Big(K-\frac{1}{l}\Big)\Big)
\ee
 JT gravity can be represented  as the statistical model of closed curves in the hyperbolic plane with a chemical potential for the area \cite{kitaev2018soft,stanford2020finitecutoffjtgravityselfavoiding}
\begin{equation}
    Z_{JT}(\beta)= \int dE\rho(E)e^{-\beta E} = \sum_{\substack{\rm curves~of \\ \rm fixed~length }}e^{\gamma\, \text{Area}}
\end{equation}
This partition function can be interpreted as the partition function of a polymer in an external field, a particle in a magnetic field, or a spinful particle moving on a hyperbolic plane. The latter interpretation is particularly natural from the holographic perspective. In this picture, the JT gravity partition function can be viewed as a transition amplitude in $AdS_2$. At the upper edge of the spectrum, $E \rightarrow +E_0$, the dual gravitational description reduces to a $dS_2$ geometry \cite{susskind2022sitter, narovlansky2023double, verlinde2024double,okuyama2025sitterjtgravitydoublescaled}.

The chord basis can also be identified in JT gravity \cite{berkooz2023quantum}. The starting point is the realization of the two-dimensional phase space of JT gravity in terms of the geodesic length between two boundaries and its conjugate momentum \cite{harlow2020factorization}. In the appropriate limit, the effective DSSYK Hamiltonian reduces to the Liouville Hamiltonian of JT gravity, while the length operator is proportional to the chord-number operator \cite{lin2022bulk}. The correspondence can be established as follows. The DSSYK Hamiltonian can be written in the form \cite{lin2022bulk}
\be
H= \frac{J}{\sqrt{\lambda(1-q)}}\left(D\sqrt{1-q^{n}} + \sqrt{1-q^{n}}D^+\right)
\ee
where $D^+=e^{-ip}$.
\be
D^+ \ket{n}=\ket{n+1}, \qquad D \ket{n}=\ket{n-1}, \qquad D\ket{0}=0
\ee
Let us introduce the rescaled length and momentum
\be
l=l_f\lambda n, \qquad k=\frac{p}{l_f\lambda}
\ee
and in the triple scaling limit: $\lambda \rightarrow 0$, $l\rightarrow \infty$, $\frac{e^{-\frac{l}{l_f}}}{2\lambda^2}=e^{-\frac{\tilde{l}}{l_f}}=\mathrm{const}$, the Liouville Hamiltonian for JT gravity is recovered
\be
H_{JT}-E_0= 2\lambda \Bigg(\frac{l_f^2k^2}{2} + 2 e^{-\frac{\tilde{l}}{l_f}}\Bigg) + O(\lambda^2)
\ee
upon the identification
\be
l_f=l_{AdS}, \qquad 2J\lambda= \frac{1}{l_{AdS}\Phi_b}
\ee
where $\Phi_b$ is the boundary value of the dilaton.

%\subsection{ 3d dS space interpretation of the bulk}

The chord basis is also identified in the 3D dS space-time. It was suggested in \cite{narovlansky2023double} that the de Sitter space in $(2+1)$ spacetime with massive observers is equivalent to the pair of DSSYK models $(R,L)$ with additional constraint for the corresponding Hamiltonians $H_R=H_L$. The (2+1) dS is described by $SL(2,C)$ CS theory with the action
\be
S=\frac{ik}{4\pi}\int \Tr\left(A_{+}dA_{+} +\tfrac{2}{3} A_{+}^3\right) - \frac{ik}{4\pi}\int \Tr\left(A_{-}dA_{-} + \tfrac{2}{3} A_{-}^3\right)
\ee
where 
\be
k=\frac{1}{2G_N}, \quad A_{+}^a= \omega^a +ie^a, \quad  A_{-}^a= \omega^a ie^a,
\ee
and $e^a$ is the triad, while $\omega^a$ is the spin connection $SO(2,1)$.

A massive observer in 3D de Sitter space is described by the gravitational Wilson line $L_A$ which yields the 3d gravitational Hamiltonian
\be
H=\frac{L_A}{\sqrt{1-q}}
\label{ds}
\ee
where $L_A$ is the proper geodesic length. The Wilson line induces the conical defect in 3d
\be
L_A\ket{\alpha}=2\sin(\pi \alpha)\ket{\alpha}
\ee
hence, comparing it with the DSSYK Hamiltonian $H_{DSSYK}= \frac{2 \cos\theta}{\sqrt{1-q}}$ the identification $2\pi\alpha=\pi -2\theta$ can be made. 

The chord basis has been identified in the dS space as follows \cite{verlinde2024double}. The Hamiltonian in de Sitter space can be presented in terms of $q$-oscillators 
\be
H_{dS}=a + a^{+}
\ee
upon introducing the Fock basis $(X,Y,Z)$ in the Teichmueller space with the skein relations
\be
XY=1+qZ^2, \qquad ZY=qYZ,\qquad ZX=q^{-1}XZ
\ee
In terms of the Fock coordinates, the $q$-oscillator operators read as
\be
\sqrt{1-q}a^+=q^{1/4}Z^{-1}Y,\qquad  \sqrt{1-q}a=q^{1/4}XZ^{-1}
\ee
The parameter $q$ at the dS side is identified as follows
\be
q=e^{-\frac{2\pi}{k}}= e^{-4\pi G_N}
\ee
The DSSYK--dS duality states that the DSSYK quantum model describes a probe propagating in dS space.

\subsection{Matrix model picture}

Let us briefly discuss the matrix-model representations underlying this class of systems. The matrix model for JT gravity/SYK was developed in \cite{saad2019jt} using the topological recursion. It correctly reproduces the all-genus expansion of the moduli space of complex structures. A generalization to the DSSYK case was proposed in \cite{jafferis2023jtgravitymattergeneralized}. For the model with additional matter, it can be formulated as a two-matrix model, or alternatively as an ETH one-matrix model with a nontrivial potential encoding the measure associated with the $q$-Hermite polynomials. The transfer matrix $T$ is identified with a random Hermitian matrix, and
\begin{equation}
    \bra{0}T^k\ket{0}= \left<\Tr M^k\right>= \frac{\int dM \Tr M^k e^{-N\Tr V(M)}}{\int dM e^{-N\Tr V(M)}}
\end{equation}
where  potential is the sum of Chebyshev polynomials
\begin{equation}
    V(M)=\sum_{k=0}^{\infty} (-1)^{k-1}q^{\frac{k(k+1)}{2}}
    \Big(\frac{1}{k}T_{2k}(M) - \frac{1}{k+1}T_{2k+2}(M)\Big)
\end{equation}
At large $N$ one can expand the moments in sum over topologies and the genus-zero term reproduces $\bra{0}T^k\ket{0}$ in the DSSYK model. Later, we shall focus on the TASEP model at $q=0$ when the potential in the matrix model is Gaussian, $V(M)= \Tr M^2$. 

The Laplace transform yields the matrix model representation for the generating function, which is the resolvent for trivial boundary conditions. If the boundary conditions are nontrivial, one has to insert the operator corresponding to the EOW branes in the matrix integral. There are two points of possible confusion:
\begin{itemize}
    \item How is the deterministic matrix $T$ identified with the random matrix $M$?
    \item How can the KPZ scaling known in the late-time asymptotics of TASEP be recognized in the Gaussian matrix integral?
\end{itemize}

Possible interpretations have been proposed in \cite{gorsky2024kpz}. The transfer matrix $T$ can be viewed as the mean-field representation of a random matrix $M$ obtained by tridiagonalization following the Edelman--Dumitriu procedure \cite{dumitriu2002matrix}. Upon tridiagonalization, which corresponds to a transition to the Krylov basis \cite{balasubramanian2023tridiagonalizing}, the matrix remains random, with a modified distribution of its matrix elements. It becomes deterministic only in the mean-field limit. Moreover, contrary to the commonly held view, the matrix $M$ in the Gaussian model is described at the spectral edge not by the Airy operator, but by the stochastic Airy operator \cite{ramirez2011beta}. On the other hand, the stochastic Airy operators at late time under proper scaling fits with the late time asymptotics of the stochastic heat equation \cite{borodin2016moments, gorin2018kpz} which is the KPZ equation under the Cole-Hopf transform. We shall discuss this point in some detail later.

\subsection{Cumulants in DSSYK and $T(1,q)$ Tutte polynomials}

A slightly different view on the partition function of DSSYK can be developed if the cumulants $k_n(q)$ are considered instead of moments of the DSSYK Hamiltonian. They are defined as follows
\begin{equation}
    \sum_{n\geq1}k_n(q) \frac{z^n}{n!}= \log  \sum_{n\geq0}m_n(q) \frac{z^n}{n!}
\end{equation}
 If we introduce for any finite set $V$ the lattice of partitions $P(V)$ and denote $P(n)=P(1,\dots,n)$, the moments and cumulants are related
 \begin{equation}
     m_{n}(q)= \sum_{\pi \in P(n)}k_{\pi}(q) \qquad 
     k_{n}(q)= \sum_{\pi \in P(n)}m_{\pi}(q)\mu(\pi,\hat{1})
 \end{equation}
where $\mu(\pi,\hat{1})=(-1)^{N(\pi) -1}(N(\pi) -1)!$ is the Mobius function and $N(\pi)$ is the number of blocks in partition $\pi$.

Remarkably, the cumulants of the $q$--Gaussian distributions are related to the Tutte polynomials of the crossing graph, which can be attributed to any chord diagram as follows. The chord corresponds to the vertex in the crossing graph, while the edge between two vertices exists if the chords intersect. Therefore, the graph $G$ is attributed to each chord diagram, and in \cite{josuat2013cumulants} it was observed that the cumulants are related to the Tutte polynomials $T(x,y)$ of the crossing graphs $G(\sigma)$
\begin{equation}
    k_n(q)=(q-1)^n\sum_{\sigma \in M(2n)} T_{G(\sigma)}(1,q)
    \label{tutte}
\end{equation}
where the sum runs over the crossing graphs corresponding to chords (pairings) of $2n$ elements. In particular, $T_{G}(1,0)$ corresponding to TASEP, and $T_{G}(1,1)$ corresponding to SSEP have clear-cut counting interpretations -- see, for instance, \cite{welsh2000potts}.

The Tutte polynomial $T(x,y)$  coincides with the partition function of the $Q$-Potts model if the pair $(x,y)$ obeys the relation 
\begin{equation}
    (1-x)(1-y)=Q
\end{equation}
see \cite{welsh2000potts} for a review, When the $Q$-Potts model is considered on the discretized fluctuating surface, a bit schematic expression for the partition function reads as 
\begin{equation}
    Z_{Potts+gravity}(\beta,Q)=\sum_{{\rm graphs}~G}T_{G}(x,y|Q,\beta)) e^{-S_{gravity}(G)}
\end{equation}
That is, we get the representation of  DSSYK partition function in terms of cumulants
\begin{equation}
    Z_{DSSYK}(\beta)= \exp \left(\sum_{n\geq 1}k_n(q) \frac{\beta^n}{n!}\right)
\end{equation}
Using the relation (\ref{tutte}) the result is similar to the Potts model on the fluctuating geometry represented by the ensemble of crossing graphs. Note that DSSYK corresponds to $Q=0$ Potts model coupled to gravity since  $x=1$ in the Tutte polynomial.

The cumulants themselves in DSSYK have the physical meaning of the correlators of the geodesic lengths or the higher Krylov complexities defined in terms of the Krylov chain for the Hartle-Hawking states as follows 
\begin{equation}
    k_n(q)=\sum_l l^n |\phi_l|^2
\end{equation}
where $k_1$ is the conventional Krylov complexity. The first  cumulants in DSSYK were perturbatively evaluated in \cite{bhattacharjee2023krylov}.

\section{ASEP versus dual gravity and TASEP as strong gravity limit}
\label{sect:4}

In this Section we compare different aspects of stochastic models with the DSSYK/dual gravity. Later, we extend the discussion initiated in \cite{watanabe2025jt, okuyama2026asepdssykdualitystrangecorrelator} to the strong gravity TASEP limit using the relation with the growth models.

\subsection{Transfer matrix vs transfer matrix}

As we outlined above, the transfer matrix in the ASEP model $T_{ASEP}=D+E$ is expressed in terms of the $q$-oscillator algebra $DE-qED=D+E$ --  see \eqref{res}. Therefore, the ASEP model, exactly like the DSSYK one, may be viewed from the perspective of a chord basis that represents the Fock states. In this notation, states $\ket{n}$ are defined as follows: a chord diagram cut at an arbitrary point can be unfolded into a straight line, along which intersecting segments corresponding to the chords are drawn. Then the state $\ket{n}$ is the state in which there are exactly $n$ open chords. With this definition, creation and annihilation operators $a$, $a^{\dagger}$ naturally arise and can be interpreted as adding and removing an open chord, respectively

In terms of the creation and annihilation operators, it reads as follows. 
\begin{equation}
T_{ASEP} = \frac{2}{1-q} I + \frac{1}{\sqrt{1-q}}(a+a^\dagger) 
\label{T}
\end{equation}
where $I$ is an identity matrix, and $a, a^\dagger$ is defined by their action on the Fock basis $\ket{n}\}_{n\geq 0}$
\begin{equation}
    \begin{array}{c}
         a\ket{n} = \begin{cases}
             \sqrt{\frac{1-q^n}{1-q}}\ket{n-1}, \quad n\geq 1\\
             0, \quad n=0
         \end{cases} \medskip \\       
a^{\dagger} \ket{n} =\sqrt{\frac{1-q^{n+1}}{1-q}} \ket{n+1}          
    \end{array}
\end{equation}
and satisfy the $q$-oscillator algebra $aa^{\dagger}-qa^{\dagger}a = 1$.

Consider a centered and scaled version of $T_{ASEP}$ (see Eq. \eq{T}) $X = \frac{1-q}{2}(T-\frac{2}{1-q}I)$, that is, in terms of $a$ and $a^\dagger$
\begin{equation}
X = \frac{\sqrt{1-q}}{2}(a+a^\dagger).    
\end{equation} 
It is obvious that all odd moments of $X$ vanish: 
\begin{equation}\label{odd}
\bra{0}X^{2k+1}\ket{0} = 0    
\end{equation}
However, even moments are not so trivial. Expand $(a+a^\dagger)$ into words of length $2k$. Only words with equal numbers of $a$ and $a^{\dagger}$ can survive between $\bra{0}$ and $\ket{0}$. Now we may normally order each word; every time one commutes $a$ past $a^{\dagger}$ it contributes with the factor $qa^{\dagger}a+1$. In vacuum expectation, the surviving scalar contributions are naturally indexed by pairings $\pi$: each occurrence of the term $1$ corresponds to a contraction of one annihilation operator with one creation operator, while each pass-through via $qa^{\dagger}a$ contributes a factor of $q$. The total number of such pass-through is exactly the number of crossings $\text{cr}(\pi)$ in the pairing. Therefore, the total contribution of the pairing $\pi$ is exactly $q^{\text{cr}(\pi)}$. Thus, we obtain the expression for even moments
\be\label{even}
\bra{0}X^{2k}\ket{0} = \left(\frac{1-q}{4}\right)^k \sum_{\pi \in \mathcal{P}_2(2k)}q^{\text{cr}(\pi)}
\ee
where the sum runs over all pairings of $\{1,2,\dots,2n\}$. 

Now, let us go from centered $X$ to initial $T_{ASEP}$. It is obvious that $T^m  = \left(\frac{2}{1-q}\right)^m(I+X)^m = \left(\frac{2}{1-q}\right)^m \sum\limits_{j=0}^m \binom{m}{j}X^j$. Putting this into vacuum expectation and substituting \eqref{even} and \eqref{odd}, we obtain the expression of the ASEP transfer matrix via the number of crossings in the chord diagrams
\begin{equation}
    \bra{0}T_{ASEP}^{m}\ket{0} = \left(\frac{2}{1-q}\right)^{m}\sum\limits_{j=0}^{\big[\frac{m}2\big]}\binom{m}{2j} \left(\frac{1-q}{4}\right)^j \sum_{\pi \in \mathcal{P}_2(2j)}q^{\text{cr}(\pi)}  
\label{DSSYK-Catalans}
\end{equation}
This expression coincides with \eq{ASEP-Catalans0}. To avoid confusion, let us note that the notation $\bra{0}...\ket{0}$ in \eq{DSSYK-Catalans} means averaging with vacuum states in terms of occupation numbers, while $\bra{e_1}...\ket{e_1}$ in \eq{ASEP-Catalans0} means averaging with boundary vectors $e_1 = (1,0,...,0)$. It is worth mention the work \cite{schouten} where the higher DSSYK correlators are constructed using the von Neumann algebraic quantum group.

Two transfer matrices $T_{DSSYK}$ and $T_{ASEP}$ are identical up to a unit matrix, and a few comments are in order. First, note that $T_{DSSYK}$ provides the shift in the Euclidean time direction, while in the ASEP the Euclidean time has the meaning of the discrete or continuum space-like coordinate in ASEP and KPZ correspondingly. Hence $T_{ASEP}$ provides the shift in the space-like coordinate in the stochastic model. For instance, the steady state in ASEP corresponds to the time-dependent matrix elements in DSSYK. The proper objects to compare are the resolvents of the transfer matrices
\begin{equation}
  G(s)=  \bra{0}\frac{1}{T_{DSSYK}-s} \ket{0} \quad \Longleftrightarrow \quad  \Omega(s)=\bra{e_1}\frac{1}{T_{ASEP}-s}\ket{e_1}
\end{equation}
These resolvents between the different coherent states allow to identify the critical behavior. 

The limit of TASEP occurs at $q\rightarrow 0$ with a hopping only in one direction. The system remains discrete in this limit and the interpretation in DSSYK is as follows. In DSSYK, the number of fermions in the interaction term  at $q\rightarrow 0$ has to be large enough
\begin{equation}
    q=\exp( - \frac{p^2}{N}) \quad \Rightarrow \quad p^2 \gg N
\end{equation}
in the dual gravity
\begin{equation}
    q=\exp(-G_N) \quad \Rightarrow \quad G_N \to \infty
\end{equation}
Therefore, we deal with a limit of strong gravity, and the sum over chords in the expressions for moments $m^k$ is dominated by chords without intersections.

Physically TASEP corresponds to strong gravity and is therefore of great interest. In following sections, we shall focus on the aspects of TASEP model as the model of growth and discuss its interrelations with other statistical systems. Such mappings will suggest the interpretation of a 2d space-time as the emergent phenomenon in the spirit of \cite{vanraamsdonk2010commentsquantumgravityentanglement, swingle2012entanglement}.

It is worth making the following remark. The partition sum in ASEP, $\bra{W}(E+D)^N\ket{V}$ (or the associated generating function) loses to some extent the stochastic origin of the model and is not informative enough. The probability $P(t,\tau_1, \dots ,\tau_N)$ which involves dynamical variables and corresponds to the matrix element of the random operators involves additional information. At the DSSYK side such a matrix elements correspond to the correlators involving the powers of Liouville field in Liouville or discrete Liouville QM.

\subsection{Liouville vs Liouville}

Let us discuss the physical origin of connections between the models. Recall that the Liouville model appears both in the triple scaling limit of DSSYK to SYK, in the bulk gravity, and in the KPZ limit of the ASEP model. Is it the same Liouville or just an accidental coincidence that occurs? 

To discuss Liouville in KPZ recall that the height function obeying ASEP in the continuum case is represented as the sum $h(i,t) = n_i +m_i$ and the evolution along the "$n$" and "$m$" directions is different. It is a free Brownian motion in the "$m$" direction and Liouville quantum mechanics along the "$n$" direction. The weights are induced by the infinite representation of the $q$-boson algebra, the parameter $q$ defines the weight for the random walk via the factor $[n_i]_q$.

As noted above, in the steady state for any initial condition $h(x,0)$, the difference $h(x,t)-h(0,t)$ converges to $H(x)$, which is the sum of two independent random (quantum) fields
\begin{equation}
    H(x)= W(x)+ X(x)
\end{equation}
The $W(x)$ corresponds to the free Brownian motion while introducing $X(x)=U(x)-U(0)$ it was found that $U(x)$ obeys the Liouville quantum mechanics on $x\in [0,L]$ \cite{barraquand2021steady}. 

We emphasize that the Liouville field emerges in the random path representation for the state space with the map
\begin{equation}
    \text{state $(\tau_1,\dots ,\tau_N)$ in the ASEP model} \quad \Longleftrightarrow \quad \text{weighted path in $Z^2$}
\end{equation}
That is, the Liouville dynamics occurs in the state space of the ASEP model and is involved in the evolution of one path to another. If we do not take the limit to KPZ, then the Liouville action is replaced by the discrete Liouville action, which plays the same role..

Turn now to the SYK model, better thought of as the triple scaling limit of DSSYK. It involves a Schwartzian action in the low-energy sector that can be presented in the form of Liouville quantum mechanics \cite{bagrets2016sachdev}. The Liouville mode parametrizes the coordinate in the Hilbert space both in the triple scaling in DSSYK in the basis of open chords and in the dual gravity in the basis of the geodesic lengths. The basis's coordinates  in boundary and bulk theories are linked by a relation $l=\lambda n$. Hence, qualitatively, the common origin of the Liouville field is provided by the identification of the space of states in three theories.

As we argued above, besides the Liouville mode direction in the state space, there is the free Brownian direction
, and the  total transfer matrix of ASEP is  
\begin{equation}
    T_{ASEP}= a^{\dagger}\otimes 1 + a\otimes 1 + 1\otimes b^{\dagger} +1\otimes b
\end{equation}
and involves the sum of the Liouville Hamiltonian and a free boson  \cite{watanabe2025jt}
\begin{equation}
    a^{\dagger}+a\rightarrow \frac{2}{\sqrt{1-q}}- (1-q)^{3/2}D_{LQM},\qquad D_{LQM}=-\frac{d^2}{d\phi^2} +e^{-\phi} + O(\epsilon^{7/4})
\end{equation}
\begin{equation}
b^{\dagger} +b  \rightarrow \frac{2}{\sqrt{1-q}}- (1-q)^{3/2}D_{free},\qquad D_{free}=-\frac{d^2}{d\chi^2} + O(\epsilon^{7/4})
\end{equation}
Here $q=e^{-\lambda}$ and the operators $b^{\dagger},b$ are the operators of the $q$-deformed oscillator in the limit $q\rightarrow 1$.
Hence, to have the second  free boson Goldstone  mode one has to consider the complex fermions in DSSYK.

Let us make a remark about the recent conjecture on the universality of the KPZ equation \cite{Daviet_2025}. It was argued that the KPZ equation describes the dynamics of the Goldstone mode which emerges upon the spontaneous breaking of the time translation symmetry (TTS). The mechanism of TTS breaking can be different, but it involves the time-dependent order parameter in the ground states of the system, while the initial system in TTS invariant. For example, the vacuum solution can be periodic like in the time crystal, or linear in time in the case of a vacuum persistent current. However, it cannot be excluded that spontaneous symmetry breaking of another global symmetry could do the same job.

In \cite{Daviet_2025} it was argued that starting with the Schwinger-Keldysh action, the following effective action can be derived for the Goldstone mode, $h(x,t)$, where the cubic interaction term is the only one allowed by symmetry:
\begin{equation}
    S(h,\tilde{h})= \int dx\, dt\, \tilde{h}(\partial_t h - \partial_{xx}h +
    g(\partial_x h)^2) - D\tilde{h}^2
\end{equation}
It immediately provides the KPZ equation for the Goldstone field. It would be important to make a precise  link between the Goldstone nature of the Schwartzian action and the general Goldstone mode interpretation of the KPZ equation.

\subsection{Towards the stringy (polymer) picture and RG interpretation}

We have already mentioned that the Euclidean time in DSSYK corresponds to the coordinate variable in ASEP steady state. Let us present some semi-quantitative arguments supporting this identification. To this aim recall the duality between the stationary states in XXZ chain and the discrete-time Neumann system describing the particle on the $S^3$ or $AdS_3$ in the oscillator potential \cite{veselov1988integrable,moser1991discrete}.
\begin{equation}
    H_{\rm Neum}= \sum_{i} (p_i^2 +\omega_i^2q_i^2), \quad \sum_{i=1}^3 q_i^2=1 ~{\rm for}~S^2
\end{equation}

Two systems are equivalent if we identify the discrete coordinate variable in XXZ and the discrete-time variable in the corresponding Neumann system $ n_{XXZ}\leftrightarrow n_{\rm Neum}$  and $q^i(t_n) \leftrightarrow S^i_n$. The XXZ has the asymmetry parameter $\Delta=1/2(q+q^{-1})$ which enters the Neumann model in two ways. First, it is related to the asymmetry in the quadratic potential
\begin{equation}
    \Delta =\frac{\omega_1}{\omega_2}, \quad \omega_2=\omega_3
\end{equation}
second, the time step in the Neumann system $t_n= h n_{\rm Neu}$ is also related to XXZ anisotropy $\Delta -1\propto O(h^2)$. Both systems have the same spectral curve.

Returning back to the steady state in ASEP interpreted as the stationary state in the XXZ chain, we therefore have the link with the discrete-time Neumann system. To explain the origin of the Neumann system we turn to the classical string propagating in the $S^3$ or $AdS_3$ geometry with some quantum numbers. In \cite{frolov2003multi} the classical dynamics of the string was considered under the simple ansatz for the embedding coordinate $ Y_a(\sigma,\tau)$
\begin{equation}
    Y_a(\sigma,\tau)= y_a(\sigma)e^{i\omega_a \tau}
\end{equation}
and has been reduced to the Neumann model with continuous worldsheet time. Hence, to derive the XXZ model required for ASEP, we can consider the classical string with discretized worldsheet time propagating in curved geometry, Such identification of XXZ model via duality \cite{veselov1988integrable} in the holographic setting has been mentioned in \cite{gorsky2005spin}.

The interpretation of stationary states of XXZ in terms of the discretized classical string in the curved space raises the question whether the non-stationary solutions to ASEP are natural in the stringy setup as well. In other words, what is the meaning of "time" in ASEP/TASEP dynamics in the dual DSSYK and gravity side. First, note 
that the partition function in ASEP $\bra{W}(E+D)^N\ket{V}$ being just the normalization constant, loses to some extent the stochastic origin of the model. The probability $P(t,\tau_1, \dots ,\tau_N)$ involves dynamical variables and corresponds to the matrix element of time-dependent random matrix product formed by the generators $(D,E)$. The time evolution in ASEP has the meaning of rearrangement of one "word" written in terms of generators $(E,D)$ into the other. We conjecture that this process is an analog of the RG operator mixing with respect to the RG time. The well known example of such a behavior in SYM is the renormalization of the vector of long operators $\mathbf{O}$ representing the ensemble of words built from two generators when in the weak coupling regime the dilatation operator is identified with the XXX Hamiltonian \cite{minahan2003bethe}, $\tfrac{d\mathbf{O}}{dt}=H_{XXX}\mathbf{O}$. 

For long word operators, the RG mixing can be linked with the stringy $\sigma$-model description \cite{kazakov2004classical}. The key point is that the RG coordinate is related to the radial coordinate in $AdS$ geometry, hence the extension of the string along this coordinate is naturally interpreted as the RG flow of the operators with the corresponding quantum numbers.

Another argument supporting the stringy picture behind TASEP is inspired by the matrix model representation of 1/2 BPS Wilson loops in SYM \cite{Erickson_2000} and the twisted Schur index interpretation of the DSSYK partition function \cite{gaiotto2024sykschurdualitydoublescaled}. As we mentioned above, the measure in the ETH matrix model in the TASEP limit $q=0$, is Gaussian. In this case, the partition function acquires the following form in the matrix model representation
\begin{equation}
   \braket{0|e^{-\beta T}|0}=  \frac{\int dM \Tr e^{-\beta M} e^{-N\
   \Tr M^2}}{\int dM e^{-N\Tr M^2}}  
\end{equation}
Hence, the DSSYK partition function in the limit $q=0$ is identical to the matrix model representation of the circular Wilson loop in the SYM theory \cite{Erickson_2000}. In that case the exact answer at strong coupling can be obtained by the summation over the planar ladder diagrams without the self-interaction of gluons inside the circular circle. This is parallel to the TASEP limit of DSSYK when only the chord ladder diagrams without intersections contribute. In the SYM the strong coupling regime reproduces the holographically dual one for the loop via the regularized area of the minimal surface embedded in AdS geometry.  The summation of the ladder diagrams also reproduced the string breaking effects in the holographic bulk \cite{Zarembo_2001}, as well as the Wilson loop correlator and the local operator in the bulk \cite{Pestun_2003}.

The disk radial interior is interpreted as the "$n$"--direction in the space of states in the stochastic model according to the wormhole length interpretation. This identification is supported by the relation between $h(i)$ and $n_i$ in the random walk picture and the identification of $n_i$ as the discrete Liouville field. The boundary circle in ASEP/TASEP is the Wilson loop itself. This also fits with the evaluation of the twisted Schur index in \cite{gaiotto2024sykschurdualitydoublescaled}.

In summary, we suggest that ASEP/TASEP is related to the dynamics of the classical string in the curved manifold with the discretized worldsheet time. Informally, this viewpoint fits with the origin of the KPZ scaling to describe the discretized polymer in a random background. The time dependence of the stochastic model governs  the RG evolution of the Wilson/t-Hooft loops. The TASEP corresponding to strong coupling gravity involves the resummation of the ladder chord diagrams, while in the ASEP case the internal vertices are taken into account.  Certainly, this conjecture concerning the RG interpretation of the ASEP/TASEP "time" evolution governed by a Markov operator requires a more detailed analysis. In next Sections, we shall use the duality between ASEP/TASEP and the ballistic deposition heap models to show how the "RG-like" behavior in the growth problem can be visualized.

\subsection{Mappings: dictionary}

Here we summarize the relation between the ASEP-like family and the DSSYK-gravity dual pair. 
\begin{table}[ht]
\centering
\caption{ASEP--DSSYK Correspondence}
\small % Reduces font size slightly for better fit
\renewcommand{\arraystretch}{1.0} % Reduces spacing between table rows (default is 1.0)
\linespread{0.95}\selectfont % Reduces line height within multi-line paragraphs

\begin{tabular}{@{} l| p{0.68\textwidth} @{}}
\toprule
\textbf{Variable / Domain} & \textbf{ASEP-KPZ $\leftrightarrow$ DSSYK Correspondence} \\
\midrule
Time \& Space Dimensions & The spatial coordinate $x$ in ASEP-KPZ maps to the Euclidean time direction in DSSYK and dual gravity. The spatial length $L$ in ASEP equals the inverse temperature $\beta = 1/T$. \\ \midrule
Boundary Conditions & Parameters of the boundary flows along the spatial coordinate in ASEP-KPZ correspond to parameters of EOW branes in Euclidean time in DSSYK. \\ \midrule
State Space \& Fields & The state space in ASEP represents an ensemble of asymmetric random walks on a 2D lattice (Liouville potential in one direction, free Brownian motion in the second). The Liouville model and free boson represent the field content of complex fermions in DSSYK. \\ \midrule
Transfer Matrices & The transfer matrices $T_{\text{DSSYK}}$ with EoW branes and $T_{\text{ASEP}}$ coincide up to a shift, both partition functions interpreted as matrix elements $\bra{W} e^{\beta T} \ket{V}$. \\ \midrule
Probability \& Gravity Limit & The state probability $P(\tau_1, \dots, \tau_N)$ in ASEP corresponds to a specific correlator in DSSYK. Since $\lambda = 4\pi G_N$, KPZ corresponds to weak gravity while TASEP corresponds to strong gravity. The KPZ solution $Z_{\text{KPZ}} = e^{h_{\text{KPZ}}} = \langle e^{\xi + \phi} \rangle_{\text{DSSYK}}$ represents Brownian motion dressed by the Liouville gravity mode $e^{\phi}$. \\
\bottomrule
\end{tabular}
\label{tab:2}
\end{table}

\section{Gaussian Random Matrices, Stochastic Airy Operator and their $q$-generalizations}
\label{sect:5}

Matrix models are the standard approach to handling low-dimensional gravity. The are essentially two types of matrix model: large--$N$ matrix models providing the effective discretization of the manifold and the Kontsevich finite--$N$ model which describes the worldvolume theory on $N$ non-compact Lagrangian branes and discretized  moduli space.  

The double scaling limit in the large matrix model $N$ is developed near the spectral edge where the matrix $M$ is replaced by the differential operator with some potential term, the Airy operator for the Gaussian model. 
It can be tridiagonalized, bringing the matrix to the effective Krylov basis \cite{balasubramanian2023tridiagonalizing}. It is usually assumed that the operator is deterministic, hence there is no place for the stochasticity of gravity in the matrix model approach. However, a more detailed analysis yields the subleading random contribution suggesting the link with the stochastic behavior.

In this Section we recall the appearance of the stochastic Airy operator in the double scaling limit and its relation with the Stochastic Heat Equation (SHE). Then, we suggest the generalization of this scaling regime for the double scaling limits in $q$-Gaussian Matrix models.

\subsection{Gaussian ensemble and Stochastic Airy Operator}

The late-time behavior $t\sim e^{2S}\sim K$ of the Krylov operator complexity in Euclidean time is governed by the descending regime of the Lanczos coefficients $b_n$. The constructive example for the Krylov chain with descending $b_n$ is provided by the  Gaussian $\beta$ ensemble. It has been shown in \cite{dumitriu2002matrix}, that random matrices from Gaussian (O)rthogonal/(U)nitary/(S)ymplectic ensemble ($\beta=1$, $2$ and $4$ respectively) can be represented in tridiagonal form: 
\be
H_\beta=\frac{1}{\sqrt{\beta K}}\left(\begin{array}{ccccc}
N(0,2) & \chi_{(K-1)\beta} & 0 & 0  & \dots  \bigskip \\
\chi_{(K-1)\beta} & N(0,2) & \chi_{(K-2)\beta} & 0  & \bigskip  \\
0 & \chi_{(K-2)\beta} & N(0,2) & \chi_{(K-3)\beta} & \bigskip \\
0 & 0 & \chi_{(K-3)\beta} & N(0,2)  & \bigskip \\
\vdots & & & &  \ddots
\end{array}
\right),
\label{b01}
\ee
where $N(\mu,\sigma)$ and $\chi_{n}$ are normal and $\chi$-distributed random variables, respectively. The matrices $H_\beta$ have the same joint law of eigenvalues as the matrices of G(O/U/S)E: 
\begin{equation}
P(\{\lambda_i\})= \prod_{i<j}  \vert\lambda_i - \lambda_j\vert^{\beta}e^{- \beta \sum\lambda_i^2}
\label{jpdf}
\end{equation}
The ensemble of matrices $H_\beta$, referred to as the Gaussian $\beta$ ensemble, provides a matrix representation of Joint Probability Density Function (JPDF) as in \eq{jpdf}) for general $\beta$.

The emerging tridiagonal matrix can be interpreted as the Hamiltonian of hopping auxiliary particle propagating on the inhomogeneous discrete chain considered as a fixed target space). In \cite{balasubramanian2023tridiagonalizing} it was found that the tridiagonalization procedure developed in \cite{dumitriu2005eigenvalues} is the transition to the Krylov basis in the initial Hermitian matrix model. The Gaussian ensemble is not unique, and tridiagonalization can be constructed for any measure in the matrix model, which produces the specific hopping $b_n$ in the Krylov chain. If the potential in the initial matrix model is even, then all $a_n=0$.

The tridiagonal matrix allows one to interpolate between the random Lanczos coefficients at finite $\beta$ and the deterministic coefficients at $\beta=\infty$. Taking the limit $\beta \rightarrow \infty$, the matrix elements of the tridiagonal matrix become deterministic with $b_n=\sqrt{1-n/K}$ and $a_n=0$, where $n=1,...,K$ and $K$ is the matrix size.

In \cite{ramirez2011beta} it has been shown that the tridiagonal matrices $H_\beta$ from the Gaussian $\beta$ ensemble at arbitrary $\beta$ near the spectral edge are well approximated by the Stochastic Airy Operator "SAO$_\beta$" defined as follows:
\begin{equation}
\mathcal{H}_\beta = -\frac{d^2}{dx^2}+x+\frac{2}{\sqrt{\beta}}B'(x)
\end{equation}
where $B(x)$ is the standard Brownian motion ($B'(x)\equiv dB(x)/dx$ is the white noise). It has been shown that the largest rescaled eigenvalues of the $\beta$ ensemble in the Krylov basis converge in distribution to the largest points of the Airy process SAO$_{\beta}$ for any $k$ \cite{ramirez2011beta}
\begin{equation}
K^{2/3}\left(2-\lambda_{\beta,l}\right)_{l=1\ldots k} \underset{K\rightarrow \infty}{\to}( \Lambda_0(\beta) > \Lambda_1 (\beta)\dots> \Lambda_{k-1}(\beta))_{Airy} 
\end{equation}
Hence, in the limit $K\to\infty$ and $T>0$ the following holds:
\be 
\exp\left[-T K^{2/3}\left(2 I-H_\beta\right)\right]\underset{K\rightarrow \infty}{\to}\exp[-T\; \text{SAO}_\beta]
\ee
Simplifying, we get
\be 
\exp\left[T K^{2/3}H_\beta\right]\underset{K\rightarrow \infty}{\to}\exp[2T K^{2/3}-T\; \text{SAO}_\beta]
\ee
The evolution operator $e^{-t H_\beta}$ in the late-time behavior $t\sim e^{2S}\sim c K$, where the spectral edge has the dominant contribution, converses to the operator   
\begin{equation}
e^{c K H_\beta} \rightarrow \exp\left(2 c K-c K^{1/3}\;\text{SAO}_{\beta}\right).
\end{equation} 

One more useful representation of $SAO_{\beta}$ found in \cite{gorin2018kpz} involves the product of tridiagonal matrices in the $N\rightarrow \infty$ limit:
\be 
\lim\limits_{K\rightarrow\infty}\frac{1}{2}\left(\left(\frac{H_\beta}{2}\right)^{\left[T N^{2/3}\right]}+\left(\frac{H_\beta}{2}\right)^{\left[T N^{2/3}\right]-1}\right)= \exp\left(-\frac{T}{2}\; \text{SAO}_{\beta}\right)
\ee
where $[...]$ designates the integer part of the variable.

By combining the connection between the asymptotics of the matrix products and SAO$_{\beta}$ with the relationship between SAO$_{\beta}$ and SHE-KPZ, the authors of \cite{gorin2018kpz} showed that the Cole–Hopf solution of the stochastic heat equation in the half-space ($\beta=1$) and in the full-space ($\beta=2$), with initial droplet state condition, has the following explicit expression:
\be
Z(0,2t^3)e^{\frac{t^3}{12}}=\lim_{K\rightarrow\infty}\frac{K}{\beta}\left[\left(\frac{H_\beta}{2}\right)^{2[tK^{2/3}]}+\left(\frac{H_\beta}{2}\right)^{2[tK^{2/3}]+1}\right]_{1,1}
\label{g-s}
\ee
Eq. \eq{g-s} is the matrix-element version of the operator correspondence discussed above.

A more general connection between KPZ solutions and SAO$_{\beta}$ has been established in \cite{borodin2016moments} where the exact relation between the Cole-Hopf solution of KPZ and the generating function of correlators for the Airy process in $\beta=2$ was obtained
\begin{equation}
E_{KPZ}\left[e^{-uZ(T,0)} e^{\frac{T}{24}}\right]=  E_{Airy}\left[\prod_{k=1}^{\infty} \frac{1}{1+u \exp (Ca_k)}\right] 
\label{eq:bor}
\end{equation}
where $\frac{T}{2}$ is identified with $C^3$. At the "KPZ side" \eq{eq:bor} reads
\begin{equation}
E_{KPZ}\left[e^{-uZ(T,0)} e^{\frac{T}{24}}\right] =1 +\sum_{k=1}^{\infty}\frac{(-1)^L}{L!}\int_{0}^{\infty} dx_1 \dots \int_{0}^{\infty} dx_L \det[K_u(x_i,x_j)]_{i,j=1}^{L}
\label{eq:bor1}
\end{equation}
where the kernel is as follows 
\begin{equation}
K_u(x,x')= \int _{-\infty}^{+\infty} \frac{dy}{1 + u^{-1} \exp((T/2)^{1/3} y)}{\rm Ai}(x-y){\rm Ai}(x'-y)
\label{eq:bor1a}
\end{equation}
At the "Airy side", the determinantal representation gives:
\begin{equation}
E_{Airy}\left[\prod_{k=1}^{\infty} \frac{1}{1+u \exp (Ca_k)}\right] = 1+ \sum_{k=1}^{\infty}\frac{(-1)^L}{L!}\int_{0}^{\infty} dy_1 \dots \int_{0}^{\infty} dy_L
\det[ K_{Ai}(y_i,y_j)]_{i,j=1}^{L}   
\label{eq:bor2}
\end{equation}
with the Airy kernel
\begin{equation}
\label{eq:bor2a}
K_{Ai}(x,y)= \int_{0}^{\infty} {\rm Ai}(x+a){\rm Ai}(y+a)
\end{equation}

\subsection{ $q$-Gaussian Ensemble and its spectral measure} 

Here we discuss the generalization of tridiagonalization to the $q$-Gaussian matrix models. First, we recall the definition of the $q$-Gaussian distribution. For $-1<q<1$, the centered $q$-Gaussian distribution $\mu_q$ is the unique probability measure on $\mathbb{R}$ with moments
\be
\mathbb{E}[X^{2n+1}] = 0, \qquad \mathbb{E}[X^{2n}] = \sum_{\pi\in\mathcal{P}_2(2n)} q^{\mathrm{cr}(\pi)}
\label{eq:01}
\ee
where $\mathcal{P}_2(2n)$ denotes the set of pairings of $\{1,\dots,2n\}$ and $\mathrm{cr}(\pi)$ is the number of crossings of the pairing $\pi$. Namely, let $X_1,\dots, X_n$ be centered  $q$-Gaussian random variables with covariance
\be
\mathbb{E}[X_i X_j] = \delta_{ij}.
\ee
All odd moments vanish, and the even moments are given by
\be
\mathbb{E}[X_{i_1} X_{i_2} \cdots X_{i_{2k}}] = \sum_{\pi \in \mathcal{P}_2(2k)} q^{\mathrm{cr}(\pi)} \prod_{(a,b)\in\pi} \mathbb{E}[X_{i_a} X_{i_b}],
\ee
Define a "crossing" as a pair of pairs $(i,j)$ and $(k,l)$ with $i < k < j < l$. Each crossing contributes with one power of $q$ to the weight of pairing. In particular:
\begin{itemize}
\item[(i)] For $q=1$ all pairings contribute equally, recovering the classical Gaussian Wick theorem;
\item[(ii)] For $q=0$ only non-crossing pairings contribute, corresponding to free (semi\-circular) probability;
\item[(iii)] For $-1<q<1$ the moments continuously interpolate between these two regimes.
\end{itemize}
This $q$-Wick expansion is applied to all linear combinations of independent $q$-Gaussian variables including the $q$-deformed $\chi$ distribution appearing in the Lanczos tridiagonalization of $q$-Gaussian random matrices, which we shall use below. The measure $\mu_q$ admits an explicit density representation due to Bozejko, K\"{u}mmerer, and Speicher \cite{bozejko1997q}. A collection of independent random variables $\{X_{ij}\}_{i\le j}$ is called $q$-Gaussian random variables if each $X_{ij}$ is distributed according to $\mu_q$, with $X_{ij}=X_{ji}$.

We consider a $N\times N$ real symmetric $q$-Gaussian random matrix $M_q$ that has independently distributed matrix elements $X_{ij}$ for $i<j$ and $X_{ii}$ for $i=j$ where $\{X_{ij}\}_{i\le j}$ are $q$-Gaussian random variables with unit variance. For $-1<q<1$, the centered $q$-Gaussian distribution has the density
\be
f_q(x) = \frac{\sqrt{1-q}}{2\pi} \sqrt{4-(1-q)x^2} \prod_{n=1}^{\infty} \frac{1-q^n}{1-\frac{1-q}{4}x^2 q^n}
\ee
This distribution is symmetric: $f_q(x) = f_q(-x)$, and is supported on the interval $x \in \left[ -\frac{2}{\sqrt{1-q}}, \, \frac{2}{\sqrt{1-q}} \right]$.

Recall that the density of the spectral measure for a general $q$-Gaussian matrix is
\be\label{1}
\rho_q(x) = \frac{\sqrt{1-q}}{2 \pi} \sqrt{4-(1-q)x^2}\prod_{n = 1}^{\infty} \frac{1-q^{n}}{1-\frac{1-q}{4}x^{2}q^{n}}, \quad |x|\leq x^{*}(q): = \frac{2}{\sqrt{1-q}} 
\ee
with the spectral radius
\be
R(q) = x^{*} = \frac{2}{\sqrt{1-q}}, \quad R(q)^2 = \frac{4}{1-q}
\ee
We will use the "radial coordinates", i.e., reparametrization $a(x) = \frac{x^2}{R(q)^{2}} \in [0,1]$. Note that $\frac{1-q}{4}x^2 = \frac{x^2}{R(q)} = a(x)$. Using the previous notation, we may decompose \eq{1} into the "semicircle" and the "deformation" parts: 
\be
4-(1-q)x^2 = (1-q)(R^2(q)-x^2),\quad \frac{\sqrt{1-q}}{2 \pi} \sqrt{4-(1-q)x^2}  = \frac{1-q}{2\pi} \sqrt{R^2(q)-x^2}
\ee
The semicircle spectral density is
\be\label{semicircle}
\rho_{\text{sc,R}}(x) = \frac{2}{\pi R^2}\sqrt{R^2-x^2} \times \mathbb{I}_{|x| \leq R}
\ee
Note that if $R = R(q)$, the prefactor is $\frac{2}{\pi R^2(q)} = \frac{1-q}{2 \pi}$. Additionally, let us substitute $a(x)$ and denote a product:
\be
\Phi_q(x) = \prod_{n=1}^{\infty} \frac{1-q^{n}}{1-a(x)q^n}.
\ee
Therefore, the spectral density has split into two parts: the first is a semicircle one (for each fixed $q$, it is a spectral density for an ordinary Gaussian matrix, but with different $R$), and the second is a $q$--dependent deformation: 
\be
\rho_{q}(x) = \rho_{\text{sc,R(q)}}(x) \Phi_q (x)    
\ee
where the support endpoints are $\pm R(q)$. Note that $\Phi_q$ interpolates between $\prod (1-q^n)$ in the center to $1$ at the edge:
\be
\begin{array}{rcl}
\disp x = 0,\; a(0) = 0 & \Longrightarrow & \disp \Phi_q(0) = \prod_{n=1}^{\infty}(1-q^n) \\
\disp x = \pm R(q),\; a = 1 & \Longrightarrow & \disp \Phi_q(\pm R) = \prod_{n=1}^{\infty} \frac{1-q^n}{1-q^n} =  1
\end{array}
\ee
In the vicinity of the spectral edge $n=N-\ell$ with $\ell=\mathcal{O}(N^{1/3})$, one has the expansion
\be
\sqrt{1-q^{N-\ell}} \simeq \sqrt{1-q^N} - \frac{q^N\log q}{2\sqrt{1-q^N}}\ell
\ee

\subsection{Mean-field theory for $q$-Gaussian Ensembles in Krylov basis}

Let $v_1\in\mathbb{R}^N$ be some unit vector. Define a set of Krylov subspaces constructed using the operator $X_q$:
\be
\mathcal{K}_k = \mathrm{span}\{v_1, X_q v_1, X_q^2 v_1,\dots, X_q^{k-1} v_1\}
\ee
Orthogonalization of this sequence produces the Lanczos basis. The few first steps of the Lanczos algorithm applied to the vector $v_1$ are explicitly described below. Define recursively:
\be
\begin{array}{ll} \disp w_k = X_q v_k - b_{k-1} v_{k-1}, & a_k = \langle v_k, w_k\rangle, \medskip \\ 
\tilde w_k = w_k - a_k v_k, & b_k = \medskip |\tilde w_k|, \\ \medskip
\disp v_{k+1} = \frac{\tilde w_k}{b_k}, & ...
\end{array}
\ee
where we started with $b_0=0$. It is known that in the basis $\{v_k\}$ the matrix $X_q$ takes a tridiagonal form
\be
X_q =
\begin{pmatrix}
a_1 & b_1 & 0 & \cdots\\
b_1 & a_2 & b_2 & \ddots\\
0 & b_2 & a_3 & \ddots\\
\vdots & \ddots & \ddots & \ddots
\end{pmatrix}
\ee
The proof is the same as for the standard $q=1$ case and follows from the symmetry of $X_q$ and Gram--Schmidt orthogonalization of the Krylov sequence, which is defined by a three-term recurrence. For Gaussian matrices ($q=1$) this procedure yields the Dumitriu-Edelman tridiagonalization of the ensemble $\beta=1$. 

For diagonal elements, each $a_k=\langle v_k, X_q v_k\rangle$ is a linear combination of independent $q$-Gaussian variables and hence is itself $q$-Gaussian with variance $2$. The distribution of sub-diagonal entries changes and follows the $q$-deformed $\chi$ distribution whose definition is given below. 

Let $x_1,\dots, x_m$ be independent $q$-Gaussian random variables with density $f_q(x)$. Define the radial variable
\be
r=\sqrt{x_1^2+\cdots+x_m^2}
\ee
The random variable $r$ is said to have the $q$-chi distribution with $m$ degrees of freedom. The probability density of $r$ is
\be
\chi_{m,q}(r) = C_{m,q}\, r^{m-1} (1-q)^{m/2} \prod_{n=1}^{\infty} \frac{(1-q^n)^m} {\left(1+\frac{1-q}{4}r^2 q^n\right)^{m/2}}, \qquad r\ge0
\ee
where $C_{m,q} = \frac{2}{\Gamma_q(m/2)} \left(\frac{1-q}{4}\right)^{m/2}$ is a normalization constant that comes from the angular integration. Since each individual coordinate $x_i$ is bounded by $\frac{2}{\sqrt{1-q}}$, the maximum available radius $r$ is reached simultaneously by all coordinates within the radius $r$ lying in the interval $0 \le r \le \frac{2\sqrt{m}}{\sqrt{1-q}}$. Collecting all expressions, we get
\be
\chi_{m,q}(r) = \frac{2}{\Gamma_q(m/2)} \left(\frac{1-q}{4}\right)^{m/2} r^{m-1} (1-q)^{m/2} \prod_{n=1}^{\infty} \frac{(1-q^n)^m} {\left(1+\frac{1-q}{4}r^2 q^n\right)^{m/2}}
\ee
Using the standard $q$-Pochhammer definition $(a;q)_\infty = \prod_{n=0}^{\infty}(1-a q^n)$ we express the infinite products (note the shift of the index $n \to n+1$) as 
\be
\chi_{m,q}(r) = \frac{2}{\Gamma_q(m/2)} \left(\frac{1-q}{2}\right)^m r^{m-1} \left[ \frac{(q; q)_\infty^2}{\left(-\frac{1-q}{4}r^2 q; \, q\right)_\infty} \right]^{m/2}
\ee
In the Lanczos tridiagonalization of a $q$-Gaussian matrix, the $k$-th off-diagonal element $b_k$ corresponds precisely to $m = 2k$. Substituting $m = 2k$ into the density function $\chi_{m,q}(r)$ gives
\be
\chi_{2k,q}(r) = \frac{2}{\Gamma_q(k)} \left(\frac{1-q}{2}\right)^{2k} r^{2k-1} \left[ \frac{(q; q)_\infty^2}{\left(-\frac{1-q}{4}r^2 q; \, q\right)_\infty} \right]^k
\ee
The expected value $\mathbb{E}[b_k]$ is given by the first moment of this radial distribution:
\be
\mathbb{E}[b_k] = \int_{0}^{\frac{2\sqrt{2k}}{\sqrt{1-q}}} r\,  \chi_{2k,q}(r) \, dr = \frac{2}{\Gamma_q(k)} \left(\frac{1-q}{2}\right)^{2k} (q; q)_\infty^{2k} \int_{0}^{\frac{2\sqrt{2k}}{\sqrt{1-q}}} \frac{r^{2k}}{\left(-\frac{1-q}{4}r^2 q; \, q\right)_\infty^k} \, dr
\ee
To compute this integral, make a change of variables to make a connection with the $q$-Beta function. Let:
\be
u = -\frac{1-q}{4}r^2 q, \quad r = \frac{2}{\sqrt{1-q}\sqrt{-q}} u^{1/2}, \quad dr = \frac{1}{\sqrt{1-q}\sqrt{-q}} u^{-1/2} \, du
\ee
After some algebra, we get
\be
\mathbb{E}[b_k] = \frac{1}{\sqrt{1-q}}\, \frac{(q^k; q)_\infty}{(q^{k+1/2}; q)_\infty}
\label{eq:q-ratio}
\ee
Taking into account the definition of  $q$-Gamma function, $\Gamma_q(x) = \frac{(q;q)_\infty}{(q^x;q)_\infty}(1-q)^{1-x}$, the ratio in \eq{eq:q-ratio} gives us
\be
\mathbb{E}[b_k] = \frac{1}{\sqrt{1-q}}\, \frac{\Gamma_q\left(k + \frac{1}{2}\right)}{\Gamma_q(k)}
\ee
Using the expansion 
\be
\frac{(q^k; q)_\infty}{(q^{k+1/2}; q)_\infty} = 1 - q^k + q^{k+1/2} + O(q^{2k})
\ee
we find the leading-order behavior of $\mathbb{E}[b_k]$ in the limit $k\gg 1$
\be
\mathbb{E}[b_k] = \sqrt{\frac{1-q^k}{1-q}} \left(1 + O(q^k) \right)
\ee
Thus, the resulting non-stochastic (averaged) tridiagonal matrix sets the three-term recurrence
\be
x H_k(x|q) = \sqrt{1-q^{k+1}} H_{k+1}(x|q) + \sqrt{1-q^k} H_{k-1}(x|q)
\ee
which is precisely the recurrence relation for continuous $q$-Hermite polynomials that emerged above in the solution of ASEP.

We conjecture that the mean-field consideration of the $q$-Gaussian matrix ensemble suggests a natural extension of the matrix-product formulation of ASEP. In the homogeneous model, the partition function is generated by repeated multiplication of the same transfer operator,
\be
Z_N=\langle W|(D+E)^N|V\rangle,
\ee
where the matrices $D$ and $E$ satisfy the algebra $DE-qED=D+E$. In the $q$-Gaussian framework one may instead consider a \emph{non-homogeneous} ASEP in which each lattice site is associated with its own pair of matrices $(D_i,E_i)$, satisfying the local algebra
\be
D_iE_i-qE_iD_i=D_i+E_i,
\ee
while the corresponding local transfer operators $X_i=D_i+E_i$ are independent realizations drawn from the $q$-Gaussian ensemble. The homogeneous product $(D+E)^N$ is then replaced by the ordered product of random matrices, $\prod_{i=1}^{N}(D_i+E_i)$. Averaging over the ensemble of $q$-Gaussian transfer matrices produces, in the Krylov basis, the deterministic tridiagonal operator derived above, whose off-diagonal elements are given asymptotically by $b_k\simeq\sqrt{\frac{1-q^k}{1-q}}$. One can speculate that the dynamics of the non-homogeneous ASEP is related to random products of matrices drawn from the $q$-Gaussian distribution, thereby establishing a connection with non-homogeneous Krylov dynamics and disordered matrix product states.

\section{TASEP/ASEP and Heaps of Pieces}
\label{sect:6}

In \cite{haug2014generalized} it has been observed that the matrix ansatz emerging in combinatorics of locally free groups coincides with one of the admissible representations of the Matrix Product Ansatz (MPA) describing the ensemble of stationary states of interacting particle systems in TASEP. Although originally formulated in terms of abstract operators or matrices, it turns out that the MPA algebra can be interpreted as a combinatorial object of topological origin generated by relations compatible with the locally free semigroup. In this interpretation, the locally free semigroup provides a natural algebraic language to describe the steady state of TASEP. 

Here we go beyond the standard locally free group and define a one-parameter deformation of its generators. Specifically, we replace generators $g_i$ with graded elements $g_i \mapsto (q')^{i} g_i$, where $q'$ is a deformation parameter. Since the commutation relations are homogeneous, this transformation preserves all locality relations, while endowing the algebra with a nontrivial graded structure. For this reason, it is natural to speak of a graded locally free algebra rather than a group in the strict sense. This deformation enriches the combinatorial and probabilistic content of the model while retaining exact solvability. As it is shown below, the graded structure of a locally free group is complementary to ASEP-like interacting particle systems possessing some duality relation.

\subsection{TASEP Matrix Ansatz and "Mikado ordering" of Heaps of Pieces}
\label{sect:5-1}

The solution of TASEP was constructed in \cite{derrida1993asep} using the operator algebra with relations $DE= D+E$. Different infinite-dimensional representations are available, among which we pay attention to two most important ones for us. The first, which constitutes the $q$--Hermite basis, we discussed at length in Section \ref{sect:2}. In the limit $q\to 0$, it generates the TASEP partition function from the transfer matrix of Motzkin paths with specific weights of the initial and final states. The second, emerging in enumerative combinatorics of partially commutative monoids (see below), generates the \L{}ukasiewicz paths and is the subject of our current discussion. 

Using the notation of previous sections, consider the boundary vectors
\be
\ket{V}=(1,0, 0, \dots), \quad  \bra{W}= (1,\alpha^{-1},\alpha^{-2}, \dots)
\label{tasep-boundaries}
\ee
and define the infinite-dimensional matrices
\be
\tilde{D} = \left(
\begin{array}{cccccccc}
\frac{1}{\beta} & \frac{1}{\beta} & \frac{1}{\beta} & \frac{1}{\beta} & \ldots \\
0 & 1 & 1 & 1 &  \ldots \\
0 & 0 & 1 & 1 &  \ldots \\
0 & 0 & 0 & 1 &  \ldots \\
\vdots & \vdots & \vdots & \vdots &  \ddots
\end{array}\right); \quad
\tilde{E} = \left(
\begin{array}{cccccccc}
0 & 0 & 0 & 0 &  \ldots \\
1 & 0 & 0 & 0 & \ldots \\
0 & 1 & 0 & 0 & \ldots \\
0 & 0 & 1 & 0 & \ldots \\
\vdots & \vdots & \vdots & \vdots & \ddots
\end{array}\right).
\label{eq:17}
\ee
In \cite{derrida1993asep} it was shown that for such a choice of operators $\tilde{D}$ and $\tilde{E}$ the relation $\tilde{D}\tilde{E} = \tilde{D}+\tilde{E}$ is satisfied and the diagonalization of $\tilde{D}+\tilde{E}$ provides the solution of the TASEP model. The advantage of the $(\tilde{D},\tilde{E})$--operator basis consists in its deep connection with the enumeration of weighted piles of pieces (HP) in the (1+1)D ballistic deposition model.

Recall that a heap of pieces is a collection of elements piled along the vertical axis. If two elements intersect or touch in their horizontal projection, the heap depends on the order in which they were deposited: the latter element is placed above the earlier one. A heap has a base, i.e., a set of positions orthogonal to the vertical axis.  The concept of heaps was first introduced in 1969 by Cartier and Foata \cite{cartier1969problemes}, who studied monoids generated by an alphabet with special commutation relations. Various heap models and their combinatorial and physical implications are reviewed in \cite{viennot-rev}. The (1+1)D HP model on square and triangular lattices has been extensively studied \cite{viennot-rev, betrema, bousquet3, hakim1983exact, dhar1, dhar2, dhar3}, both combinatorially and via transfer matrix diagonalization or Bethe Ansatz. The combinatorics of (2+1)D heaps above the 2D rectangular base in connection with the domino tilling has been discussed in \cite{strehl}.  

Beyond enumeration, HP models are related to several mathematical problems, such as the combinatorics of partitions \cite{hakim1983exact, gouyou1988equivalence, bousquet3, bousquet2002lattice} and the asymptotic theory of directed 2D lattice paths \cite{banderier2002basic}. Particular attention deserves the work \cite{difrancesco} which establishes connection between heaps of pieces, paths and cluster algebras. The link between heap statistics and (1+1)D growing braids as been discussed in \cite{vershik2000statistical}. By removing the Yang-Baxter relations from the braid group $B_n\{\sigma_1,\dots,\sigma_n\}$ and passing to the locally-free group $\mathcal{LF}_n(g_1,\dots,g_n)$,
\be
B_n: 
\begin{cases}
\sigma_j \sigma_{j+1} \sigma_j = \sigma_{j+1} \sigma_j \sigma_{j+1},\\
\sigma_j \sigma_k = \sigma_k \sigma_j \quad \text{for } |j-k|\ge 2
\end{cases}
\;\;\Longrightarrow\;\;
\mathcal{LF}_n: 
g_j g_k = g_k g_j \quad \text{for } |j-k|\ge 2,
\label{eq:braid} 
\ee
one obtains an algebraic definition of heaps: each equivalence class of words in the locally free semigroup generated by $g_i$ corresponds to a unique heap configuration. Here, we consider square pieces that cannot touch along side faces, while top and bottom faces may touch. The relation between the partition function of (1+1)D HPs without touching vertical faces and the steady state of TASEP with open boundaries was discussed in \cite{haug2014generalized, tamm2022steady}.

\subsubsection{TASEP and enumeration of heaps in "Mikado frame" by \L{}ukasiewicz paths}
\label{sect:5-1-1}

\begin{figure}[ht]
\center{\includegraphics[width=12cm]{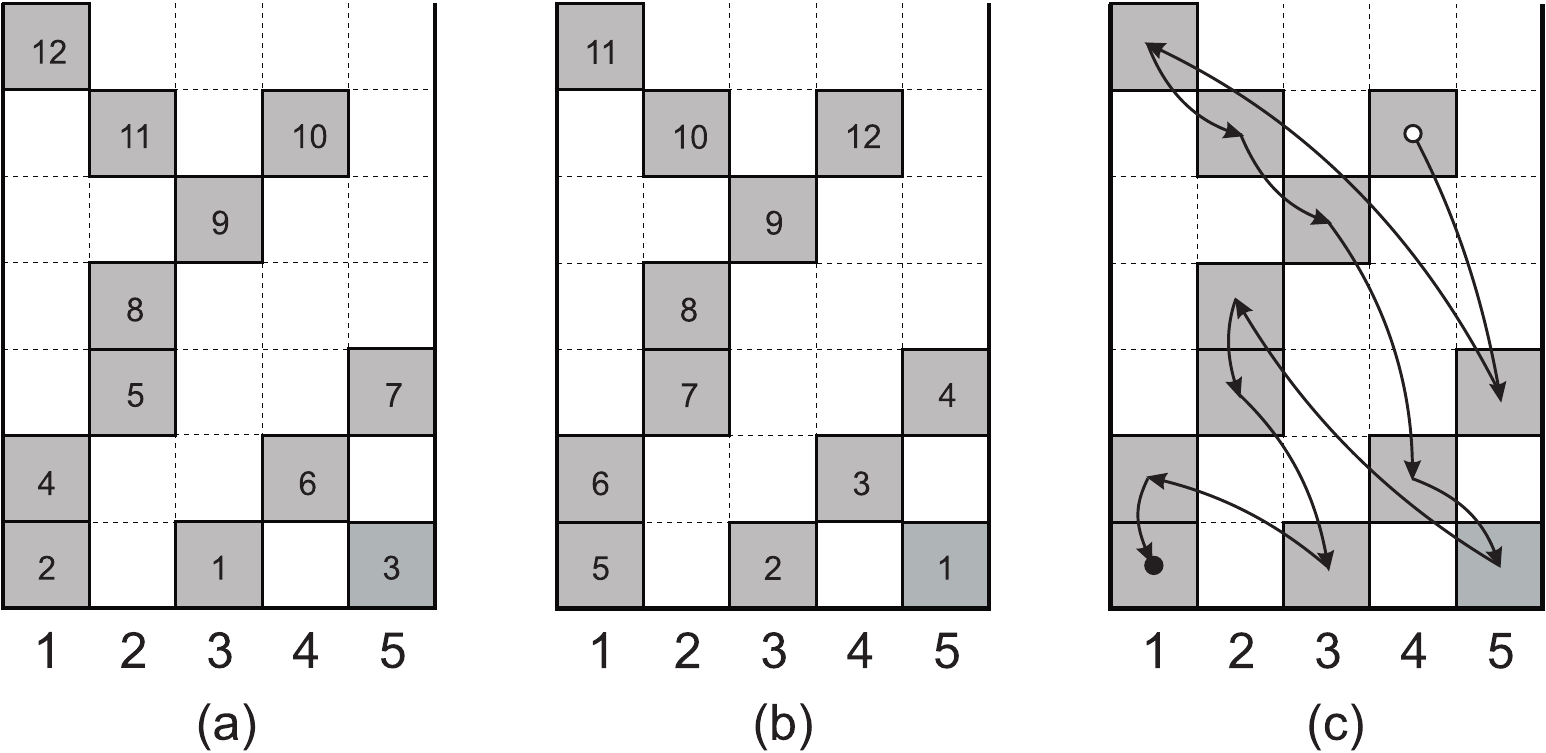}}
\caption{A particular heap. (a) Heap obtained by sequentially dropping bricks corresponding to the word $W_a=g_3g_1g_5g_1g_2g_4g_5g_2g_3g_4g_2g_1$; (b) same heap with a different sequence $W_b=g_5g_3g_4g_5g_1g_1g_2g_2g_3g_2g_1g_4$; (c) unique ``Mikado ordering'' of pieces.}
\label{fig:01}
\end{figure}

Let us outline the computation of the partition function $Z_N(n)$ of (1+1)D heaps of pieces as shown in \fig{fig:01}. By introducing a unique enumeration of heaps, a transfer matrix equation can be written for $Z_N(n)$, which coincides with that of the steady state of the open-boundary TASEP \cite{haug2014generalized, tamm2022steady} for $\beta=1$.

Each heap results from a deposition process, but different deposition sequences can yield the same geometric heap (\fig{fig:01}a,b). To enumerate pieces uniquely, consider the roof of the heap: pieces that, if removed, leave a valid heap and, if redeposited, reconstruct the original heap. We iteratively remove the rightmost piece of the current roof until the heap is empty. This procedure provides the "Mikado representation" of heaps using partially ordered sequences of generators $g_i$. By definition, the sequential normal ordering of words in the locally free monoid is referred to as the "Mikado ordering" and a Markovian transition rule $s_{k+1} \ge s_k - 1$ on generator indices. This name comes from the physical analogy with the classic pick-up sticks game (Mikado): to systematically order a 2D heap of pieces, one peels off pieces one by one from top to bottom (or left to right) without disturbing the resting pieces below.

\begin{figure}[ht]
\center{\includegraphics[width=14cm]{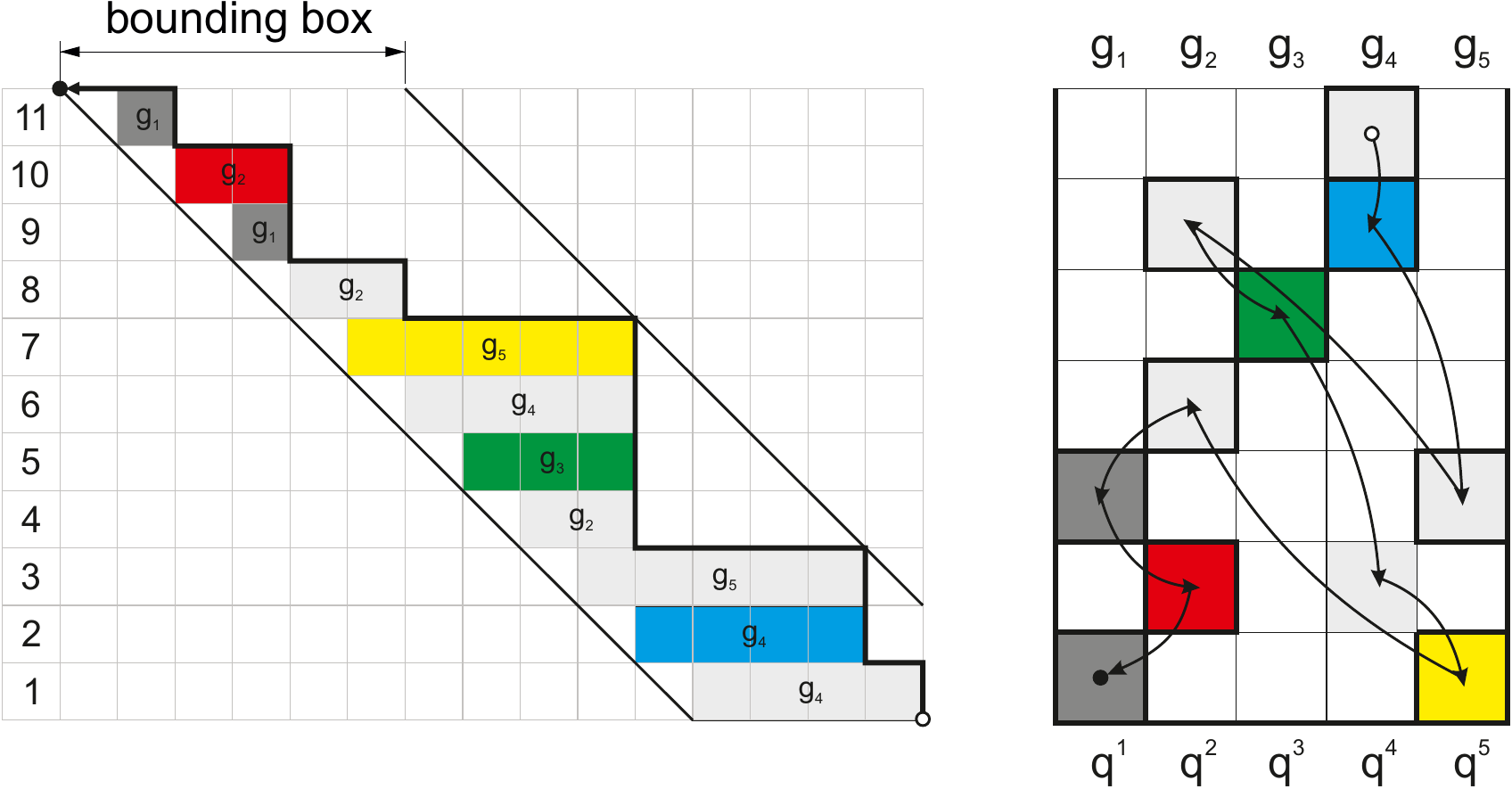}}
\caption{Left: Young tableau representation of a heap shown in the right panel; Right: particular configuration of a heap where arrows show Mikado enumeration.}
\label{fig:02}
\end{figure}

Mikado orderings can be represented by Young tableaux as shown in \fig{fig:02}. The corresponding paths are the so-called \L{}ukasiewicz paths: for a row of length $j$ at level $k$, the next row $k+1$ can have length between $j-1$ and $n$. Rotating the tableau by $\pi/4$ establishes a bijection between \L{}ukasiewicz and the standard Dyck path.

Let $Z_N(x,x_0)$ denotes the number of heaps with Mikado ordering starting at $x_0$ and ending at $x$ ($1\le (x_0,x)\le n$), it satisfies the recursion
\be
\begin{cases}
\disp Z_{N+1}(x,x_0) = \sum_{x'=1}^{x+1} Z_N(x',x_0), & x=1,\dots,n,\\
Z_0(x,x_0) = \delta_{x,x_0}.
\end{cases}
\label{eq:05}
\ee
which is convenient to rewrite as
\be
Z_N(x,x_0) = \bra{W} \tilde{T}^N(n)\ket{V},
\label{eq:07}
\ee
where $\ket{V}$ and $\bra{W}$ are some initial and final vectors, and the transfer matrix takes the form
\be
\tilde{T}(n)=\left(
\begin{array}{cccccccc}
1 & 1 & 1 & 1 & \ldots & 1 \\
1 & 1 & 1 & 1 & \ldots & 1 \\
0 & 1 & 1 & 1 & \ldots & 1 \\
0 & 0 & 1 & 1 & \ldots & 1 \\
\vdots & \vdots & \vdots & \ddots & \ddots & \vdots \\
0 & 0 & 0 & 0 & \ldots & 1
\end{array}\right);
\label{eq:06}
\ee
In what follows, we consider $\ket{V} = (1,0,...,0)$ and $\bra{W} = (1,\alpha^{-1},\alpha^{-2}, \dots)$.

In particular, \eq{eq:06} and \eq{eq:17} are strikingly similar: in the limit $n \to \infty$, $\tilde{T}(n) \to \tilde{D}+\tilde{E}$ with $\beta = 1$. Hence,
\be
Z_N^{\text{TASEP}} (\alpha, \beta = 1) = \alpha \lim_{n\to\infty} \sum_{y=1}^{n} \alpha^{-y} Z_{N}^{\text{HP}}(x=1,y)
\label{beta1}
\ee
which allows to express the TASEP partition function as a weighted sum over heaps with the topmost piece at $x=1$ ("pyramids" in Viennot's terminology \cite{viennot-rev}).

\begin{figure}[ht]
    \centering
    \includegraphics[width=15cm]{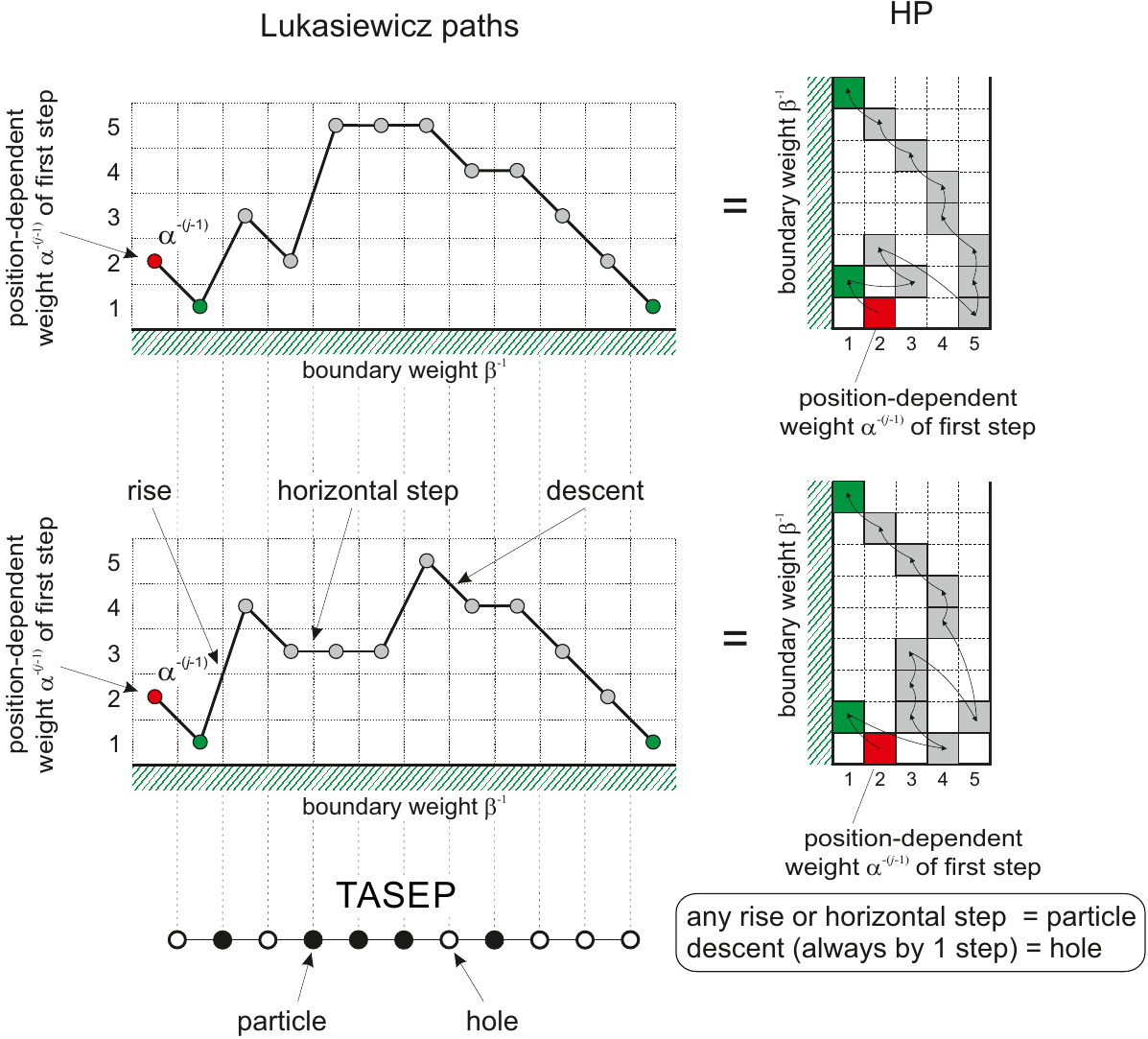}
    \caption{Example of two distinct \L{}ukasiewicz paths (corresponding to two distinct pyramids) mapped to the single TASEP configuration. Particularly, the first step carries the weight $\alpha^{-(j+1)}\big|_{j=2}= \alpha^{-1}$ and the steps 2 and 12 carry the boundary weight $\beta^{-1}$.}
    \label{fig:tasep-hp01}
\end{figure}

Equation \eq{beta1} can be generalized to arbitrary $\beta$ by assigning an additional weight $\beta^{-1}$ to pieces in the first column, which is equivalent to placing an "adsorbing" vertical wall in $x=0$. In the \L{}ukasiewicz path picture, this multiplies the path by $\beta^{-1}$ each time it touches the diagonal, similar to adsorption in a 1D potential well. Consequently,
\be
Z_N^{\text{TASEP}} (\alpha, \beta) = \lim_{n\to\infty}\bra{W}\tilde{T}^N_{\beta}(n)\ket{V}= \alpha \beta \sum_{\text{all pyramids of size $N$}} \alpha^{-y} \beta^{-[\#(x=1)]},
\label{mapping_main}
\ee
where the boundary vectors are defined in \eq{tasep-boundaries}, $y$ is the coordinate of the last piece in the CF order, and $\#(x=1)$ counts the pieces in $x=1$. The mapping of TASEP configuration with given weights of $\alpha$ and $\beta$ onto ensemble of \L{}ukasiewicz paths is schematically depicted in \fig{fig:tasep-hp01}. 

We would like to emphasize that the TASEP$\to$HP mapping is not a bijection: one particular TASEP configuration of particles and holes is mapped onto the \emph{subset} of different \L{}ukasiewicz paths. The example of two distinct \L{}ukasiewicz paths (and hence of two distinct configurations of pyramids) corresponding to one TASEP configuration is shown in \fig{fig:tasep-hp01}.

Now, the whole machinery of generating functions discussed in Section \ref{sect:2} can be repeated. The difference consists only in changing the basis of the transfer matrix. We reproduce the final answer for the stationary TASEP generating function interpreted as CF summation over pyramids in the heap:
\be
\Omega_n (s,\alpha, \beta) = \sum_{N=0}^{\infty} \bra{W}\tilde{T}^N_{\beta}(n)\ket{V} s^N = -\frac{1}{s} \frac{\frac{(p_1+1)(p_1^n - (-\alpha)^{-n})}{p_1+\alpha^{-1}}-\frac{(p_2+1)(p_2^n - (-\alpha)^{-n})}{p_2+\alpha^{-1}}}{(p_1+\beta^{-1})p_1^n - (p_2+\beta^{-1})p_2^n}
\label{Omega_n}
\ee
where $p_{1,2} = \frac{-1\pm \sqrt{1-4s}}{2s}$. For $s\to 0$ and $n\to \infty$, 
\be
\lim_{n \to \infty} \Omega_n (s,\alpha, \beta) = \\
\frac{2(\sqrt{1-4s}+1-2s)}{\left(\sqrt{1-4s}+1-2s\alpha^{-1}\right)\left(\sqrt{1-4s}+1-2s\beta^{-1}\right)}.
\label{TASEP_Omega}
\ee
Note that in the limit $n\to \infty$ the symmetry $\alpha \leftrightarrow \beta$ is restored. 

The view on TASEP in terms of heap statistics allows one to inscribe the TASEP phase transitions in the standard classification of pinning--depinning transitions in wetting phenomena \cite{forgacs1986wetting}. Since the transitions are  determined by the smallest singularity of $\Omega(s,\alpha, \beta)$, we have: (i) $s_1 = 1/4$ (square root singularity) -- maximal flow phase; (ii) $s_2(\beta) = \beta (1-\beta)$ (pole) -- high-density phase; (iii) $s_3(\alpha) = \alpha (1-\alpha)$ (pole) -- low-density phase. Phase transitions occur when $s_1 = s_2(\beta) \to \beta = 1/2, \quad s_1 = s_3(\alpha) \to \alpha = 1/2, \quad s_2(\beta) = s_3(\alpha) \to \beta = \alpha$.

In terms of \L{}ukasiewicz paths, these phases correspond to (i) freely diffusing paths, (ii) paths pinned to the wall, and (iii) fully elongated paths. The diffusive-to-pinned transition occurs at pinning weight $\beta^{-1}=2$ \cite{ternovsky1992statistics}. A path confined between two weighted adsorbing walls $\alpha^{-1}, \beta^{-1}$ is analogous to open-boundary TASEP \cite{krug1994disorder}. As noted in \cite{tamm2022steady}, this correspondence clarifies the symmetry $\alpha \leftrightarrow \beta$: the attractive potential $U(x) = \delta_{1,x}\log \beta$ at the boundary $x=1$ mirrors the path's stretching in the potential $V(x) = x\log \alpha$ acting at the end point of the trajectory.

\subsubsection{Cartier-Foata-Viennot duality relation between TASEP/HP and hardcore gas on a segment}
\label{sect:5-1-2}

Here we demonstrate the bijection between partition functions of heaps of pieces on a segment $1\le j\le n$ with a sticky boundary at $j=0$ and a 1D hardcore gas of particles on a segment $1\le j\le n$ with an "exceptional" particle at $j=1$. This connection is based on the Cartier-Foata theorem which proves that for any partial commutation graph $G = (V, E)$, the growth generating function $W(z) = \sum_{N \ge 0} a_N z^N$ for the number of traces (classes of equivalent words) is given by
$$
W(z) = \frac{1}{\sum\limits_{I \text{ clique of } \bar{G}} (-1)^{|I|} z^{|I|}}
$$
where $\bar{G}$ is the complement of $G$, and $I$ is a set of generators that are pairwise non-adjacent in $G$ (i.e. form cliques in $\bar{G}$), $|I|$ is the number of such generators. This abstract theorem was extended by G.X. Viennot \cite{viennot-rev} to heaps of pieces, showing that the generating functions of a whole heap of pieces is uniquely defined by a single layer of a heap. We first recall the statement of the theorem following G.X. Viennot \cite{viennot-rev} and then apply it to our particular case.

Let $Z_N$ be a partition function of a heap constructed over some graph ${\cal G}$ serving as a base, where the vertices of ${\cal G}$ designate possible locations of elementary pieces, and the edges connect vertices that cannot be simultaneously occupied in a single layer (in our case, the graph is just a segment of $n$ sites). Let $\Theta (s)$ be the corresponding generating function:
\be
\Omega(s) = \sum_{N=1}^{\infty} Z_N s^N \equiv \sum_{\text{allowed configurations}} s^{\# \text{~of pieces}}\, ,
\label{eq:34}
\ee
Define another partition function $Q_k$ of all distinct configurations of $k$ elementary pieces in a single layer, i.e., all possible subsets of $k$ vertices of $\cal G$, such that no edge has both ends included in the subset. This generating function is 
\be
\Theta(t) = 1 + \sum_{k=1}^{k_{\max}} Q_k (t)^k,
\label{eq:35}
\ee
where $k_{\max}$ is the maximum possible size of this subset. In this formulation $\Omega(t)$ is the partition function of a hard-core lattice gas on a graph $\cal G$ with fugacity $s$. The theorem of G.X. Viennot \cite{viennot-rev} states that
\be
\Omega(s) = \frac{1}{\Theta(t=-s)}
\label{eq:37}
\ee
This duality coincides with the duality of partition functions of \emph{elementary} and \emph{full} symmetric polynomials \cite{smirnov2024symmetric} and seems to be closely related to the construction of supersymmetric partition functions of lattice spin systems.

Generalizing \eq{eq:37}, we can establish a correspondence between the partition functions of heaps with a  sticky left boundary on the $n$-site segment and a combination of weighted hardcore gasses on segments of  1,2,... $n$ sites. To proceed, define the Hamiltonian of the standard hardcore gas on the $n$-site segment:
\be
H(\sigma_1,...,\sigma_n|t) = \ln t \sum_{j=1}^{n} \sigma_j  + J \sum_{j=1}^{n-1} \sigma_j \sigma_{j+1}
\label{eq:gas1}
\ee
where $\sigma_j=(0,1)$ is the occupation number at site $j$, $\ln t$ is the chemical potential of a particle at site $j$, and $J$ in the limit $J \to -\infty$ enforces the hard-core exclusion of neighboring particles. The corresponding partition function reads:
\be
W(t,n)=\lim_{J\to-\infty} \sum_{\{\sigma_1,...,\sigma_n\}} e^{H(\sigma_1,...,\sigma_n|t)}
\label{eq:gas2}
\ee
To include in the description the TASEP rates $\alpha$ and $\beta$, we consider two combinations of weighted partition functions of hard-core gasses: 
\begin{itemize}
\item[(i)] $\Theta^{(1)}_{\beta}(t,n)$, which is the $\beta$--weighted difference of two partition functions of hard-core gasses on the $n$--site and $(n-1)$--site segments: 
\be
\Theta^{(1)}_{\beta}(t,n) = (1-\beta)W(t,n-1)-W(t,n)
\label{eq:Theta1}
\ee
\item[(ii)] $\Theta^{(2)}_{\alpha}(t,n)$, which is the $\alpha$--weighted sum of partition functions of hard-core gasses on $1-, 2-,...,n$--site segments. The generic expression of $\Theta^{(2)}_{\alpha}(t,n)$ consistent with the TASEP boundaries $\ket{V} = (1,0,...,0)$ and $\bra{W} = (1,\alpha^{-1},\alpha^{-2}, \dots)$, is
\be
\Theta^{(2)}_{\alpha}(t,n) = \sum_{k=1}^{n-1}(-t\alpha^{-1})^k W(t,n-k) + (-t\alpha^{-1})^n
\label{eq:Theta2}
\ee
\end{itemize}
The partition function $(\alpha,\beta)$-TASEP, $\Omega_{\alpha,\beta}(n,s)$ (see \eq{Omega_n}) can now be expressed as the quotient of weighted partition functions $\Theta^{(1)}_{\beta}(t,k)$ and $\Theta^{(1)}_{\alpha}(t,k)$ of hard-core gasses: 
\be
\Omega_{\alpha,\beta}(n,s) = \left.\frac{\Theta^{(2)}_{\alpha}(t,n)}{\Theta^{(1)}_{\beta}(t,n)}\right|_{t=-s}= \left.\alpha\beta \frac{\sum_{k=1}^{n-1}(-t\alpha^{-1})^k W(t,n-k) + (-t\alpha^{-1})^n}{(1-\beta)W(t,n-1)-W(t,n)}\right|_{t=-s}
\label{eq:Theta3}
\ee
where $W(t,n)$ is the partition function of the standard hard-core gas on a $n$-site segment defined in \eq{eq:gas2}. In the following, we demonstrate on a particular example $n=6$ the validity of \eq{eq:Theta3}. From \eq{Omega_n} we have:
\begin{multline}
\Omega_{\alpha,\beta}(s,n=6)= \\ \frac{b s \left(a^5+(1-5 a) a^4 s-(a-1)^3 a^2 s^3+\left(6 a^2-4 a+1\right) a^3 s^2-(a-1) s^5+(a-1)^2 a s^4\right)}{a^5 \left(b ((s-5) (s-1) s-1)+(3 s-4) s^2+s\right)}
\label{tasep-s}
\end{multline}
From \eq{eq:Theta1}--\eq{eq:Theta2} we compute the quotient $\Theta^{(2)}_{\alpha}(t,n=6)$ and $\Theta^{(1)}_{\beta}(t,n=6)$:
\begin{multline}
\frac{\Theta^{(2)}_{\alpha}(t,n=6)}{\Theta^{(1)}_{\beta}(t,n=6)}= \\ \frac{b t \left(a^5+(5 a-1) a^4 t+(a-1)^3 a^2 t^3+\left(6 a^2-4 a+1\right) a^3 t^2+(a-1) t^5+(a-1)^2 a
   t^4\right)}{a^5 (b t (t+1) (t+5)+b+t (t+1) (3 t+1))}
\label{gas-t}
\end{multline}
It can be checked that $\Omega_{\alpha,\beta}(s,n=6) = \Theta^{(2)}_{\alpha}(t=-s,n=6)/\Theta^{(1)}_{\beta}(t=-s,n=6)$. In the limit $n\to\infty$ phase transitions in TASEP can be interpreted as phase transitions in ensembles of weighted hard-core gasses. 

\subsection{Weighted heaps of pieces and $q'$-graded locally free algebra $q'\mathcal{LF}_n$}
\label{sect:5-2}

In forthcoming Sections \ref{sect:5-2-1}--\ref{sect:5-2-3} we introduce and describe the $q'$-deformation of locally-free group, $\mathcal{LF}_n$, which allows to enumerate $q$--weighted heaps and establish a connection with "magnetic $q$--Dyck paths". We show below that the graded algebra $q'\mathcal{LF}_n$ has connections with the $q$-deformation of $(D,E)$--Matrix Product Ansatz for ASEP, however is not equivalent to it (we use the notation $q'$ for heaps of pieces to make a distinction with the parameter $q$ used in ASEP). In Section \ref{sect:5-3} we discuss the connection between $q$ and $q'$ in more detail.

\subsubsection{Definition of a $q'$-graded locally free algebra}
\label{sect:5-2-1}
   
We introduce a generalization of the locally-free group $\mathcal{LF}_n$ that incorporates a grading associated with a weight $(q')^j$ ($0<q<1$) for each generator $g_j$. This graded version allows us to enumerate heaps of pieces with weighted generators (”$q'$--heaps”), naturally leading to the notion of magnetic Dyck paths. To define a graded structure, we assign a degree to each generator $\deg(g_j)=j$ and apply this definition multiplicatively to words. That is, for $w = g_{j_1} g_{j_2} \cdots g_{j_r}$ we set $\deg(w) = j_1 + j_2 + \cdots + j_r$.

The corresponding graded locally-free algebra $q'\mathcal{LF}_n$ over a base field is the associative algebra generated by $\{g_j\}$, subject to the same commutation relations as $\mathcal{LF}_n$, but equipped with a grading defined by the degree. We further define a weight:
\be
w_{q'}(g_{j_1}\cdots g_{j_r}) = (q')^{\,j_1+\cdots+j_r} = (q')^{\deg(w)},
\ee
which extends linearly to all elements of the algebra. The algebra is non-commutative for nearest neighboring pieces of the heap, while is commutative for non-interfering (i.e. "distant" pieces), in agreement with Viennot’s theory of heaps \cite{viennot-rev}. 

A normal form basis of $q'\mathcal{LF}_n$ can be chosen by selecting, for each equivalence class of words, a canonical Mikado representative -- as for $q'=1$. Each normally ordered word in the $q'$--graded basis corresponds uniquely to a $q'$-deformed heap of $N$ pieces. Each heap of pieces can be assigned to a path of length $2N$ represented in the form of a Young tableau in \fig{fig:02}, where a horizontal row corresponds to the generator $g_j$ that preserves the Mikado order. Under this bijection, the weight $(q')^j$ of a generator $g_j$ is naturally interpreted as a contribution $(q')^h$ for a row of length $h$. The total weight of the Dyck path is then
\be
w_{q'}(\text{path}) = \prod_{\text{All $h_k$ in  Young tableau}}^{2N} (q')^{h_k} = (q')^{\text{area(path)}},
\ee
where $h_k$ is the particular length of the $k$th line in the Young tableau, and $\text{area(path)}$ denotes the total area under the Dyck path rotated by $\pi/4$. This is exactly the total $q'$-weight associated with the heap, and we can define the partition function for all heaps of $N$ pieces (or, equivalently, Dyck paths of length $2N$) as the sum over all weighted Dyck paths,
\be
Z_N(q,n) = \sum_{\text{Dyck paths in a strip $n$}} (q')^{\text{area(path)}}.
\ee
Here, the restriction on the maximal height $n$ corresponds to the finite number of generators $\{ g_1,\dots,g_n\}$. In this formulation, $Z_N(q',n)$ is  the generating function of heaps of pieces in a graded locally free algebra $q'\mathcal{LF}_n$ which bijectively coincides with the number of $q'$-Dyck paths weighted with the area in the width strip $n$.

At this point, it is important to distinguish between two different, but equivalent, transfer-matrix representations of $q'$-weighted heaps. These representations correspond to two different geometric frames: (i) the Young tableau (or \L{}ukasiewicz path) frame, and (ii) the Dyck path frame obtained by rotation of the Young tableau by $\pi/4$. We now describe these two frames separately and establish their precise correspondence.

\subsubsection{Young tableau and $q'$-Dyck path frames for enumeration of $q'$-heaps}
\label{sect:5-2-2}

We first work on Young's tableau representation, where each heap (or Cartier-Foata-ordered word) is represented as a tableau whose rows have lengths $j_1, j_2, \dots, j_N$, ($1 \le j_k \le n$), with the admissibility condition $j_{k+1} \le j_k + 1$. The length $j_k$ of the $k$-th row corresponds to the generator $g_{j_k}$.

Let $Z_N(j,n)$ denote the partition function of tableaux with $N$ rows whose last row has length $j$. Adding a new row of length $j$ contributes a magnetic weight $(q')^j$, coming from the area interpretation discussed above. The recursion relation therefore reads
\be
Z_{N+1}(j,n)= (q')^j \sum_{j'=j-1}^{n} Z_N(j',n), \qquad j=1,\dots,n 
\label{eq:YTrec}
\ee
with the convention $Z_N(0,n)=0$ and initial condition $Z_0(j,n)=\delta_{j,1}$. Equation \eq{eq:YTrec} can be written in a matrix form as
\be
Z_{N+1} = T^{\text{YT}}_{q'}(n)\, Z_N,
\label{eq:recYT}
\ee
where the $n\times n$ transfer matrix that describes the Young tableau is
\be
T^{\text{YT}}_{q'}(n)=
\left(
\begin{array}{cccccc}
(q')^1 & (q')^1 & (q')^1 & (q')^1 & \ldots & (q')^1 \\
(q')^2 & (q')^2 & (q')^2 & (q')^2 & \ldots & (q')^2 \\
0   & (q')^3 & (q')^3 & (q')^3 & \ldots & (q')^3 \\
0   & 0   & (q')^4 & (q')^4 & \ldots & (q')^4 \\
\vdots & \vdots & \vdots & \ddots & \ddots & \vdots \\
0 & 0 & 0 & 0 & \ldots & (q')^n
\end{array}
\right)
\label{eq:TrYT}
\ee
Thus, in the Young tableau frame in which the upper-triangular structure encodes the constraint $j_{k+1} \le j_k+1$ and all nonzero matrix elements carry "magnetic weights" which depend on the number of the row. This transfer matrix naturally enumerates the weighted heaps and is directly adapted to the graded locally-free algebra $q'\mathcal{LF}_n$.

Instead of describing the growth of heaps in terms of successive rows of a Young tableau, one may equivalently formulate the problem in the $q'$-Dyck path frame, obtained by viewing the same configurations as paths evolving in discrete time with a local height variable. We rotate the Young tableau by an angle $\pi/4$, mapping rows of the tableau to diagonal steps of a $q'$-Dyck path -- see \fig{fig:02}. Under this transformation: (i) adding a row corresponds to adding a local step of the Dyck path; (ii) the row index $j$ becomes the height of the path; (iii) the constraint $j_{k+1} \le j_k+1$ becomes the Dyck path constraint of unit up-steps and arbitrary down-steps.  

Let $\mathcal{Z}_N(j,n)$ denote the partition function of magnetic Dyck paths of length $N$ whose height at step $N$ equals $j$, with $1\le j\le n$. In magnetic interpretation, an upward step $j\to j+1$ carries unit weight, while a downward step $j\to j-1$ carries a magnetic weight $(q')^{\,j-1}$, reflecting the contribution to the area under the path. This formulation makes explicit the interpretation of weighted heaps as magnetic Dyck paths: the grading in the locally-free algebra manifests itself as a height-dependent magnetic field acting on downward steps. The recursion relation in this rotated frame takes the local form
\be
\mathcal{Z}_{N+1}(j,n)= \mathcal{Z}_N(j-1,n)+ (q')^{j}\,\mathcal{Z}_N(j+1,n) \qquad j=1,\dots,n ,
\label{eq:Dyck-rec}
\ee
with the boundary conditions $\mathcal{Z}_N(0,n)=\mathcal{Z}_N(n+1,n)=0$. Equation \eq{eq:Dyck-rec} can be written in transfer-matrix form as
\be
\mathcal{Z}_{N+1}=T^{\text{Dyck}}_{q'}(n)\,\mathcal{Z}_N,
\label{q-Dyck}
\ee
where the Dyck path transfer matrix is strictly tri-diagonal:
\be
T^{\text{Dyck}}_{q'}(n)=\left(
\begin{array}{cccccc}
0 & 1 & 0 & 0 & \ldots & 0 \\
(q')^{0} & 0 & 1 & 0 & \ldots & 0 \\
0 & (q')^{1} & 0 & 1 & \ldots & 0 \\
0 & 0 & (q')^{2} & 0 & \ddots & 0 \\
\vdots & \vdots & \vdots & \ddots & \ddots & 1 \\
0 & 0 & 0 & 0 & (q')^{n-1} & 0
\end{array}
\right)
\label{eq:T-Dyck}
\ee
Although the transfer matrix in this frame differs structurally from the Young tableau transfer matrix, both descriptions generate the same partition function $Z_N(q',n)$ and encode the same ensemble of weighted heaps.

In what follows, we pay most attention to statistics of "$q'$-Dyck Brownian bridges", which are $2N$--step magnetic Dyck paths with coinciding initial and final points located at $x=1$. The corresponding partition function ${\cal Z}_{2N}(q')$ is the solution of \eq{q-Dyck} with the transfer matrix \eq{eq:T-Dyck} and the boundary vectors $\bra{W}=(1,0,0,...,0)$, $\ket{V}=(1,0,0,...,0)^{\top}$: 
\be
{\cal Z}_{2N}(q') = \bra{W} (T^{\text{Dyck}}_{q'})^{2N}(n)\ket{V} = C_n(q')
\label{q-brbr}
\ee
where $C_n(q')$ are the Carlitz-Riordan $q'$-Catalan numbers that satisfy the recursion
\be
C_n(q') = \sum_{k=0}^{n-1}(q')^k C_k(q') C_{n-k-1}(q')
\ee
The function $C_n(q')$ is the $q'$-extension of the Catalan number $C_n(1) = \frac{1}{n+1} \binom{2n}{n}$. 

The $q'$-Catalan Karlitz-Riordan number $C_N(q')$ counts the ensemble of standard Dyck paths of length $2N$ weighted by the area $A$ and, simultaneously, the number of nonequivalent $q'$-graded heaps of $N$ pieces. The area-weighted generating function is defined as:
\begin{equation}
F(s,q') = \sum_{N=0}^{\infty} C_N(q') s^N
\end{equation}
By decomposing an arbitrary path at its first return to the origin, the generating function satisfies the non-linear functional recurrence relation:
\begin{equation}
F(s,q') = \frac{1}{1 - s F(sq', q')}
\end{equation}
This relation provides expansion into a continued fraction. By introducing an auxiliary function $\psi(s) = \text{Ai}_{q'}(s)$, where $\text{Ai}_{q'}(s)$ is the Ramanujan $q'$-Airy hypergeometric series, one can linearize the system via the algebraic ansatz:
\begin{equation}
F(s,q') = \frac{\psi(s)}{\psi(s/q')}
\end{equation}
Substituting this ratio back into the recurrence directly yields the linear second-order $q'$-difference equation:
\begin{equation}
\psi(s) - \psi(qs) + s\psi\left((q')^2s\right) = 0
\end{equation}
The unique solution is:
\begin{equation}
F(s,q') = \frac{\text{Ai}_{q'}(s)}{\text{Ai}_{q'}(s/q')}
\label{eq:genfunction}
\end{equation}
where the Ramanujan $\text{Ai}_{q'}(s)$ function is defined as follows
\be
\text{Ai}
_{q'}(s)=\sum_{k=0}^{\infty}\frac{(q')^{k^2}(-s)^k}{(q';q')_k}; \quad (t;q')_k=\prod_{k=0}^{k-1} \left(1-t (q')^k\right)
\ee

Near the critical edge $s_c = 1/4$ one can write $q' = e^{-\epsilon}$ ($0 < \epsilon \ll 1$) and introduce the continuous spatial coordinate $x$:
\begin{equation}
s(x) = \frac{1}{4}(1 - \epsilon^{2/3} x)
\end{equation}
Such a scaling in $\epsilon$ makes the continuous equation homogeneous. Evaluating the terms $qs$ near $s_c$ in the first-order approximation with $q' \approx 1 - \epsilon$, we get the following.
\begin{equation}
sq' \approx \left( \tfrac{1}{4}(1 - \epsilon^{2/3} x) \right) (1 - \epsilon) = \tfrac{1}{4} \left( 1 - \epsilon^{2/3} \left(x + \epsilon^{1/3} \right) \right)
\end{equation}
Compared with the continuous step definition $q's = \frac{1}{4}(1 - \epsilon^{2/3}(x + \Delta x))$, we have $\Delta x = \epsilon^{1/3}$. Now we can expand $\psi(q's)$ and $\psi((q')^2 s)$ in a Taylor series centered at $x$ with step size $\Delta x = \epsilon^{1/3}$. By evaluating the generating function $F(s,q') = \psi(s)/\psi(q's)$ within the scaling window, we match the local boundary values. Since the classical Catalan function satisfies $C(1/4) = 2$, the expansion of the ratio of the $\psi(s)$ and $\psi(q's)$ yields:
\begin{equation}
F(s,q') \approx 2 - 2\epsilon^{1/3} \frac{\text{Ai}'(x)}{\text{Ai}(x)}; \quad x = \frac{1-4s}{\epsilon^{2/3}}, \quad \epsilon = -\ln (q') \approx 1-q'
\label{eq:airy1}
\end{equation}

To verify that this expression recovers the classical Catalan square-root singularity as $q' \to 1^-$, take the limit $\epsilon \to 0$ for a fixed value of $s < 1/4$, which sends $x$ to $+\infty$. The well-known asymptotic profiles of the Airy functions for large positive arguments read:
\begin{equation}
\text{Ai}(x) \sim \frac{1}{2\sqrt{\pi}x^{1/4}} e^{-\frac{2}{3}x^{3/2}}, \quad \text{Ai}'(x) \sim -\sqrt{x}\,\text{Ai}(x)\quad \Rightarrow\quad \frac{\text{Ai}'(x)}{\text{Ai}(x)} \sim -\sqrt{x}
\end{equation}
Substituting this asymptotic ratio back into $F(s,q')$ taken at $q'=1$ we get:
\begin{equation}
F(s,q'=1) = \lim_{\epsilon \to 0} \left[ 2 - 2\epsilon^{1/3} \left(-\sqrt{\frac{1-4s}{\epsilon^{2/3}}}\right) \right] = 2 + 2\sqrt{1-4s}
\end{equation}
This reproduces the classical algebraic branching singularity of standard Catalans.

The straightforward check shows that the partition functions of $q'$-heaps of $N$ pieces, 
\be
Z_N(q') = \bra{W}(T_{q'}^{YT})^N(n)\ket{V}
\ee
and the partition function of $2N$-step $q'$-Dyck Brownian bridge, ${\cal Z}_{2N}(q')$, defined in \eq{q-brbr}, coincide:
\be
Z_N(q') = {\cal Z}_{2N}(q') = C_n(q')
\label{heap-brbr}
\ee
Let us point out the difference between the Carlitz-Riordan $C_n(q')$ Catalan numbers, considered here, and the Touchard-Riordan $C_n^T(q)$ discussed in ASEP.consideration.

\subsubsection{Burgers equation, KPZ, and rational Painlev\'e II}
\label{sect:5-2-3}

Let us show that the generating function $F(s,q')$ of the $q'$-weighted heaps of pieces in the limit $q'\to 1$ (see \eq{eq:airy1}) reproduces the regularized solution of the Burgers equation with small diffusivity $\nu$. Consider the velocity field $u(x,t)$ governed by the stochastic Burgers equation with $0<\nu\ll 1$ and space-time white noise $\eta(x,t)$:
\begin{equation}
\frac{\partial u}{\partial t} + u \frac{\partial u}{\partial x} = \nu \frac{\partial^2 u}{\partial x^2} + \sqrt{2D}\frac{\partial \eta}{\partial x}
\end{equation}
Introduce a surface height profile field $h(x,t)$ defined as:
\begin{equation}
u(x,t) = -\frac{\partial h}{\partial x}
\end{equation}
Integrating the Burgers equation over the spatial coordinate $x$ transforms the velocity dynamics directly into the canonical KPZ equation for non-linear surface growth:
\begin{equation}
\frac{\partial h}{\partial t} = \nu \frac{\partial^2 h}{\partial x^2} - \frac{1}{2}\left(\frac{\partial h}{\partial x}\right)^2 + \sqrt{2D}\eta(x,t)
\end{equation}

The structure of the profile $h(x,t)$ governed by the non-perturbed inviscid Burgers equation (Hopf) ($\nu=0$, $D=0$) is:
\begin{equation}
\frac{\partial u_0}{\partial t} + u_0 \frac{\partial u_0}{\partial x} = 0
\label{eq:hopf}
\end{equation}
The solution of \eq{eq:hopf} is obtained by integrating along the characteristics. The boundary profile satisfies the equation $t u_0^2 - x u_0 + 1 = 0$ and has two possible branches. Selecting the physical branch that remains stable under short-time evolution, we have the explicit expression for the velocity field:
\begin{equation}
u_0(x,t) = \frac{x - \sqrt{x^2 - 4t}}{2t}
\end{equation}
This solution exhibits a vertical gradient divergence ($\partial_x u_0 \to \infty$) when the square root vanishes, defining the critical shock at the edge $x_c(t) = 2t^{1/2}$. Evaluating the velocity field at this critical edge, we get $u_c(t) = u_0(x_c, t) = t^{-1/2}$.

Let us analyze the localized profile near the edge $x_s(t)$ in the weak-diffusivity and zero-noise limit ($0<\nu\ll 1$, $D=0$). Consider the following multiscale ansatz:
\begin{equation}
x = x_s(t) + \nu^{2/3} y, \quad u(x,t) = u_0(x_c,t) + \nu^{1/3} w(y,t)
\end{equation}
where $y$ is the  rescaled and shifted $x$-variable, and $w(y,t)$ is the rescaled and shifted velocity field $u(x,t)$. In the vanishing diffusivity limit ($\nu \to 0$), we arrive at the following equation:
\begin{equation}
w_{yy} = w w_y - \frac{1}{2}t^{-3/2}
\label{eq:w(y,t)}
\end{equation}
Taking into account that $w w_y = \frac{\partial}{\partial y}\left(\frac{1}{2}w^2\right)$, we integrate the equation \eqref{eq:w(y,t)} with respect to $y$. Setting the far-field boundary integration constant to zero yields the non-autonomous Riccati equation:
\begin{equation}
w_y = \frac{1}{2}w^2 - \frac{y}{2t^{3/2}}
\label{eq:ricc}
\end{equation}
To linearize the Riccati equation, we express the localized velocity profile $w(y,t)$ as the logarithmic derivative of an auxiliary field $\phi(y,t)$:
\begin{equation}
w(y,t) = -2 \frac{\partial}{\partial y} \ln \phi(y,t) = -2 \frac{\psi_y}{\psi}
\end{equation}
Differentiating this substitution with respect to $y$ gives us:
\begin{equation}
w_y = -2 \frac{\phi_{yy}}{\phi} + 2 \left( \frac{\phi_y}{\phi} \right)^2
\end{equation}
Substituting both $w$ and $w_y$ back into the Riccati equation and multiplying the remaining terms by $-\frac{1}{2}\phi$ provides the linear second-order ordinary differential equation:
\begin{equation}
\phi_{yy} - \frac{y}{4t^{3/2}}  \phi = 0
\end{equation}
Introducing the rescaled coordinate $\zeta = (4t^{3/2})^{-1/3}y$, we arrive at the classical Airy differential equation:
\begin{equation}
\phi(y,t) = \text{Ai}\left( \frac{y}{(4t^{3/2})^{1/3}} \right)
\end{equation}
Thus, the smooth regularized solution inside the localized shock layer is as follows:
\begin{equation}
w(y,t) = -2 \frac{\partial}{\partial y} \ln \left(\text{Ai}\left( \frac{y}{(4t^{3/2})^{1/3}} \right) \right) = -\frac{2^{1/3}}{\sqrt{t}} \frac{\text{Ai}'\left( \frac{y}{2^{2/3}\sqrt{t}} \right)}{\text{Ai}\left( \frac{y}{2^{2/3}\sqrt{t}} \right)}
\label{eq:airy2}
\end{equation}
Comparing  \eqref{eq:airy1} to \eqref{eq:airy2} we establish the following dictionary between the growth of $q'$-graded heaps described by the Carlitz-Riordan $q'$-difference equation in the limit $q'\to 1$ and the Burgers equation with small diffusivity $\nu$:

\begin{table}[h]
\centering
\caption{Dictionary between $q'$-combinatorics and fluid dynamics}
\begin{tabular}{l|l|l}
\toprule
{\bf Identity} & {\bf $q'$-Catalan Paths} & {\bf Burgers equation} \\ \midrule
Parameter   & $\epsilon = 1-q'$  & Diffusivity coefficient $\nu$ \\ \midrule
Variable  & Distance to edge $(1-4s)$           & Spatial coordinate $y$ \\ \midrule
Discretization parameter  & $\Delta X = \epsilon^{1/3}$         & $dy = \nu^{1/3} dz$ \\ \midrule
Fluctuation field     & $\partial_x \ln \text{Ai}(x)$       & $\partial_z \ln \text{Ai}(z)$ \\ \bottomrule
\end{tabular}
\end{table}

For completeness, recall that the standard Painlevé II equation is a non-linear second-order ordinary differential equation defined by:
\begin{equation}
u''(s) = 2u(s)^3 + s u(s) + a
\label{eq:pain}
\end{equation}
where $a$ is a complex parameter. Generically for $a \neq 0$, \eqref{eq:pain} defines transcendental functions. However, when $a = \pm \frac{1}{2}$, the equation admits a special non-linear first-order reduction. Consider a function $w(s)$ satisfying the non-autonomous Riccati equation, which is another way of representing the noiseless Burgers equation \eq{eq:ricc}:
\begin{equation}
w' = \frac{1}{2}w^2 - \frac{1}{2}s
\label{eq:riccati}
\end{equation}
Let us define a field $u(s) = -\frac{1}{2}w(s)$. Differentiating $u(s)$ twice with respect to $s$ yields:
\begin{equation}
\begin{array}{l}
\disp u' = -\frac{1}{2}w' = -\frac{1}{4}w^2 + \frac{1}{4}s = -u^2 + \frac{1}{4}s \medskip \\ \disp q'' = -2qq' + \frac{1}{4} = -2q\left(-q^2 + \frac{1}{4}s\right) + \frac{1}{4} = 2q^3 - \frac{1}{2}sq + \frac{1}{4}
\end{array}
\end{equation}
By rescaling of coordinates: $s = -2^{-1/3}x$ and $u(s) = 2^{-1/3}\tilde{u}(x)$, equation \eqref{eq:riccati} gets mapped to:
\begin{equation}
\tilde{u}''(x) = 2\tilde{u}(x)^3 + x\tilde{u}(x) - \frac{1}{2}
\end{equation}
This shows that the deterministic Riccati equation that governs the mean viscous shock layer is an exact linearizable solution of the Painlev\'{e} II equation for $a = -1/2$. Recall that the solution to \eqref{eq:riccati} is given by log-derivative of the Airy function:
\begin{equation}
w(s) = -2\frac{\partial}{\partial s} \ln \text{Ai}\left(2^{-1/3}s\right)
\end{equation}
When stochastic fluctuations are introduced, the symmetry of the underlying system undergoes a structural change encoded in the replacement $a = -1/2$ by $a = 0$ which corresponds to the transcendental solution. Under the influence of white noise, the system leaves the deterministic profile and moves into the domain of the Tracy--Widom distribution for Painlev\'e II at $a=0$:
\begin{equation}
u''(s) = 2u^3(s) + s u(s), 
\end{equation}
where $u(s) \to \text{Ai}(s)$ as $s \to \infty$.

\subsection{ASEP/HP correspondence}
\label{sect:5-3}

In Section \ref{sect:5-1} we have connected the steady-state solution of TASEP (i.e. $q=0$--ASEP) in the $N$-site segment with the enumeration of the topologically nonequivalent configuration of heaps of $N$ pieces (HP) in a box with neighboring interactions. The solution of the HP problem we have found by diagonalization of the transfer matrix $T_n$ which encodes the local rules which fix the unique Mikado enumeration of heaps by words in the locally-free group $\mathcal{LF}_n$. It turned out that the transfer matrix $T_n$ defines the hopping rules of the $N$-step \L{}ukasiewich paths on a $n$-site segment. The size of the box, $n$, in the HP model is a dummy parameter for TASEP which just sets the size of the matrix in the algebra $DE-ED=D+E$ and to have an exact correspondence between the enumeration of heaps and the number of configurations in the steady-state TASEP, we should take a limit $n\to\infty$ keeping $N$ constant. 

The correspondence between TASEP and HP is set as follows: the horizontal step or the step up of the \L{}ukasiewich path corresponds to a particle in TASEP, while the step down corresponds to a hole. We emphasize that the TASEP/HP--correspondence is not bijective. From the configuration of \L{}ukasiewich path we can uniquely reconstruct the configuration of TASEP, however, the reverse is not true: the TASEP configuration does not fix the unique configuration of \L{}ukasiewich path.  

\subsubsection{ASEP, Young tableaux and weighted \L{}ukasiewicz paths}
\label{sect:5-3-1}

It is legitimate to ask whether there exists a correspondence between ASEP and $q$-heaps and whether one can enumerate configurations in the steady-state ASEP by counting topologically nonequivalent $q$-weighted heaps. The answer to this question is positive, and in the following we describe the corresponding construction.

In complete analogy with the TASEP/HP--correspondence, we introduce a representation of the ASEP algebra ($DE-qED=D+E$) that has the triangular structure of the transfer matrix \eq{eq:06}. This triangular structure encodes the Cartier-Foata enumeration of words and sets the topology of \L{}ukasiewich paths, while the absolute values of the matrix elements give the weights of the corresponding jumps.

First, we define the operator $E$ as in the TASEP algebra: 
\begin{equation}
\begin{array}{cc}
E\ket{n}=\ket{n+1}, \quad \text{i.e.}\quad 
E = \left(\begin{array}{ccccc}
0 & 0 & 0 & 0 & \cdots \\
1 & 0 & 0 & 0 & \\
0 & 1 & 0 & 0 & \\
0 & 0 & 1 & 0 & \\
\vdots & & & & \ddots
\end{array}\right). 
\end{array}
\end{equation}
Assume now that $D$ is upper triangular, i.e., $D_{ij} = 0$ if $j<i$. Therefore,
\begin{equation}
\begin{array}{cc}
 (DE)_{ij}  = \sum_{k\geq0}D_{i,k}E_{k,j} = \sum_{k\geq0}D_{i,k}\delta_{k,j+1} = D_{i,j+1}, \\
 \\
 (ED)_{ij} = \sum_{k\geq0}E_{i,k}D_{k,j} = \sum_{k\geq0}\delta_{i-1,k}D_{k,j} = D_{i-1,j} \quad \text{for} \quad i\geq1,\\
 \\
 (ED)_{0,j}=0
\end{array}
\end{equation}
We immediately obtain:
\begin{equation}
    D_{i,j+1}-qD_{i-1,j}=D_{i,j}+\delta_{i,j+1}.
\end{equation}
Note that $D_{0,j+1}-q(ED)_{0,j}=D_{0,j}+\delta_{0,j+1}$, therefore, for all $j\geq 0 \quad D_{0,j+1}=D_{0,j}$. Consequently, the whole top row $D$ is constant: $D_{0,j} = b, \quad (j=0,1,2,\dots)$. \\

Let $j=i+k, \quad k\geq 0 ; \quad a_{i,k}:=D_{i,i+k}$. Therefore, 
\begin{equation}\label{sys1}
    D_{i,j} = a_{i,k} = a_{i,k+1} - qa_{i-1,k+1}.
\end{equation}
On the other hand, if $j=i-1$
\begin{equation}
  a_{i,0}=1+qa_{i-1,0} 
\label{sys2}
\end{equation}
From the above derivation we get
\begin{equation}\label{sys3}
    a_{0,k}=b \quad k\geq 0
\end{equation}
The solution of three equations above \eqref{sys1}--\eqref{sys3} gives by the following ansatz:
\begin{equation}
    a_{i,k} = \sum_{r=0}^{i-1} q^{r} \binom{k+r}{r} + q^{i}b\binom{k+i}{i}, \quad i\geq0,k\geq0
\label{answ}
\end{equation}
It is easy to check that \eqref{answ} satisfies each of \eqref{sys1}--\eqref{sys3}.
Therefore,
\begin{equation}
    D_{i,j} = \begin{cases}
\disp        0, \quad j<i \\ \medskip 
\disp        \sum_{r=0}^{i-1} q^{r}\binom{j-i+r}{r} + q^{i}b\binom{j}{i}, \quad j\geq i
    \end{cases}
\end{equation}

Now, let us take into account the boundary conditions:
\begin{equation}
\begin{array}{l}
\disp    -\alpha \bra{W}E +\gamma \bra{W}D = - \bra{W}, \medskip \\
\disp    -\delta E \ket{V}+\beta D \ket{V} = \ket{V}.
\end{array}
\end{equation}
As before, we use the decomposition:
\begin{equation}
    \bra{W} = \sum_{n\geq0} w_n \bra{n}, \quad \ket{V} = \sum_{n\geq0} v_n \ket{n}.
\end{equation}
Therefore, the equation for the left boundary is as follows:
\begin{equation}
    \alpha w_{n+1} =  w_{n} + \gamma \sum_{m=0}^n w_m D_{m,n}, \quad n\geq 0. 
\end{equation}
For the right boundary,
\begin{equation}
    \beta \sum_{m=n}^{\infty} D_{n,m}v_m =  v_n+\delta v_{n-1}, \quad n\geq0, \quad v_{-1} = 0.
\end{equation}
Note that if we take a TASEP limit (i.e. $\gamma=\delta=0$), we will immediately get
\begin{equation}\label{bound}
    \bra{W} = \Big(1, \frac{1}{\alpha}, \Big(\frac{1}{\alpha}\Big)^2, \dots\Big),
    \quad \ket{V}=\ket{0} = (1,0,0,\dots).  
\end{equation}
as expected (see \eqref{tasep-boundaries}).\\

Finally, in matrix form:
\begin{equation}\label{ED}
E=\left(\begin{array}{ccccc}
0 & 0 & 0 & 0 & \cdots \\
1 & 0 & 0 & 0 & \\
0 & 1 & 0 & 0 & \\
0 & 0 & 1 & 0 & \\
\vdots & & & & \ddots
\end{array}\right), \quad D = \left(\begin{array}{ccccc}
    b & b & b & b & \cdots\\
    0 & 1+qb & 1+2qb & 1+3qb & \\
    0 & 0 & 1+q+q^2b & 1+2q+3q^2b & \\
    0 & 0 & 0 & 1+q+q^2+q^3 \\
    \vdots & & & & \ddots
\end{array}  \right). 
\end{equation}
For the left boundary, it is possible to 
write a vector explicitly:
\begin{equation}
\begin{array}{l}
\disp w_0 = 1,\\ \medskip
\disp w_1 = \frac{1+\gamma b}{\alpha}w_0,\\ \medskip     
\disp w_2=\frac{1+\gamma(1+qb)}{\alpha} w_1 + \frac{\gamma b}{\alpha} w_0,\\ \medskip    
\disp w_3 = \frac{1+\gamma(1+q+q^2b)}{\alpha}w_2 + \frac{\gamma(1+2qb)}{\alpha}w_1 + \frac{\gamma b }{\alpha}w_0.
\end{array}
\end{equation}
And so on. Therefore:
\begin{equation}
    \bra{W}=\Big(1, \frac{1+\gamma b}{\alpha}, \frac{(1+\gamma b)(1+\gamma(1+qb))}{\alpha^2} + \frac{\gamma b}{\alpha}, \dots\Big)
\end{equation}
Due to the infinite sum in the recursion equation at the right boundary, an explicit form cannot be obtained. However, if we require the two boundary rates to be trivial—specifically, $\gamma=\delta=0$—then the equations for the boundary vectors are uniquely resolved, giving \eqref{bound}. At the same time, under this assumption, we can set $b = \frac{1}{\beta}$ in \eqref{ED}. 

To summarize, we have constructed a representation of the quadratic ASEP algebra $DE-qED=D+E$ in which the matrix $D$ remains upper triangular while $E$ is the same shift operator as in the TASEP case. This preserves the Cartier--Foata ordering of words and, consequently, the underlying topology of the associated \L{}ukasiewicz paths. The deformation parameter $q$ does not modify the allowed moves of the paths; instead, it changes the weights assigned to these moves through the matrix elements of $D$. Thus, the ASEP steady state can be interpreted as an ensemble of reweighted \L{}ukasiewicz paths.

Since the triangular structure of the transfer matrix is precisely the one generated by the local free algebra graded $q'\mathcal{LF}_n$, this construction simultaneously establishes a correspondence between the steady-state ASEP and the combinatorics of heaps weighted with $q'$. The parameter $q'$, which grades the Cartier--Foata normal forms by the number of commutations of neighboring generators, coincides with the ASEP asymmetry parameter $q'=q$. Therefore, the stationary measure of ASEP can be viewed as the generating function of topologically distinct $q$-weighted heaps, or equivalently, weighted \L{}ukasiewicz paths, providing a direct combinatorial interpretation of the matrix-product solution of the ASEP steady state.

It should be emphasized that the matrix representation obtained here coincides with the matrices $D_1$ and $E_1$ introduced by Williams in \cite{williams2022combinatorics}. There, the representation was derived from a different perspective. Its purpose was to provide an explicit infinite-dimensional solution of the Matrix Ansatz for the stationary distribution of the finite ASEP and to establish positivity properties of the corresponding Markov chain. In particular, the matrices $D_1$ and $E_1$ were used to connect the stationary probabilities of the ASEP with the weighted generating functions of the staircase tableaux and related combinatorial objects. Thus, the asymmetry parameter of the exclusion process is naturally interpreted as the grading parameter of the locally-free algebra.

\subsubsection{ASEP, $q^{-1}$--Hermite polynomials and magnetic Dyck paths}
\label{sect:5-3-2}

In our analysis of ASEP, there exists a formal duality between the Touchard–Riordan Catalan regime, describing weighted crossing partitions, and the Carlitz–Riordan Catalan regime, describing magnetic Dyck paths. Specifically, the Touchard–Riordan matching corresponds to the partition function of standard continuous $q$-Hermite polynomials when the system is evaluated with the left and right boundary vectors set to $\bra{W^{(n)}}= \ket{V^{(n)}}^{\top}=(1, 0, \dots, 0, 0)$. By changing the boundary vectors to $\bra{W^{\prime (n)}}= \ket{V^{\prime(n)}}^{\top}= (0, 0, \dots, 0, 1)$ and applying the transformation $q \to 1/q$, the Carlitz–Riordan Catalans naturally emerge in the $1/q$--ASEP model. 

The Touchard--Riordan and Carlitz--Riordan models present two different families of generic Al-Salam–Chihara orthogonal polynomial systems. The parameter inversion $q\to 1/q$ transfers the underlying orthogonal measure from the continuous $q$--Hermite family to the discrete $q^{-1}$--Hermite family, implying that the asymptotics of generating functions at the spectral edge are ultimately governed by quotient of two $q$-Airy functions.

The generating function of ASEP with boundary vectors $\bra{W^{(n)}} \equiv \bra{e_n}=(0,0,\dots,0,1)$, $\ket{V^{(n)}} \equiv \ket{e_n} = (0,0,\dots,0,1)^{\top}$ is given by 
\begin{equation}\label{eq:ASEP-bridge}
	\Omega_n(s)=\sum_{N=0}^{\infty} \bra{W^{(n)}}T^N(n) \ket{V^{(n)}}s^{N} = -\frac{1-q}{s} \frac{H_{n-1}(x|q)}{H_n(x|q)}, \quad x=1-\frac{1-q}{2s}
\end{equation} 
where $H_n(x|q)$ are the standard continuous $q$-Hermite polynomials which satisfy the three-term recurrence relation:
\begin{equation}
\begin{array}{l}
H_{n+1}(x|q) = 2x H_n(x|q) - (1-q^n) H_{n-1}(x|q) 
\end{array}
\end{equation}
The substitution $q \to 1/q$ converts these recurrence relations into:
\begin{equation}
\begin{array}{l}
H_{n+1}(x|1/q) = 2x H_n(x|1/q) + q^{-n}(1-q^n) H_{n-1}(x|1/q) 
\end{array}
\end{equation}

To align these equations with the $q^{-1}$-Hermite polynomials (Ismail-Masson), we perform the complex rotation $x = iy$. This yields the exact identity mapping:
\begin{equation}
\begin{array}{l}
H_n(x|1/q) = i^n h_n(-ix|q) 
\end{array}
\end{equation}

Applying the transformation $q \to 1/q$ to the quotient \eqref{eq:ASEP-bridge} that defines the generating function, we get:
\begin{equation}
\Omega_n(s)\Big|_{q \to 1/q} = \frac{1-q}{qs} \frac{H_{n-1}(x_{\text{new}}|1/q)}{H_n(x_{\text{new}}|1/q)}, \quad x_{\text{new}} = 1 + \frac{1-q}{2qs}
\end{equation}
Using the complex mapping definitions, the ratio of polynomials can be rewritten as follows:
\begin{equation}
\frac{H_{n-1}(x_{\text{new}}|1/q)}{H_n(x_{\text{new}}|1/q)} = \frac{i^{n-1} h_{n-1}(-ix_{\text{new}}|q)}{i^n h_n(-ix_{\text{new}}|q)} = \frac{1}{i} \frac{h_{n-1}(-ix_{\text{new}}|q)}{h_n(-ix_{\text{new}}|q)}
\end{equation}

To study the behavior in the vicinity of the spectral boundary, we expand the Plancherel-Rotach scaling for $n \to \infty$ with $0 < q < 1$. We parameterize the coordinate using a local edge variable $t$:
\begin{equation}
-ix_{\text{new}} = \sinh \xi_n = \frac{t q^{-n/2} - t^{-1} q^{n/2}}{2}
\end{equation}
As in $n \to \infty$, the term $q^{-n/2}$ dominates:
\begin{equation}
-ix_{\text{new}} \approx \frac{t q^{-n/2}}{2}
\end{equation}
Equating this asymptotic behavior with our expression for $x_{\text{new}}$, we can determine the scaling of the parameter $s$ close to the boundary:
\begin{equation}
-i\left(1 + \frac{1-q}{2qs}\right) \approx \frac{t q^{-n/2}}{2} \implies qs \approx -i \frac{1-q}{t} q^{n/2}
\end{equation}

The $q$-Airy function (Ramanujan function) is defined as:
\begin{equation}
\text{Ai}_q(z) = \sum_{k=0}^\infty \frac{q^{k^2}}{(q;q)_k} (-z)^k
\end{equation}
At the spectral edge, the $q^{-1}$-Hermite polynomial scales as:
\begin{equation}\label{asymp}
h_n(\sinh \xi_n | q) \sim q^{-n^2/2} t^n \text{Ai}_q\left(\frac{1}{t^2}\right)
\end{equation}
For the polynomial $h_{n-1}(\sinh \xi_n | q)$, using the relation $\sinh \xi_n (t) = \sinh \xi_{n-1}(tq^{-\frac{1}{2}})$ and applying the asymptotic \eqref{asymp} to $h_{n-1}(\sinh \xi_n (t)|q) = h_{n-1}(\sinh \xi_{n-1}(tq^{-\frac{1}{2}})|q)$ with the variable $u = tq^{-\frac{1}{2}}$ we obtain the following.
\begin{equation*}
h_{n-1}(\sinh \xi_n | q) \sim q^{-n^2/2 + n/2} t^{n-1} \text{Ai}_q\left(\frac{q}{t^2}\right)
\end{equation*}

Taking the ratio of these two asymptotic limits, we get:
\begin{equation}
\frac{h_{n-1}(-ix_{\text{new}}|q)}{h_n(-ix_{\text{new}}|q)} \sim \frac{q^{-n^2/2 + n/2} t^{n-1} \text{Ai}_q\left(\frac{q}{t^2}\right)}{q^{-n^2/2} t^n \text{Ai}_q\left(\frac{1}{t^2}\right)} = \frac{q^{n/2}}{t} \frac{\text{Ai}_q\left(\frac{q}{t^2}\right)}{\text{Ai}_q\left(\frac{1}{t^2}\right)}
\end{equation}
Now we can combine the transformed prefactor with the asymptotic ratio:
\begin{equation}
\Omega_n(s)\Big|_{q \to 1/q} \sim \left( \frac{1-q}{qs} \right) \frac{1}{i} \left( \frac{q^{n/2}}{t} \frac{\text{Ai}_q\left(\frac{q}{t^2}\right)}{\text{Ai}_q\left(\frac{1}{t^2}\right)} \right) = \frac{(1-q) q^{n/2}}{i \, qs \, t} \frac{\text{Ai}_q\left(\frac{q}{t^2}\right)}{\text{Ai}_q\left(\frac{1}{t^2}\right)}
\end{equation}
Substituting the edge scaling relation ($qs \approx -i \frac{1-q}{t} q^{n/2}$) into the denominator provides a cancellation of the background.
\begin{equation}
\Omega_n(s)\Big|_{q \to 1/q} \sim \frac{(1-q) q^{n/2}}{i \left( -i \frac{1-q}{t} q^{n/2} \right) t} \frac{\text{Ai}_q\left(\frac{q}{t^2}\right)}{\text{Ai}_q\left(\frac{1}{t^2}\right)} = \frac{\text{Ai}_q\left(\frac{q}{t^2}\right)}{\text{Ai}_q\left(\frac{1}{t^2}\right)}
\end{equation}

From the Plancherel-Rotach scaling relation, we found the asymptotic equation for the spatial coordinate at the spectral boundary:
\begin{equation}
qs \approx -i \frac{1-q}{t} q^{n/2}
\end{equation}
The parameter $t$ can be explicitly expressed in terms of $s$
\begin{equation}
t = -i \frac{1-q}{qs} q^{n/2}
\end{equation}
The fugacity variable, $s$ used in the Carlitz $q$-Catalan and $q$-Airy (see \eqref{eq:genfunction}) gets mapped onto $s$ in the Plancherel-Rotach scaling as $s \to - \frac{q^3 s^2}{(1-q)^2 q^n}$. The behavior of a system considered near the edge directly mirrors the statistics of chord diagrams via the Touchard-Riordan $q$-Catalan polynomials -- see \eqref{C^T_q}. In summary, we arrive at the following conclusion.

\begin{itemize}
\item In the ``regular'' $0 < q < 1$ regime, for boundary vectors $\bra{W}= \ket{V}^{\top}=(1, 0, \dots, 0, 0)$ the $2N$-th moment of the continuous $q$-Hermite polynomials is given by expression:
\begin{equation}
m_{2N}(q) = 4^{-N}\int (2x)^{2N} d\mu_q(x) = 4^{-N} \bra{W} T_0^{2N} \ket{V}
\end{equation}
A walk of length $2N$ starting and ending at $0$ with up-step weights $1$ and down-step weights $\lambda_n = 1-q^n$ enumerates chord diagrams on the  $2N$ vertices. This can be expressed through the Touchard-Riordan polynomial $C_N^T(q)$, which counts perfect matches ordered by their crossing number $\text{cr}(M)$:
\be
m_{2N}(q) = \Big(\frac{1-q}{4}\Big)^N \sum_{\pi \in \mathcal{P}(2N)} q^{\text{cr}(\pi)} = \frac{1}{4^N} \sum_{k\geq 0} (-1)^{k} \Bigg(\binom{2N}{N-k}-\binom{2N}{N-k-1}\Bigg)q^{\binom{k+1}{2}}
\ee

\item Transformation $q \to 1/q$, together with the change of boundary vectors to $\bra{W'}= \ket{V'}^{\top}= (0, 0, \dots, 0, 1)$, parameterizes a regime where configurations with maximal crossing numbers are exponentially favored ($q^{-1} > 1$). Algebraically, this maps the system from the continuous family to the discrete $q^{-1}$-Hermite family (Ismail-Masson). The following intuitive interpretation behind the transform $q\to 1/q$ could be useful. Touchard’s original expression for chord crossings can be written as a continued fraction:
\begin{equation}
\sum_{N=0}^\infty C_N^T(q) z^N = \cfrac{1}{1 - \cfrac{[1]_q z}{1 - \cfrac{[2]_q z}{1 - \dots}}}
\end{equation}
It looks like, under Plancherel-Rotach edge scaling, this continued fraction collapses into the functional equation for the $q$-Airy quotient $C(z) = \frac{1}{1-zqC(q z)}$ where the poles of $C(z)$ are exactly the discrete roots of $\text{Ai}_q(z) = 0$. Thus, the poles $\text{Ai}_q(1/t^2) = 0$ represent the asymptotic condensation threshold, maximizing the number of chord intersections.
\end{itemize}

Let us complete this Section summarizing the main results related to the ASEP/HP correspondence.

1. {\it The $q'$-heaps correspond to the $q'$-Dyck paths and the $q'$-Catalan numbers}. We have shown that the combinatorial structure of $q'$-heaps described by the $q'$-graded locally-free algebra, introduced in this work as the natural extension of the "Mikado" enumeration of partially-commutative monoids, can be naturally encoded by $q'$-Dyck paths. As a consequence, the partition function of $q'$-heaps is given by $q'$-Catalan numbers.

2. {\it Cartier-Foata-Viennot duality maps the $q'$-heap partition function to a ratio of partition functions of weighted sum of 1D hard-core gases}. Using this duality, the partition function of heaps and of TASEP with arbitrary boundary conditions can be rewritten in terms of a ratio of partition functions of the Ising-type hard gas model. This construction allows for generalization to $q'$-weighted heaps.  

3. {\it The tridiagonalization of $q$-Gaussian random matrices provides a link between $q$-Gaussian ensembles and ASEP-like systems}. The fact that the expectation values of the coefficients of a $q$-Gaussian random matrix after tridiagonalization generate the transfer matrix associated with $q$-Hermite polynomials is not surprising. However, we have emphasized that the very same structure arises in connection with the $q$-deformed $(D,E)$ algebra of ASEP. This relation appears to be original and suggests a nontrivial link between deformed random matrix ensembles and the steady-state behavior of multiparticle ASEP-like systems.

4. {\it ASEP is mapped onto heaps via modified $q'$-weighted \L{}ukasiewicz paths}. A direct correspondence between ASEP configurations and $q'$-heaps is established using a modified version of Lukasiewicz paths. This construction maps ASEP steady states onto path combinatorics of $q'$-heaps. 

5. {\it Simultaneous transformation $q\to 1/q$ and change of boundary vectors $(1,0,...,0,0)\to (0,0,...,0,1)$ in ASEP model maps the statistics of chord diagram crossings described by combinatorics of Touchard-Riordan $q$-Catalan numbers onto the statistics of weighted partitions described by the $q'=1/q$-Carlitz-Riordan Catalan numbers for magnetic Dyck paths}. This duality seems to have deep connections with the construction of HOMFLY-type algebraic invariants for some series of torus knots. 

\section{Discussion}
\label{sect:disc}

Here, we highlight some directions for further research.

1. {\bf Beyond the steady-state solution for single-species TASEP}. The first class of questions concerns familiar generalizations of stochastic systems. We have discussed steady-state solutions for Markov process, which correspond to the ground states of associated XXZ spin chains (for ASEP) and Ising model (for TASEP). Moreover, we have focused on the simplest observable, namely, the resolvent of the transfer matrix. It is therefore natural to ask about the gravity interpretation of the non-stationary solutions of a Markov process, or, equivalently, of  excited states in the spin-chain representation. Consideration of observables other than the resolvent seems also desirable.

Next, it is natural to regard the $SU(N)$ versions of the stochastic system, for example, the multispecies TASEP model. The corresponding question has been studied using the Schur index in \cite{lewis2025schur} and presumably corresponds to higher-spin gravity in terms of the CS representation of $SL(N,C)$. The corresponding generalization of DSSYK is yet hidden; however, there is a natural framework in the stochastic approach, multispecies ASEP and TASEP. Another view on multispecies SSEP model has been proposed in \cite{popkov} where the family of higher critical exponents has been described. It would be interesting to connect these seemingly different approaches.

It is also useful to consider the higher-level stochastic systems associated with two well-known families: the Macdonald process and the higher-spin vertex model. From Schur-index counting and from the gravity model with EOW branes, we learn that these stochastic models correspond, respectively, to 5d SYM with adjoint matter and to 5d SQCD with $N_f=4$. Conversely, the SSEP model with symmetric hopping naively corresponds to a vanishing gravitational constant, yet it still exhibits nontrivial stochastic dynamics. Natural generalizations of EOW branes also involve representations of the full Sklyanin algebra, since the Askey--Wilson polynomials are related to a degeneration of this algebra \cite{gorsky1993degenerations}.

2. {\bf Inhomogeneous TASEP and the Goldfish model}. Another direction of research concerns developing of possible counterparts of the dualities in stochastic models \cite{gorsky2022dualities} within the DSSYK--gravity framework. To fully exploit these dualities, one can consider multispecies inhomogeneous TASEP \cite{arita2013matrix, kim2023schubert}, for which the steady state on a ring is expressed in terms of Schubert polynomials.

This setting involves two types of dualities. First, the analog of bispectral duality is realized through the interchange of inhomogeneities and eigenvalues associated with the boundary conditions (twists). Second, inhomogeneous TASEP is QQ-dual to the Goldfish model, with the number of sites in TASEP identified with the number of particles in the Goldfish model \cite{gorsky2022dualities}. The Goldfish model itself is bispectrally dual to the open relativistic Toda system, whose wave function governs the Whittaker process \cite{borodin2014macdonald}. For $N=2$, this reduces to the discrete Liouville Hamiltonian, which describes the DSSYK Hamiltonian in the chord basis, or, equivalently, the bulk gravity Hamiltonian.

The possible physical origin of the inhomogeneities in ASEP or TASEP may be related to some measurement protocols, such as those observed for SSEP \cite{russotto2026universality}. In such an interpretation, the inhomogeneities correspond to the local strengths of the measurements. In the dual Calogero--Ruijsenaars--Toda systems, the interpretation can be made more transparent since such systems emerge in many situations. We hope to discuss duality issues elsewhere.

3. {\bf ASEP/TASEP Markov Equation as a holographic RG}. We have conjectured that the Markov dynamics can be interpreted as the RG evolution of a non-local Wilson loop operator, with the XXZ spin-chain Markov operator for ASEP generating the dynamics along the radial holographic direction. However, the relationship between this holographic RG interpretation and the growth interpretation of ASEP/TASEP in terms of ballistic deposition remains unclear. In particular, the emergent growth geometry in TASEP, which is well visualized in the heap model, involves a complicated two-dimensional structure filled with clusters and percolating holes \cite{khanin2010ballistic}. The possible relation between this asymptotic geometry and the polymer picture of strongly coupled gravity deserves further investigation.

4. {\bf Transition amplitudes on the Krylov chain}. In \cite{gorsky2024kpz}, the KPZ scaling was identified in the formal limit $q=1$, where the Hermite polynomials provide the natural basis. To study the emergence of KPZ scaling, the autocorrelator was considered in the Krylov basis, and the Krylov complexity was analyzed both for an artificially truncated Krylov chain and for systems with a finite-dimensional Hilbert space. It was found that the transition from the Gaussian scaling regime to the KPZ scaling regime happens as a third-order phase transition in the presence of an artificial cutoff, while it becomes a smooth crossover in a finite-dimensional system. A similar analytical and numerical analysis for general $q$ will be presented elsewhere \cite{gorsky26}.

5. {\bf Closed TASEP and the Riemann surface}. In our work, we have focused on the open TASEP model, meanwhile TASEP on the ring also has a rich behavior \cite{ayyer2014inhomogeneous,baik2016tasep}. In this case, the auxiliary Riemann surface captures the information on interesting observables \cite{prolhac2020riemann}. It is worth considering the mapping of the closed TASEP to the dual gravity model.

6. {\bf Crossing graphs and relation to the $Q=0$ Potts model}. In the paper, we briefly discuss the crossing graph representation of the DSSYK partition function, which provides a link to the $Q=0$ Potts model coupled to gravity. The $Q=0$ Potts model on a graph is interesting in its own, as it is related to dilute random-loop and percolation models. Combinatorially, it counts spanning trees or forests on the graph. It belongs to the class of critical models and, when coupled to quantum gravity, provides a candidate for a topological theory with $c=0$. It would be interesting to investigate all aspects of its relation to DSSYK and develop the explicit realization of the summation over crossing graphs as some version of discrete quantum gravity.

7. {\bf Partition functions and knot/link invariants}. The chord diagrams can be interpreted as the braid graphs for singular knots, and the dual crossing graphs provide  the weight systems and the Vasiliev invariants of knots \cite{kazaryan2023weight}. Another connection to knot polynomials is provided by the relation between the Tutte polynomial and the Jones invariants of alternating knots \cite{welsh2000potts}. We plan to discuss these issues in a separate publication \cite{kovalenko}.

8. {\bf Towards different reincarnations of $q$}. The parameter $q$ has different interpretations depending on the problem under consideration: (i) it is the weight assigned to crossings of chord diagrams, (ii) it is related to the gravitational coupling,  (iii) it characterizes the asymmetry of hopping in ASEP, (iv) it is the equivariant parameter in the half-Schur index. Could one propose a transparent explanation for the equivalences of these seemingly different notions apart from their relation to representations of $U_q(SL(2))$ (as discussed in \cite{schouten})? 

One possibility is that this equivalence emerges from the first quantization of a spinful particle using path integrals in 2d and in 3d. Indeed, it was shown explicitly in \cite{kitaev2019statistical} that the JT partition function can be written in terms of a spinful particle on the hyperbolic disk. A similar representation in terms of a spinful observer in 3d was discussed in \cite{verlinde2024double}. Let us first recall that, in the first-quantized picture, the spin factor in 2d and in 3d \cite{polyakov1988fermi} can be represented by assigning a weight to self-intersections of particle trajectories. 

Secondly, recall  that a chord diagram involves two basic ingredients: the number of chords and the pattern of their intersections. If we interpret the chord diagram as a braid graph for singular knot , the number of chords can be associated with the number of  self-intersections of the singular knot, i.e. with number of loops attached to the particle trajectory, with the spin serving as fugacity for number of chords. The parameter $q$, in turn, weights the intersections between chords in the canonical ensemble. In the language of singular knots, these intersections might be interpreted as linking interactions between the attached loops: although the number of intersections is fixed, the loops can still interact with one another. Such an interaction could naturally be attributed to chirality coupled to gravity. This interpretation is at least qualitatively consistent with the role of $q$ as a measure of chiral asymmetry in ASEP hopping rates. In this way, the  seemingly different interpretations of $q$-crossing weights in chord diagrams, gravitational coupling, and hopping asymmetry in ASEP—may have a common origin in the spin and topology of the first-quantized particle path integral. This viewpoint does not contradict the interpretation of q as the twisted spin fugacity for counting 1/4 BPS states specified by the sum of the angular momentum $j_3$ along the symmetry axis of the half-three sphere and the $R$-charge in \cite{gaiotto2024sykschurdualitydoublescaled}.

\section{Conclusion}
\label{sect:conc}

In our study, we have discussed a stochastic--model interpretation of low-dimensional gravity and DSSYK within the KPZ universality class, extending the discussion in \cite{pelliconi, watanabe2025jt}. We have explored the role of the TASEP model, together with its ASEP and $q$-TASEP generalizations. In particular, we have argued that TASEP corresponds to the strong-coupling regime of 2d $\sin$-dilaton gravity.

We mainly focused on the generating function for the steady state of open TASEP and ASEP, which is the resolvent of the TASEP/ASEP transfer matrix. It coincides with the resolvents of the DSSYK transfer matrix and the Green function for the 2d gravity. We have used the chord basis in stochastic models and argued that the dual crossing graph representation of partition function in terms of cumulants provides a promising link to the $Q=0$ Potts model coupled to some version of discrete 2d gravity.

One of the main motivations of our study is the attempt of exploring the connection with stochastic growth models to justify low-dimensional space-time as an emerging phenomenon in the spirit of works \cite{swingle2012entanglement, vanraamsdonk2010commentsquantumgravityentanglement}. To this end, we focus on various interpretations of ASEP/TASEP as growth models. One approach exploits the fact that the space of states of the ASEP can be represented by the random walks on the 2d discrete space with the Liouville process in one direction and the Browmian motion in the other, while for the TASEP the Liouville process is substituted by the Brownian bridge. Also, using the mapping of ASEP/TASEP to the heap models based on Cartier-Foata duality, we are able to establish a connection with the structure of Lee-Yang zeros. The new generalization of the TASEP--heap duality to the ASEP--$q^{-1}$-heap duality has been proposed. 

The 2d JT gravity and its $\sin$--dilaton generalization can be described via the matrix model, hence there is a question where the stochasticity comes in from in the matrix model approach. We have argued using mathematically rigorous results that the careful double scaling limit of the matrix model involves the random term in the Airy operator. Exploring the connection between the Stochastic Heat equation and the Stochastic Airy equation, we suggest the origin of KPZ stochasticity at the gravity side in the matrix model language.

\begin{acknowledgments}

The authors thank A. Valov for valuable discussions at different stages of this work, K. Khanin for helpful comments, and A. Povolotsky for reading the manuscript and for important comments. A.G. thanks IHES for the hospitality and support.

\end{acknowledgments}

\begin{appendix}

\section{ASEP generating function via Cramer's method} 
\label{sect:app1}

Using Cramer's method for \eq{gen}, we can write: 
\be
Y(s,k) = \frac{\text{det}B(k)}{\text{det}(T(n)-\frac{1}{s} I)} = \frac{v_{n,k}}{u_n}
\label{W}
\ee
where $B(k)$ is the matrix obtained from $\left(T(n)-\frac{1}{s} I\right)$ by replacing the $k$-th column by the column $-\frac{1}{s}\ket{V^{(n)}}$. Define:
\be
T(n)-\frac{1}{s} I = \left(\begin{array}{cccccc}
         a & b_1 & 0 & 0& \dots & 0 \\
         b_1 & a & b_2& 0& \dots & 0 \\
         0 & b_2 & a & b_3& \dots & 0 \\
         0 & 0 & b_3 & a& \dots & 0\\
         \vdots & \vdots & \vdots& \vdots & \ddots & b_{n-1} \\
         0 & 0& 0& 0& \dots  b_{n-1} & a
\end{array}\right); \qquad \begin{cases} \disp  a = \frac{2}{1-q}-\frac1s, \medskip \\ \disp b_k = \frac{\sqrt{1-q^k}}{1-q} \end{cases}
\label{t-a-b}
\ee
The denominator, $u_n= \text{det}\left(T(n)-\frac{1}{s} I\right)$, satisfies the recursion:
\be
\begin{cases}
\disp u_0(s)=1 \medskip \\
\disp u_1(s)=\frac{2}{1-q}-\frac{1}{s} \medskip \\
\disp u_{n+1}(s) = \left(\frac{2}{1-q}-\frac{1}{s}\right)u_n(s) - \frac{1-q^n}{(1-q)^2}u_{n-1}(s),\quad n\geq 1
\end{cases}
\label{recursion}
\ee
Simplifying \eq{recursion}, we can express its solution in terms of $q$-Hermite polynomials. Define $\widehat{u}_n(s)=(1-q)^nu_n(s)$. This substitution deforms a recursion as
\be
\begin{cases}
\disp \widehat{u}_0(s)=1 \medskip \\
\disp \widehat{u}_1(s)=2-\frac{1-q}{s} \medskip \\
\disp \widehat{u}_{n+1}(s) = \left(2-\frac{1-q}{s}\right)\widehat{u}_n(s) - (1-q^n)\widehat{u}_{n-1}(s),\quad n\geq1
\end{cases}
\label{recursion2}
\ee
Denoting $x\equiv x(s)=1-\frac{1-q}{2s}$, rewrite \eq{recursion2} in the form
\be
\begin{cases}
\disp \widehat{u}_0(s)=1 \medskip \\
\disp \widehat{u}_1(s)=2x \medskip \\
\disp \widehat{u}_{n+1}(s)=2x\widehat{u}_{n}(s) - (1-q^n)\widehat{u}_{n-1}(s)
\end{cases}    
\ee
The last equation defines continuous $q$-Hermite polynomials. Thus, we arrive at the following expression for $u_n(s)$:
\be
u_{n}(s)= \frac{1}{(1-q)^n}H_n\left(1-\frac{1-q}{2s} \Bigg| q\right)
\ee

Recursion for $v_{n,k}(s)$ requires more involved computations. Note that the determinant is multilinear in columns, so here we will drop the $-\frac1s$ factor for a while and then simply multiply the answer by $-\frac1s$. In what follows, we use the shortened notation (see \eq{braket})
\be
\ket{V^{(n)}} = \left(v_0, v_1, \dots, v_{n-1}\right)^{\top}; \qquad v_0=1, \quad v_m = \frac{Q_m}{\sqrt{(q;q)_m}} \quad (m=0,...,n-1) 
\label{ket2}
\ee
where $(q;q)_m=\prod_{j=1}^m (1-q^j)$. Expanding $v_{n,k}=\text{det}B(k)$ by the column $k$, we get: 
\be
v_{n,k}(s)=\sum_{j=1}^n (-1)^{j+k} x_{j-1} M_{j,k}(s)
\label{expand-v}
\ee
In \eq{expand-v} $M_{j,k}(s) = \det A_n(s)[j|k]$, where $A_n[j|k]=\left(T(n)-\frac{1}{s}I\right)$ and by $A_n(s)[j|k]$ we understand the matrix $A_n$ with deleted row $j$ and column $k$. Depending on the relative positions of $j$ and $k$ the following options are available:
\begin{enumerate}
\item For $j=k$ one has
\be
A_n(s)[j|k] = \left(\begin{array}{cc}
A_{k-1}^{(0)} & 0 \\
0 & A_{n-k}^{(k)}
\end{array}\right)
\ee
where
\be
    A_m^{(r)}(s)=\left(\begin{array}{ccccc}
        a &b_{r+1} & 0 & \dots & 0   \\
        b_{r+1} & a & b_{r+2} & \dots & 0\\
        0 & b_{r+2} & a & \dots & 0\\
        \vdots & \vdots & \vdots & \ddots& b_{r+m-1}\\
        0 & 0 & 0 & \dots b_{r+m-1}& a       
    \end{array}\right)
\ee
and the coefficients $a$ and $b_k$ are defined in \eq{t-a-b}. Denoting $u_m^{(r)}=\det A_m^{(r)}$, the contribution of this summand can be written as $v_{k-1}\, u_{k-1}(s)\, u_{n-k}^{(k)}(s)$.

\item For $j<k$ we group the row/column indices into three consecutive domains: $\{1,\dots,j-1\},\, \{j,\dots,k-1\}, \, \{k+1, \dots, n\}$. Then $A_n(s)[j|k]$ takes the upper block--diagonal form:
\be
    A_n(s)[j|k] = \left(\begin{array}{ccc}
        A_{j-1}^{(0)} & * & 0 \\
        0 & R_{j,k} & * \\
        0 & 0 & A_{n-k}^{(k)}
    \end{array}\right)
\ee
Therefore, $M_{j,k} = u_{j-1}(s)\, \det R_{j,k}\, u_{n-k}^{(k)}(s)$, where the matrix
\be
R_{j,k}=\left(\begin{array}{cccccc}
        b_j & a & b_{j+1} & 0 & \dots & 0   \\
        0 & b_{j+1} & a & b_{j+2} & \dots & 0\\
        0 & 0 & b_{j+2} & a & \dots & 0\\
        \vdots & \vdots & \vdots & \vdots& \ddots& a\\
        0 & 0 & 0 & 0& \dots  & b_{k-1}    \end{array}\right) 
\ee
is the upper diagonal, i.e., $\det R_{j,k} = \prod_{r=1}^{k-1}b_r$. Thus, the total contribution of this summand is $(-1)^{j+k}v_{j-1}u_{j-1}(s)\left(\prod_{r=j}^{k-1}b_r\right) u_{n-k}^{(k)}(s)$.

\item For $j>k$ we group the row/column indices into three consecutive domains: $\{1,\dots,k-1\},\, \{k+1,\dots,j\},\, \{j+1, \dots, n\}$. Then $A_n(s)[j|k]$ has the following lower block--diagonal form:
\be
    A_n(s)[j|k] = \left(\begin{array}{ccc}
        A_{k-1}^{(0)} & 0 & 0 \\
        * & L_{k,j} & 0 \\
        0 & * & A_{n-j}^{(j)}
    \end{array}\right)
\ee
Therefore, $M_{j,k} = u_{k-1}(s)\, \det L_{k,j}\, u_{n-j}^{(j)}(s)$ where $L_{k,j}$ is the lower diagonal of $\text{diag}(L_{k,j})=b_k,b_{k+1}, \dots,b_{j-1}$, i.e., $\det R_{j,k} = \prod_{r=k}^{j-1}b_r$. The total contribution of this summand is $(-1)^{j+k}v_{j-1}u_{k-1}(s)\left(\prod_{r=k}^{j-1}b_r\right) u_{n-j}^{(j)}(s)$.
\end{enumerate}

Collecting items 1--3 together and taking into account that $\prod_{r=k}^{k-1} b_r=1$, we get
\be
v_{n,k}(s) = u_{n-k}^{(k)}(s)\sum_{j=1}^k (-1)^{j+k}v_{j-1}u_{j-1}(s) \prod_{r=j}^{k-1}b_r+u_{k-1}(s)\sum_{j=k+1}^n (-1)^{j+k}v_{j-1}u_{n-j}^{(j)}(s) \prod_{r=k}^{j-1}b_r
\label{v(s)}
\ee
Substituting $b_{r} = \frac{\sqrt{1-q^r}}{1-q}$ and the components of $\ket{V^{(n)}}$ in \eq{v(s)}, we obtain the following:
\begin{multline}
v_{n,k}(s) = \sqrt{(q;q)_{k-1}} u_{n-k}^{(k)}(s)\sum_{j=1}^k (-1)^{j+k}\frac{Q_{j-1}}{(q;q)_{j-1}}\frac{u_{j-1}(s)}{(1-q)^{k-j}}+ \\ \frac{u_{k-1}(s)}{\sqrt{(q;q)_{k-1}}}\sum_{j=k+1}^n (-1)^{j+k} Q_{j-1} \frac{u_{n-j}^{(j)}(s)}{(1-q)^{j-k}} 
\label{x}
\end{multline}
Now we can rewrite \eq{x} in terms of $q$--Hermite polynomials. Defining "shifted" $q$--Hermite polynomials,
\be
H_{n+1}^{(r)}(x|q) = 2xH_{n}^{(r)}(x|q) - (1-q^{r+n})H_{n-1}^{(r)}(x|q),
\ee
we can write the functions $u_n^{(r)}$ in terms of $H_{n}^{(r)}(x|q)$:
\be
u_n^{(r)}=\frac{1}{(1-q)^n}H_n^{(r)}\left(1-\frac{1-q}{2s}\Bigg|q\right)
\ee
Simplifying all expressions, we arrive at the following equation
\begin{multline}
 v_{n,k}(s) = \frac{1}{(1-q)^{n-1}}\bigg[\sqrt{(q;q)_{k-1}} H_{n-k}^{(k)}(x|q)\sum_{j=1}^k (-1)^{j+k}\frac{Q_{j-1}}{(q;q)_{j-1}} H_{j-1}(x|q)+ \\ \frac{H_{k-1}(x|q)}{\sqrt{(q;q)_{k-1}}}\sum_{j=k+1}^n (-1)^{j+k} Q_{j-1} H_{n-j}^{(j)}(x|q) \bigg]   
\label{v_nk}
\end{multline}
where $x=1-\frac{1-q}{2s}$. Thanks to \eq{PQ} the coefficients $Q_j$ could also be expressed through continuous $q$-Hermite polynomials:
\be
Q_n = (i\sqrt{D})^n H_n \left(\frac{C}{2 i \sqrt{D}}\Bigg|q \right)
\ee
with $C$ and $D$ defined in \eq{CD}. Therefore, we may rewrite \eq{v_nk} as 
\begin{multline}
v_{n,k}(s) = \frac{-1}{s(1-q)^{n-1}}\Big[\sqrt{(q;q)_{k-1}}  H_{n-k}^{(k)}(x|q)\sum_{j=1}^{k} (-1)^{j+k}\frac{(i\sqrt{D})^{j-1}}{(q;q)_{j-1}}H_{j-1}(y|q) H_{j-1}(x|q)+\\ \frac{H_{k-1}(x|q)}{\sqrt{(q;q)_{k-1}}}\sum_{j=k+1}^{n} (-1)^{j+k} (i\sqrt{D})^{j+1} H_{j+1}(y|q) H_{n-j}^{(j)}(x|q) \Big] 
\end{multline}
where $y = \frac{C}{2i\sqrt{D}}$, and the $-\frac1s$ factor appears in the definition of $v_{n,k}$.

Finally, the ASEP partition function $Y(s,k)$ in \eq{W} takes the following explicit form
\begin{multline}
Y(s,k) = \frac{v_{n,k}(s)}{u_n(s)} = \frac{(1-q)^n}{H_n(x|q)}v_{n,k}(s) = \\ \frac{(-1)^{k+1}}{s}\frac{1-q}{H_n(x|q)} \bigg[\sqrt{(q;q)_{k-1}} H_{n-k}^{(k)}(x|q)\sum_{j=1}^{k} \frac{(-i\sqrt{D})^{j-1}}{(q;q)_{j-1}}H_{j-1}(y|q) H_{j-1}(x|q)+ \\ \frac{H_{k-1}(x|q)}{\sqrt{(q;q)_{k-1}}}\sum_{j=k+1}^{n} (-i\sqrt{D})^{j-1} H_{j-1}(y|q) H_{n-j}^{(j)}(x|q) \bigg] 
\label{ASEP}
\end{multline}

Define the ASEP generating function, $\Omega_n^{ASEP}(s)$ 
\be
\Omega_n^{ASEP}(s) =\sum_{N=0}^{\infty} \bra{W^{(n)}}T^N(n)\ket{V^{(n)}}s^N 
\label{gen-ASEP}
\ee
Writing \eq{gen-ASEP} in a form $\Omega_n^{ASEP}(s) = \braket{W^{(n)}| \textbf{Y}(s) } =\sum_{k=1}^nw_{k-1}Y(s,k)$, where $ \textbf{Y}(s)$ and $Y(s,k)$ are given by \eq{W'} and \eq{ASEP} respectively; $w_{k-1}$ is the $(k-1)$ component of the vector $\bra{W^{(n)}}$, given by \eq{braket}, we get
\begin{multline}
\Omega_n^{ASEP}(s) = \frac{(-1)}{s} \frac{1-q}{H_n(x|q)} \sum_{k=1}^{n}(-1)^k \frac{(i\sqrt{B})^{k-1}H_{k-1}(a|q)}{\sqrt{(q;q)_{k-1}}} \times \\ \bigg[\sqrt{(q;q)_{k-1}}  H_{n-k}^{(k)}(x|q)\sum_{j=1}^{k} \frac{(-i\sqrt{D})^{j-1}}{(q;q)_{j-1}}H_{j-1}(y|q) H_{j-1}(x|q)+ \\ \frac{H_{k-1}(x|q)}{\sqrt{(q;q)_{k-1}}}\sum_{j=k+1}^{n} (-i\sqrt{D})^{j-1} H_{j-1}(y|q) H_{n-j}^{(j)}(x|q) \bigg] 
\end{multline}
Recall for completeness that $x=1-\frac{1-q}{2s}, \quad y = \frac{C}{2i\sqrt{D}}, \quad a=\frac{A}{2i\sqrt{B}}$. Defining $m=\min(k,j), \quad M=\max{(k,j)}$ and combining two regions into one sum, we obtain the following expression for the generating function of ASEP in the most general form:
\begin{multline}
\Omega_n^{ASEP}(s) = \frac{1-q}{s} \frac{1}{H_n(x|q)} \times \\ \sum_{k=1}^n \sum_{j=1}^n (-1)^{k+j+1} \frac{(i\sqrt{{B}})^{k-1}(i\sqrt{D})^{j-1}}{(q;q)_{m-1}} H_{k-1}(a|q)H_{j-1}(y|q)H_{m-1}(x|q)H^{(M)}_{n-M}(x|q)
\label{Omega ASEP}
\end{multline}

\section{TASEP limit: $q\to 0$}
\label{sect:app2}

Our starting point is equation \eq{gen-TASEP0}
\be
\Omega_n^{(TASEP)}(s) = \frac{1}{sU_n(x)} \sum_{k=1}^n\sum_{j=1}^n(-1)^{k+j+1}A^{k-1}C^{j-1}U_{m-1}(x)U_{n-M}(x)
\label{gen-TASEP}
\ee
The convergence radii of $s$ and $\eta$ are: $0<s<\frac{1}{4}, \quad 0<\eta<\frac{1}{2} \Rightarrow |\rho| <1$. The kernel 
\be
K_n(k,j) = \frac{U_{m-1}(x)U_{n-M}(x)}{U_n(x)}.
\ee
can be written in terms of $\rho$ as
\be
K_n(k,j) = \frac{\rho^{|k-j|+1} - \rho^{k+j+1}-\rho^{2n-k-j+3}+\rho^{2n-|k-j|+3}}{(1-\rho^2)(1-\rho^{2n+2})}
\ee
Thus, the generating function for TASEP becomes
\be
\Omega_n^{TASEP}(s) = -\frac{1}{s}\sum_{k=1}^n\sum_{j=1}^n (-A)^{k-1}(-C)^{j-1} \frac{\rho^{|k-j|+1} - \rho^{k+j+1}-\rho^{2n-k-j+3}+\rho^{2n-|k-j|+3}}{(1-\rho^2)(1-\rho^{2n+2})}.
\ee
Since the denominator of $\Omega_n^{TASEP}(s)$ does not depend on $k,j$, it can be taken out of the sum. With the help of four functions
\be
\begin{array}{ll}   
\disp S_1(n)=\sum_{k,j=1}^n (-A)^{k-1}(-C)^{j-1}\rho^{|k-j|+1}, & \disp S_3(n)=\sum_{k,j=1}^n (-A)^{k-1}(-C)^{j-1}\rho^{2n-k-j+3} \medskip \\
\disp S_2(n)=\sum_{k,j=1}^n (-A)^{k-1}(-C)^{j-1}\rho^{k+j+1}, & \disp S_4(n)=\sum_{k,j=1}^n (-A)^{k-1}(-C)^{j-1}\rho^{2n-|k-j|+3} 
\end{array} 
\ee
we rewrite $\Omega_n^{TASEP}(s)$ as follows 
\be
\Omega_n^{TASEP}(s) = -\frac{S_1(n) + S_2(n) + S_3(n) + S_4 (n)}{s(1-\rho^2)(1-\rho^{2n+2})}
\ee
Thus, the problem is reduced to calculating limits of $S_i(n), \quad i=1,...,4$. Let us begin with $S_2$, which is separable:
\be
S_2(n) = \sum_{k=1}^n (-A)^{k-1}\rho^k \sum_{j=1}^n (-C)^{j-1}\rho^{j+1} = \rho^3 \left(\sum_{k=1}^n (-A\rho)^{k-1}\sum_{j=1}^n (-C\rho)^{j-1}\right)
\ee
For $S_1$ we split the summation regions into two parts: for $k\geq j$ we take $k=j+d$; for $j>d$ we take $j=k+d$:
\be
S_1(n) = \rho \sum_{j=1}^n(AC)^{j-1}\sum_{d=0}^{n-j} (-A\rho)^{d} + \rho \sum_{k=1}^{n-1}(AC)^{k-1}\sum_{d=1}^{n-k}(-C\rho)^d
\ee
Assuming $|AC|<1, \quad |A\rho|<1, \quad |C\rho|<1$, we take the limit $n \rightarrow \infty$ and get 
\be
\left[\begin{array}{l}
\disp S_1^{\infty} =\frac{\rho(1-AC\rho^2)}{(1+AC)(1+A\rho)(1+C\rho)} \medskip \\
\disp S_2^{\infty} = \frac{\rho^3}{(1+A\rho)(1+C\rho)}
\end{array}\right] \Rightarrow S_1^{\infty}-S_2^{\infty} = \frac{\rho(1-\rho^2)}{(1-AC)(1+A\rho)(1+C\rho)}.
\ee
Similarly, 
\be
\left[\begin{array}{ll}
\disp S_3(n) = & \disp \rho^{2n+1}\sum_{k=1}^{n}\left(\frac{-A}{\rho}\right)^{k-1} \sum_{j=1}^{n}\left(\frac{-C}{\rho}\right)^{j-1}  \medskip \\
\disp S_4(n) = & \disp \rho^{2n+1}\sum_{j=1}^{n}(AC)^{j-1} \sum_{d=0}^{n-j}\left(\frac{-A}{\rho}\right)^{d} + \medskip \\  
& \disp \rho^{2n+1}\sum_{k=1}^{n-1}(AC)^{k-1} \sum_{d=1}^{n-k}\left(\frac{-C}{\rho}\right)^{d} 
\end{array}
\right] 
\Rightarrow 
\begin{array}{l} S_3^{\infty} = 0 \medskip \\ S_4^{\infty} = 0 \end{array}
\ee
So, finally, in the $n \rightarrow \infty$ limit the generating function $\Omega_n^{TASEP}(s)$ takes the well-known TASEP form
\be
\Omega^{TASEP}(s) = -\frac{\rho}{s}\frac{1}{(1-AC)(1+A\rho)(1+C\rho)}= \frac{\alpha \beta}{\alpha + \beta -1} \frac{1}{(1-\frac{\eta}{\alpha})(1-\frac{\eta}{\beta})}
\ee
where $ \eta = \frac{1-\sqrt{1-4s}}{2}$.

\section{Phase transitions in ASEP with full set of rates}
\label{sect:app3}

The singularities of $\Omega(s)$ could originate from: (i) the denominator $1-s\lambda(\theta)$ of the resolvent in \eq{spec}; (ii) poles of $\widehat{X}(\theta)$; (iii) poles of $\widehat{Y}(\theta)$. Consider all of these possibilities separately.

1. The resolvent ("bulk") singularity is located at $\lambda(\theta) = \frac{2(1-\cos(\theta))}{1-q}$ where $0\leq\theta\leq \pi$. In this case $\Omega(s)$ is a power series in $s$ around $s=0$, so the dominant asymptotics are controlled by the smallest $s=\frac{1}{\lambda(\theta)}$. Therefore the desired singularity is 
\be
s_{0} = \frac{1-q}{4}
\label{s_0}
\ee    

2. To investigate the singularity coming from the left boundary factor $\widehat{W}(\theta)$ recall that
\be
w_n = \frac{P_n}{\sqrt{(q;q)_n}}, \quad P_n = (i\sqrt{B})^n H_n\left(\frac{A}{2i \sqrt{B}}\Bigg|q\right)  
\ee
Thus we get
\be
\widehat{W}(\theta) = \sum_{n\geq0} \frac{(-i\sqrt{B})^n}{(q;q)_n} H_n\left(\frac{A}{2i \sqrt{B}}\Bigg|q\right)  H_n(\cos(\theta)|q)
\label{hatW}    
\ee
The analysis of \eq{hatW} becomes straightforward with the help of the Poisson kernel for the continuous $q$-Hermite polynomials:
\be
\sum_{n\geq 0 } \frac{t^n}{(q;q)_n}H_{n}(x|q)H_n(y|q) = \frac{(t^2;q)_{\infty}}{(te^{i(\theta+\phi)}, t e^{i(\theta-\phi)}, t e^{-i(\theta-\phi)}, te^{-i(\theta+\phi)}; q)_{\infty}}
\label{Pois}
\ee
where $x=\cos(\theta),\quad y=\cos(\phi)$. Setting in \eq{hatW} $t = -i\sqrt{B}, \cos(\phi) = \frac{A}{2i \sqrt{B}}$ we immediately have: $r_{L_{\pm}}=te^{\pm i \phi}$. Thus, $r_{L_{+}}+ r_{L_{-}} = 2t \cos(\phi) = -A, \quad r_{L_{+}}r_{L_{-}} = t^2 = -B$, and hence, $r_{L_{\pm}}$ are roots of an equation $r^2+Ar-B=0$. In terms of $r_{L_{\pm}}$ using \eq{Pois}, we can rewrite \eq{hatW} as follows:
\be
\widehat{W}(\theta) = \frac{(-B;q)_{\infty}}{(r_{L_{+}}e^{i\theta},r_{L_{-}}e^{i\theta},r_{L_{+}}e^{-i\theta},r_{L_{-}}e^{-i\theta};q)_{\infty}}
\ee
Therefore, factors that produce the pole are exactly $(r_{L_{\pm}}e^{i\theta};q)_{\infty}$ and $(r_{L_{\pm}}e^{-i\theta};q)_{\infty}$. A $q$-Pochhammer symbol $(z;q)_{\infty}$ vanishes at $z=q^{-m}, \quad m = 0,1,2,\dots$, so the nearest singularity comes from $m=0$: $r_{L_{\pm}}e^{i\theta}=1$ and/or $r_{L_{\pm}}e^{-i\theta}=1$.
The solutions of these two equations are indistinguishable, since $x = \cos(\theta) = \frac12 (r_{L_{\pm}} + r_{L_{\pm}}^{-1})$. 

As in the case of the "bulk" singularity, the dominant asymptotics are controlled by the smallest $s=\frac{1}{\lambda(\theta)}$. Since $s=\frac{1}{\lambda(\theta)} = \frac{1-q}{2(1-x)}$, the candidates for the dominant singularity are
\be
s_L = \frac{1-q}{2-r_{L_{\pm}}-r_{L_{\pm}}^{-1}}
\ee
Thus, we should keep the one of $r_L\in \{r_{L_{+}}, r_{L_{-}}\}$, for which $|s(r)|$ is smaller. Recalling equation $r^2+Ar-B=0$ for $r=r_L$, we conclude that the desired $r_L$ is 
\be
r_L = \frac{-A-\sqrt{{A^2}+4B}}{2}.
\ee
Substituting $A$ and $B$, we finally obtain
\be
s_L = \frac{1-q}{2-(r_L+r_L^{-1})}, \quad r_L = \frac{-(1-q-\alpha+\gamma) - \sqrt{(1-q-\alpha+\gamma)^2+4\alpha \gamma}}{2\alpha}.
\label{s_L}
\ee

3. For the right boundary factor $\widehat{V}(\theta)$ we proceed exactly the same way as for $\widehat{W}(\theta)$. So, we skip the derivation and provide the final answer for the dominant singularity
\be
s_{R}=\frac{1-q}{2-(r_R+r_R^{-1})}, \quad r_R = \frac{-(1-q-\beta+\delta) - \sqrt{(1-q-\beta+\delta)^2+4\beta \delta}}{2\beta}.
\label{s_R}
\ee

By definition, the stationary flow is $J = \lim_{N \rightarrow  \infty} \frac{Z_{N-1}}{Z_N} = \min\{s_0,s_L,s_R\}$. The phase behavior of ASEP is determined by $J$: at $J = s_0$ the system is in the maximal flow phase; at $J = s_L$ -- in the low-density phase; at $J = s_R$ -- in the high-density phase. Transitions between phases occur when the corresponding singularities merge. There are the following options:
\begin{enumerate}
\item $s_L = s_0$, i.e., $\alpha = \frac{1-q}{2}+\gamma$. Since $q\in [0,1]$, this condition implies:
\be
0\leq (\alpha-\gamma)\leq \frac{1}{2}
\ee
\item $s_R = s_0$, i.e., $\beta = \frac{1-q}{2}+\delta$ gives us
\be
0 \leq (\beta- \delta) \leq \frac{1}{2}
\ee
\item $s_R = s_L$ suggests different options, for example $\alpha = \beta, \quad \gamma = \delta$ (this condition is sufficient but not necessary). There are two branches of $s_L=s_R$: $r_R=r_L$, and $r_R = \frac{1}{r_L}$, defined by \eq{s_L} and \eq{s_R}. Begin with the case $r_L=r_R$. By definition, the root of this equality should satisfy both equations
\be
\begin{array}{l}
\alpha r^2 + (1-q- \alpha + \gamma )r - \gamma = 0 \medskip \\
\beta r^2 + (1-q- \beta + \delta )r - \delta = 0
\end{array}
\label{trans-r}
\ee
which gives
\be
r=\frac{\delta-\gamma}{\alpha-\beta}
\label{trans-2}
\ee
Substituting this value of $r$ back into \eq{trans-2} we obtain the first-branch condition for a transition point
\be
\begin{array}{l}
\disp \alpha \neq \beta, \quad \gamma \neq \delta, \quad \frac{\delta-\gamma}{\alpha - \beta} < 0 \medskip \\ 
\disp 0\leq 1-\alpha + \gamma + \alpha \frac{\delta-\gamma}{\alpha - \beta} - \gamma \frac{\alpha-\beta}{\delta - \gamma}\leq 1
\end{array}
\ee
where $\frac{\delta-\gamma}{\alpha - \beta}< 0$ follows from the fact that we chose the negative root. Another option is: $\alpha = \beta, \quad \gamma = \delta$ (see below). The second root is obtained in a similar way. The simultaneous solution of quadratic equations: 
\be
\begin{array}{l}
\disp \alpha r^2 + (1-q- \alpha + \gamma )r - \gamma = 0, \medskip \\
\disp \beta \frac{1}{r^2} + (1-q- \beta + \delta )\frac1r - \delta = 0   
\end{array}
\label{trans-22}
\ee
gives 
\be
r = - \frac{\beta + \gamma}{\alpha + \delta}
\ee
which is always negative. Substituting this value of $r$ back into \eq{trans-22} gives the second-branch condition:
\be
0\leq 1- \alpha + \gamma - \alpha \frac{\beta + \gamma}{\alpha + \delta} + \gamma \frac{\alpha+ \delta}{\beta + \gamma} \leq 1
\ee
\end{enumerate}

\section{Wave function as correlator of height functions in KPZ-like models. Polymer picture}
\label{sect:app4}

Integrable probabilities are interested in life in a realm of
the non-equilibrium statistical physics and belong to the KPZ
universality class involving the different versions of
stochastic growth problems.
The continuum KPZ stochastic equation
for the height function $h(x,t)$  in $(1+1)$ reads as
\be
\frac{\partial h(x,t)}{\partial t}=\frac{1}{2} \frac{\partial^2 h(x,t)}{\partial^2 x} +
\frac{1}{2} \Big(\frac{\partial h(x,t)}{\partial x}\Big)^2 + \eta(x,t)\,,
\ee
where $\eta(x,t)$ is the Gaussian noise.
Upon the Cole-Hopf transform $Z(x,t)=\exp( h(x,t))$ the KPZ equation
becomes mapped into a stochastic equation of Fokker-Planck type
\be\label{rr1}
\frac{\partial Z}{\partial t} = \frac{1}{2}\frac{\partial^2 Z}{\partial x^2} + \eta Z\,.
\ee
The function $Z(x,t)$ can be treated
as a partition function of a polymer of length $t$ in a random environment.

The evaluation of $Z(x,t)$ can be performed via the replica trick
when instead of the single polymer $n$ polymers are considered, hence
one introduces the joint probability
averaged over disorder:
$$\Psi(x_1,\dots x_n,t)= \left<Z(x_1,t)\dots Z(x_n,t)\right>,
$$
It turns out (see \cite{spohn2012stochastic, corwin2012kardar} for reviews)  that the
joint probability obeys the non-stationary equation for the
attractive Lieb-Lineger many-body system with $n$-particles
:
\be\label{ll}
\displaystyle{
\frac{d\Psi}{dt}= H_n \Psi\,, \qquad H_n= -
\frac{1}{2} \sum^n \frac{\partial^2}{\partial x_j^2} -\frac{1}{2} \sum_{i\neq j}^n \delta(x_i -x_j)\,.
%\label{ll}
}
\ee
This example explains one of the roles that an integrable many-body system
plays in the context of non-equilibrium dynamics. The Bethe ansatz equations
now come into the game and yield the spectrum of eigenvalues of (\ref{ll}).
Similarly, one can consider the polymer in the semi-discrete (1+1) space-time
when a space coordinate is discretized. If one introduces
$N$ directed polymers in semi-discrete space-time in the random medium  starting from one point, the joint polymer partition function
obeys the non-stationary Schrodinger equation for the N-particle
open Toda chain \cite{oconnell2012directed}.

The trigonometric Calogero-Moser system appears in a polymer setting similar to the following \cite{cardy2003stochastic}.
Consider the multiple radial SLE stochastic process which describes the growth
of N interacting polymers in a random environment on the disk.
 Polymers grow from points $x_i(t)$ in the boundary circle
and obey the process
\be\label{rr3}
dx_i(t)= \sum_{i\neq j} \cot \frac{x_i(t) -x_j(t)}{2} dt 
- \sqrt{k} \sum_{k\neq i} dB_k\,,
\ee
where $B_k$ are independent  Brownian motions starting from the origin.

The joint probability up to conjugation coincides
with the wave function of the Calogero-Moser Hamiltonian at energy $E$:
\be\label{rr4}
H_{CM}\Psi(x_1,\dots x_n)= E \Psi(x_1,\dots x_n),
\ee

\be\label{rr5}
H_{CM}= -\frac{k}{2}\sum \frac{\partial^2}{\partial x_i^2} + \frac{2-k}{2k} \sum_{i\neq j}
\frac{1}{\sin^2(x_i -x_j)/2} - \frac{N(N-1)}{2k}.
\ee
where  $E=\frac{k^2 -16}{32}$.

\end{appendix}

\bibliographystyle{unsrt}
\bibliography{ref.bib}

% --- BibTeX ---
%\begin{thebibliography}{9}
%\bibitem{Gorsky2010qtCatalanHOMFLY}
%E.~Gorsky,
%\newblock \emph{$q,t$-Catalan numbers and knot homology},
%\newblock arXiv:1003.0916.

%\bibitem{BirmanLin1993KnotPolynomialsVassiliev}
%J.~Birman and X.-S.~Lin,
%\newblock \emph{Knot polynomials and Vassiliev's invariants},
%\newblock Inventiones Mathematicae \textbf{111} (1993), 225--270.
%\end{thebibliography} 

\end{document}